\documentclass[prd, preprint,amscd,amsmath,amssymb,verbatim,showpacs,nofootinbib]{revtex4}
\usepackage{graphicx} % for figures
\usepackage{epstopdf} % so can use EPS or PDF figures
\usepackage{tensor}
\usepackage{amsfonts}
\newcommand{\be}{\begin{equation}}
\newcommand{\ee}{\end{equation}}
\newcommand{\bea}{\begin{eqnarray}}
\newcommand{\eea}{\end{eqnarray}}
\newcommand{\beaa}{\begin{eqnarray*}} 
\newcommand{\eeaa}{\end{eqnarray*}} 

\newcommand{\na}{\nabla}
\newcommand{\dis}{\displaystyle} 
\newcommand{\bsube}{\begin{subequations}}
\newcommand{\esube}{\end{subequations}}
\begin{document}

% %%%%%%%%%%%%%%%%%%%%%%%%%%%%%%
\title{Self force on an accelerated particle}
\author{Thomas M. Linz}
\email{tmlinz@uwm.edu}
\affiliation{ Leonard E Parker Center for Gravitation, Cosmology and Astrophysics, Department of Physics,
University of Wisconsin--Milwaukee, P.O. Box 413, Milwaukee, Wisconsin 53201,
USA}

\author{John L. Friedman}
\email{friedman@uwm.edu}
\affiliation{ Leonard E Parker Center for Gravitation, Cosmology and Astrophysics, Department of Physics,
University of Wisconsin--Milwaukee, P.O. Box 413, Milwaukee, Wisconsin 53201,
USA}

\author{Alan G. Wiseman}
\email{agw@gravity.phys.uwm.edu}
\affiliation{ Leonard E Parker Center for Gravitation, Cosmology and Astrophysics, Department of Physics,
University of Wisconsin--Milwaukee, P.O. Box 413, Milwaukee, Wisconsin 53201,
USA}
%\author{Thomas Linz, John L. Friedman, Alan G. Wiseman}
%\affiliation{Department of Physics, University of Wisconsin Milwaukee, 53201, USA}

\date{December 11, 2013}

\begin{abstract}

We calculate the singular field of an accelerated point particle
(scalar charge, electric charge or small gravitating mass) moving
on an accelerated (non-geodesic) trajectory in a generic
background spacetime.  Using a mode-sum regularization scheme,
we obtain explicit expressions for the self-force regularization
parameters.  In the electromagnetic and gravitational case, we
use a Lorenz gauge.  This work extends the work of Barack and Ori
\cite{BO1} who demonstrated that the regularization parameters
for a point particle in geodesic motion in a Schwarzschild
spacetime can be described solely by the leading and subleading
terms in the mode-sum (commonly known as the $A$ and $B$ terms)
and that all terms of higher order in $\ell$ vanish upon
summation (later they  showed the same behavior for geodesic motion in 
Kerr \cite{bo03}, \cite{barack09}).  We demonstrate that these 
properties are universal to point particles moving through any 
smooth spacetime along arbitrary (accelerated) trajectories.  
Our renormalization scheme is based on, but not identical to, the Quinn-Wald axioms.
As we develop our approach, we review and extend work showing that 
that different definitions of the singular field used in the literature are
equivalent to our approach.  Because our approach does  not
assume geodesic motion of the perturbing particle, we are able
use our mode-sum formalism to explicitly recover a well-known
result: The self-force on static scalar charges near a
Schwarzschild black hole vanishes.

\end{abstract}

\maketitle % generate title, including abstract

\section{Introduction}
\label{Introduction}

The likelihood that the gravitational radiation from stellar-size
black holes spiraling in to a supermassive galactic black will be
observable has spurred work on the extreme-mass-ratio inspiral
problem and on the analogous problem of a point particle with
scalar or electric charge moving in a curved spacetime.  
The trajectory of a small body moving in a curved spacetime 
deviates from the geodesic motion of a point particle 
at linear order in the charge or mass. Derivations of the 
trajectory use matched asymptotic expansions and a point-particle 
limit of a family of finite bodies whose charge, mass and radius 
simultaneously shrink to zero; they show that one can describe 
this first-order trajectory by a renormalized self-force.% 
\footnote{The most recent and rigorous of these are by Gralla, Harte, and 
Wald \cite{gw08,ghw09} (with a formal proof for an electromagnetic charge), 
by Pound \cite{pound10}, and by Poisson, Pound and Vega \cite{Poisson}, 
who also review the history and give a comprehensive bibliography.}   
% 
% Derivations of the 
% deviation from geodesic motion based on matched asymptotic 
% expansions show 
% Quinn \cite{Quinn} briefly discusses the self-force on a scalar 
% charge in this framework, and a rigorous derivation for an electromagnetic 
% charge is given by Gralla, Harte and Wald \cite{ghw09}. Gralla and Wald \cite{gw08} give a corresponding argument for a massive particle, and Poisson et al. \cite{Poisson} review derivations using matched asymptotic expansions.  
In an initial {\em MiSaTaQuWa} form developed for the metric perturbation of a
massive particle by Mino, Sasaki and Tanaka \cite{MST} and by
Quinn and Wald \cite{QW} -- and for scalar fields by Quinn
\cite{Quinn} -- one uses the Hadamard expansion of the retarded 
Green's function to identify a singular part of the field and 
a corresponding singular part $f^{sing}_\alpha$ of the expression 
$f^{ret}_\alpha$ for the particle's self-force written in terms 
of the retarded field.    

To subtract the singular from the retarded expression for the self-force, 
one first regulates each.  This can in principle be done, as 
described in the MiSaTaQuWa papers, by evaluating them at a 
finite proper distance $\rho$ from the trajectory and then taking a limit 
of their difference as $\rho\rightarrow 0$.  Nearly all explicit 
calculations of the self-force on particles moving in
Kerr or Schwarzschild geometries, however, have used a mode-sum form 
of the renormalization introduced by Barack and Ori \cite{BO00,BO1}, with 
early development and first applications by them, Mino, Nakano, and Sasaki and 
Burko \cite{BMNOS,BarackBurko,Burko}.  Its subsequent development and applications 
by a number of researchers are reviewed 
by Barack \cite{barack09} and Poisson et al. \cite{Poisson}.  In mode-sum regularization, 
one writes $f^{sing}_\alpha$ and $f^{ret}_\alpha$ as sums of angular 
harmonics on a sphere through the particle, replacing the short-distance 
cutoff $\rho$ by a cutoff $\ell_{max}$ in the $\ell,m$  harmonics, 
and expressing the renormalized self-force as a limit 
$\dis\lim_{\ell_{\rm max}\rightarrow \infty} 
	\left(\sum_{\ell=0}^{\ell_{max}}f^{ret,\ell}_\alpha - \sum_{\ell=0}^{\ell_{max}}f^{sing,\ell}_\alpha\right)$ or, equivalently, as the convergent sum 
$\dis{ \sum_{\ell=0}^{\infty}}(f^{ret,\ell}_\alpha - f^{sing,\ell}_\alpha)$.

In this paper we generalize the results of Barack and Ori \cite{BO1, bo03, barack09} 
for geodesic motion in a Schwarzschild or Kerr background to accelerated 
trajectories in generic spacetimes, checking the method by computing the 
known (vanishing) self-force on a scalar particle at rest in a Schwarzschild 
background \cite{Wiseman}.
A striking freature of mode-sum regularization is that only the leading 
and subleading terms in $\ell^{-1}$ give nonzero contributions to the singular 
expression for the self-force:  For a point particle with scalar charge, and, 
in a Lorenz gauge, for an electric charge and a point mass, 
$f^{sing,\ell}_\alpha$ has the form 
\be
    f^{sing,\ell\pm}_\alpha = \pm A_\alpha L + B_\alpha,
\label{eq:fsingAB}\ee 
where $L=\ell+1/2$, $A_\alpha$ and $B_\alpha$ are independent of $\ell$,
and the sign $\pm$ refers to a limit of the direction-dependent 
singular expression taken as one approaches the sphere through the particle 
from the outside or inside.\footnote{ Although the mode-sum expansion of the 
retarded field has terms of higher powers in $\ell^{-1}$, we show that 
the sum of these terms vanishes. In computing the self-force, however, 
one improves convergence by explicitly including higher-order 
terms in the singular field.}   

The form of Eq.~(\ref{eq:fsingAB}) describes the large $\ell$ behavior 
of $f^{sing}_\alpha$ and depends only on the short distance behavior 
of the retarded field.    
The values of the vectors $A_\alpha$ and $B_\alpha$ also depend on 
the choice of spherical coordinates in a neighborhood of the particle.  
For the electromagnetic and gravitational cases, the definition of 
$f^{ret}_\alpha$ involves, in addition to the retarded field (the vector 
potential ${\texttt A}_\alpha^{ret}$ or the perturbed gravitational field 
$h^{ret}_{\alpha\beta}$), the background metric $g_{\alpha\beta}$ and 
the particle's 4-velocity $u^\alpha$, each evaluated at the particle's 
position $z(0)$.
An expression for $f^{ret}_\alpha$ in the neighborhood of the particle  
then depends on how one extends $g_{\alpha\beta}[z(0)] $ and $u^\alpha[z(0)]$ 
to the neighborhood.  Different smooth extensions leave the form 
of Eq.~(\ref{eq:fsingAB}) and the value of the vector $A_\alpha$ unchanged,
but they change the value of the subleading term, $B_\alpha$. 

To obtain Eq.~(\ref{eq:fsingAB}) and the values of $A_\alpha$ and $B_\alpha$, 
one arbitrarily extends $f^{sing}_\alpha$ from a normal neighborhood 
of the particle to a thick sphere spanned by spherical coordinates, 
but the values of $A_\alpha$ and $B_\alpha$ do not depend on that extension. 
Although the coefficient of any finite angular harmonic does depend on the extension, 
two different extensions that are smooth outside the normal 
neighborhood differ only by a smooth function; coefficients of 
the angular harmonics of a smooth function on the sphere fall 
off faster than any power of $\ell$. 
     
In the next section, we also show the equivalence of several renormalization methods:  
renormalization using mode-sum regularization, using regularization 
based on a short-distance cutoff, and related versions of renormalization 
that involve an angle average over a sphere of radius $\rho$ about the particle.  
We use the Detweiler-Whiting form of the singular field \cite{DW} as 
a smooth, locally defined solution to simplify the analysis. 
This section is primarily a review, but the discussions of equivalence do not 
appear in one place in the literature and are often restricted to geodesic motion.  

Generalizing the mode-sum formalism to accelerated trajectories allows one 
to use the method to find the self-force on charged, massive particles in 
spacetimes with background scalar or electromagnetic fields.  
The extension also allows one to consider flat-space limits of 
bound orbits in which the size of the orbit remains finite. 
It simplifies the derivation of the self-force on a static particle, 
as we show in in Sec.~\ref{RadialForce}.  For the gravitational case, a consistent 
treatment of an accelerated particle of mass $m$ must 
include the matter responsible for the acceleration. Because 
our computation of the gravitational self-force includes only 
the contribution from the particle, it does not by itself 
describe the correction to the orbit of an accelerated mass at 
order $m$.  The gravitational
and electromagnetic contributions to the singular part of the 
self-force, however, have a natural form as a sum of the contributions 
we obtain here and contributions arising at subleading order 
from terms that couple the two fields \cite{lfw14,pz14}.  
We mention in Sec.~\ref{gravitational} our work in progress 
on the self-force on a charged, massive particle in an 
electrovac spacetime. The study may be useful in deciding whether 
including the self-force prevents one from overcharging a 
near-extremal black hole \cite{hubeny,ist,bck,zvph}.  

\subsection*{Our renormalization scheme and connection with
previous results}

Our renormalization scheme is a slight modification of the method
used by Quinn \cite{Quinn}.  However, in the process of carrying
out the calculation, we can show the equivalence of our technique
with that of Detweiler and Whiting \cite{DW}.  

Quinn uses the DeWitt-Brehme \cite{DewittBrehme} formalism to
expand the gradient of the retarded field in close proximity to
the scalar charge. This gives a highly divergent Coulomb field
({\it i.e.} a term of  $O(\epsilon^{-2})$, where $\epsilon$ is a
measure of the distance from the particle) and another divergent
term proportional to the acceleration  which scales as
$O(\epsilon^{-1})$. Some non-divergent terms proportional to the
square of the acceleration and the curvature tensors appear at
$O(\epsilon^{0})$.  Also appearing at $O(\epsilon^{0})$  are the
terms that actually produce the self-force: the $\dddot x$ 
terms which give the standard Abraham-Lorentz-Dirac force, as
well as an integral over the past history ({\it i.e.} the tail
term).  Quinn notes that the first terms (the Coulomb, the
acceleration and acceleration-squared terms) are the same as
those that are present in flat spacetime for the half-advanced plus
half-retarded field.  

Quinn's  second axiom asserts that a charge moving in flat spacetime
with a half-advanced plus half-retarded field will experience no
self-force.  His first axiom allows him to subtract the
flat spacetime terms from the curved spacetime field without modifying
the resulting force. The subtraction removes all singular terms
because the singular terms are the same in flat spacetime as they are in
curved space.  The subtraction leaves only the curvature terms
and those terms actually responsible for the force. His second
axiom also requires an angle average over a sphere near the
charge. This angle average removes the  curvature terms and
leaves only those terms which contribute to the self-force.

This is an elegant procedure; however the final step of angle-averaging over a small ball centered on the charge is difficult
to carry out when the fields are computed using a mode-sum with
coordinates centered on the black hole. 

To get past this technical difficulty, we modify Quinn's
prescription.  When we do the subtraction, we also subtract the
curvature terms. This eliminates all the divergent terms and it
eliminates the need for angle-averaging. The only terms that
remain are those that contribute to the self-force.

Detweiler and Whiting arrive at a similar prescription, but through
a somewhat different argument. The connection between the methods
can be explained as follows: The quantity they subtract from the
full retarded field gradient is  the gradient of their
$\Psi^{S}(x)$ ({\it i.e.} Eq.~(17) of \cite{DW}).  By expanding 
this vector and keeping terms of $O(\epsilon^{-2}$),
$O(\epsilon^{-1}$) and  $O(\epsilon^{0}$), one arrives at
precisely the same quantity we subtract from the retarded field
gradient: the Coulomb field, the acceleration terms, the
acceleration-squared terms, and the Riemann curvature terms.

Though many of the steps in this paper are closely related to
other work, the over-riding intent of this paper is to give a
self-contained discussion of the self-force problem: We start
with the field equations and end with the regularization
parameters (the $A$ and $B$ terms) needed for a mode-sum
calculation of the finite part of the self-force for an
accelerated point particle moving in an arbitrary spacetime.

%As mentioned above, our renormalization scheme is patterned after
%Quinn \cite{Quinn}. We will also demonstrate
%the previously noted equivalence of this method with
%the scheme presented in Detweiler and Whiting \cite{DW} and
%Poisson, Pound and Vega \cite{Poisson}.
%However the emphasis in these other papers  is conceptual:
%we need to develop a formalism that will give us practical
%way to compute the renormalizaion parameters.

Why do we take the time to derive the field
gradient expressions anew? Why don't we shorten the paper and
simply take the expressions from Quinn \cite{Quinn}  (Or
Poisson, Pound and Vega \cite{Poisson} see their
Eqs.~(17.39) and (18.20))? There are two (related) reasons why we
do not do this. First,  we wish to exploit the techniques of
Barack and Ori to derive the regularization parameters;
therefore we choose to express the field gradient with a
notation and quasi-Cartesian coordinate system that is
essentially identical to theirs, allowing us to use their
methods directly.  At the very least, using previously obtained
results for the field gradients would require a messy
notation or coordinate conversion in order to write them in a
usable form for our calculations, or, alternatively, recasting the techniques of
Barack and Ori into a Quinn-like notation.  Simply rederiving the field gradients 
in our notation is the most straight-forward and illuminating path.

A second reason for not simply lifting the expressions for the
field gradients from the previous literature is that such
expressions really do not contain sufficient information for our
calculation. In Quinn, for example, the expression of the field
gradient is given at a field point that lies along a spatial
geodesic orthogonal to the world line; Quinn's
expression relates the field gradient on this orthogonal spatial
slice to the state of motion (acceleration and jerk) of the
particle on this same spatial slice.  In order to compute the
mode-sums, we need the field gradient at points on a
slice of constant coordinate time.  Our gradient is expressed in
term of of the state of motion of the particle at that same
instant of coordinate time.  Thus the results given in \cite{Quinn}
and  \cite{Poisson} don't give us this necessary starting point
for our calculation.\footnote{Of course one could, for example, use 
Quinn's expression to obtain  the field gradient on a slice of
constant time.  Each point on our slice of constant time
corresponds to a point on a particular slice that is orthogonal
to the world line.  However, this slice would intersect the world
line at a different coordinate time.  One then has to
relate the motion of the particle as it passes through this
orthogonal slice to the motion at the desired moment in
coordinate time.}

We begin in Sect. \ref{SelfForceSingField} with the singular
field of a scalar charge, using the results of Quinn \cite{Quinn}
and Detweiler and Whiting \cite{DW}, to express $f^{sing}_\alpha$
explicitly in Riemann normal coordinates (RNCs). In Sect.
\ref{ModeSumReg}, we show that $f^{sing,\ell}_\alpha$ has the
form given in Eq. (\ref{eq:fsingAB}), and we find explicit expressions for the
regularization parameters $A_\alpha$ and $B_\alpha$. Next, in
Sect. \ref{HigherSpins}, we generalize this analysis to electric
charges and masses in a Lorenz gauge.  In each case, the mode-sum
analysis is closely patterned on the work of Barack and Ori
\cite{BO2}.  Finally, in Sect. \ref{RadialForce}, we check the
validity of the method by recovering the result of Wiseman
\cite{Wiseman} for a static charge in a Schwarzschild spacetime. 

In this paper we will always use the conventions of Misner, Thorne and Wheeler,
\cite{MTW} unless otherwise noted.

\section{Self Force and the Singular Field.}
\label{SelfForceSingField}

We consider a point particle (a scalar charge $q$, electric
charge $e$, or mass $m$) traveling on an accelerated trajectory
$z(\tau)$ in a smooth spacetime $(M,g_{\alpha\beta})$, where 
$\tau$ is proper time.  We will use RNCs about a point  
$\tau =0$ of the trajectory and, for mode-sum regularization, 
spherical coordinates $(t,r,\theta,\phi)$ associated with an 
arbitrary smooth Cartesian chart.  For brevity of notation,
we assume that $t=0$ at $\tau=0$.    

We consider a field point $x$ that lies on the spacelike $t=0$ slice 
and is in a convex normal neighborhood $C$ of $z(0)$. 
We denote by $\epsilon$ the geodesic distance from the particle's 
position at an arbitrary time $\tau$ to $x$; that is, $\epsilon$ is the length of 
the unique geodesic from an arbitrary point on the trajectory $z(\tau)$. After performing the various derivative operations to get to an expression for the singular field and singular force, we will choose our arbitrary point to be $z(0)$. In particular, when we reach the mode-sum section, we will consider $\epsilon$ to be the length of the unique geodesic from  $z(0)$ to $x$ (see Figure 1).
\begin{figure}[h!]
\begin{center}
\includegraphics[width=3in]{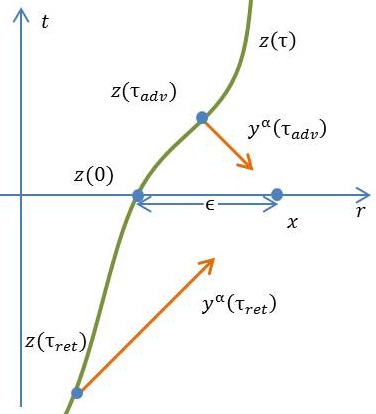}
\caption{The particle trajectory $z(\tau)$.  Two null vectors  
$y^\alpha(\tau_{ret})$ and $y_{\alpha}(\tau_{adv})$ are tangent 
to future- and past-directed null geodesics from points along 
the trajectory to a field point $x$. A geodesic from $z(0)$ 
to $x$ has length $\epsilon$.}
\label{Fig1}
\end{center}
\end{figure}

In this section we focus on the self-force on a scalar
charge, but we will subsequently use the same mathematical framework 
to calculate the self-force for both electric charges and
point masses.  We begin with the MiSaTaQuWa renormalization of 
a scalar field, using the axiomatic 
renormalization description given by Quinn \cite{Quinn}. We 
use the Hadamard expansion of the advanced and retarded fields 
to show that this description is equivalent to subtracting a singular 
part $f^{sing}_\alpha$ of the retarded expression $f^{ret}_\alpha$ 
for the self-force, regularized by a short-distance cutoff.  
From the explicit form of $f^{sing}_{\alpha}$, we show that, for 
geodesic motion, the renormalization is equivalent 
to an angle average of $f^{ret}_\alpha$. We end 
the section with a check of the equivalence of 
the singular field seen in a Hadamard expansion 
and Detweiler-Whiting form.  

% The following section concludes our 
% demonstration 
% of equivalent renormalizations with an analysis of the mode-sum 
% prescription. 

\subsection{Self-Force on a Scalar Charge}

The self-force correction to the equation of motion of a particle with 
charge $q$ is $O(q^2)$.  
For matter with scalar charge density $\rho$, 
a zero-rest-mass scalar field $\Phi$ satisfies
\begin{equation}
\nabla_{\alpha}\nabla^{\alpha}\Phi=-4\pi\rho,
\label{ScalarFieldEquation}
\end{equation}
where, for a point-particle, the density $\rho$ is given by the distribution 
\be
  \rho(x) = q \int d\tau \delta^4(x,z(\tau)),
\label{rho_pp}\ee
and the formal expression for the self-force, 
\begin{equation}
f^\alpha=q\nabla^\alpha\Phi,
\label{ScalarForceDefinition}
\end{equation}
diverges at the position of the particle.  
%%BEGIN HERE 10/11

Quinn's description of the renormalized self-force is stated as two axioms, 
based on the Quinn-Wald axioms \cite{QW} for higher spins. 
To present the axioms, we introduce 
 a set of RNCs whose origin is the position $z(0)$ of 
the particle at proper time $\tau=0$.  Coordinates and components 
in these coordinates will be denoted by hatted indices. The coordinates of 
the particle's position are then $z^{\hat{\alpha}}(\tau)$, with 
$z^{\hat{\alpha}}(0)=0$, and a field point $x$ has coordinates 
$x^{\hat{\alpha}}$.  

Quinn allows $\Phi$ to be any scalar field satisfying Eq.~(\ref{ScalarFieldEquation}), 
with $\rho$ given by Eq.~(\ref{rho_pp}).  The field $\Phi$ can then
then include an arbitrary homogeneous background scalar field as well 
as the retarded field of the particle (if, in fact, a retarded field is 
well defined on the spacetime); the renormalized force, which we denote by 
$f^{ren}_{Q\,\alpha}$, then includes the force due to the homogeneous background 
field as well as the self-force.  
We initially state the axioms with this generality and then restrict 
consideration to the retarded field.  With this restriction, the force 
$f_{Q\,\alpha}^{ren}$ is the self-force $f^{ren}_\alpha$.  
  
Quinn's first axiom, the comparison axiom can be stated as follows:
\begin{verse}

\hspace{1mm} 
\phantom{xx}Consider two point particles in two possibly different spacetimes,
each particle having scalar charge $q$. Suppose that, at points $z(0)$ and 
$\tilde z(0)$ on their respective trajectories, the magnitude of the particles' 
4-accelerations coincide.  We may then choose RNC systems  
about $z(0)$ and about $\tilde z(0)$ for which the components of the 
4-velocities and 4-accelerations coincide:  
\be
u^{\hat\alpha} = \tilde u^{\hat\alpha}, \qquad a^{\hat\alpha} = \tilde a^{\hat\alpha}.
\ee  
Let $\Phi$ and $\tilde{\Phi}$ be the retarded scalar fields
of the particles.  With the RNC systems used to identify 
neighborhoods of $z(0)$ and $\tilde z(0)$, the 
difference between the renormalized scalar forces, $f_{Q\,\alpha}^{ren}$ and
$\tilde{f}_{Q\,\alpha}^{ren}$ is given by the limit as $r\rightarrow0$
of the gradients of the fields averaged over a sphere of geodesic
distance $r$ about $z(0)$.%
\footnote{With $S_r$ the set of points that lie a geodesic distance $r$ 
from $z(0)$ along a geodesic perpendicular to the trajectory, the average 
of a function $f$ is  
$\dis\langle f\rangle_r := |S_r|^{-1}\int_{S_r} f dS$, where $|S_r|$ is the 
area of $S_r$.}
\end{verse}
\begin{equation}
f_Q^{ren,\hat\alpha}-\tilde{f}_Q^{ren,\hat\alpha}=q\lim_{r\rightarrow0}\langle\nabla^{\hat\alpha}\Phi-\nabla^{\hat\alpha}\tilde\Phi
\rangle_r .
\label{Quinn'sComparisonAxiom}
\end{equation}

Quinn's second axiom simply states that the renormalized scalar force vanishes for the 
half-advanced + half-retarded field of a uniformly accelerated charge in 
flat space:
\begin{verse}

\hspace{5mm} If, for a uniformly accelerated scalar charge in flat space,  
$\tilde{\Phi}=\frac12(\tilde{\Phi}^{ret}+\tilde{\Phi}^{adv})$, 
then $\tilde{f}_Q^{ren,\alpha}=0$. 
\end{verse} 

To define the self-force, we assume that the spacetime of the field $\Phi$ is 
globally hyperbolic so that retarded and advanced fields are well defined, 
and we set $\Phi=\Phi^{ret}$.  With this restriction, the axioms imply that the 
self-force is given by 
\be
f^{ren,\hat\alpha} =q\lim_{r\rightarrow0}\langle\nabla^{\hat\alpha}\Phi^{ret}-\nabla^{\hat\alpha}\tilde\Phi
\rangle_r.
\label{RegularizedField}\ee
As in this equation, we will henceforth use the RNC identification 
of normal neighborhoods of the flat and curved spacetimes to regard 
$\tilde\Phi$ as a field on $C$.

% The renormalized field, $\Phi^{ren}$ is smooth and regular at the
% particle and throughout the entire spacetime, which necessarily
% requires that the singular field has the same singular structure
% s the retarded solution.

% Therefore, to calculate $f^{ren,\alpha}$ we need three things:
% first we need to find the retarded and advanced solutions to the
% field equations, second we need to define an appropriate singular
% field, and third we must find a consistent manner to perform the
% subtraction of these two divergent quantities (for example if we
% used Quinn's flat spacetime half advanced plus half retarded
% solution, we would need to include an angle average in our
% subtraction). 

\subsection{Short-Distance Expansion of the Retarded and Advanced Fields}

While Quinn's description makes no explicit mention of a singular part of 
the scalar force, we will see that the result of the subtraction and angle 
average is equivalent to identifying and subtracting such a singular 
part: a vector field $f^{sing}_\alpha$ defined in a neighborhood of the 
particle.  To do so, we partly follow Quinn, who in turn uses the DeWitt-Brehme 
formalism, to find the singular behavior of the advanced and retarded 
fields near the particle.  Quinn obtains an expression for the field 
at points $x$ lying on a geodesic on a spacelike hypersurface orthogonal 
to the trajectory at a point $z(0)$ in terms of the particle's 
velocity, acceleration and jerk at $z(0)$.  For the mode-sum regularization 
of Sect.~\ref{ModeSumReg}, however, we need an expression for the field on a 
hypersurface that is not orthogonal to the trajectory.  For this reason, we 
obtain in this section the field at an arbitrary nearby point $x$ and 
use it to identify $f^{sing}_\alpha$.

% Eq.~(\ref{RegularizedField}) for $f^{ren,\alpha}$ involves a 
% limit as $x\rightarrow z(0)$, or, using Fig.~1, as
% $\epsilon\rightarrow 0$. 
To rewrite $f^{ren}_\alpha$ of Eq.~(\ref{RegularizedField}) as a 
difference of the form 
\be
  f^{ren}_\alpha = \lim_{x\rightarrow z(0)}[ f^{ret}_\alpha(x) - f^{sing}_\alpha(x)],
\label{eq:fsing}\ee 
we restrict $x$ to lie in the normal neighborhood $C$ of $z(0)$. Because 
we have chosen $C$ to be convex, any points $x,x' \in C$ are joined by a geodesic, 
and the advanced and retarded Green's functions 
have the Hadamard forms,
\begin{equation}
G^{adv/ret}(x,x')=\Theta_{\pm}(x,x')\left[U(x,x')\delta(\sigma(x,x'))-V(x,x')\theta(-\sigma(x,x'))\right].
\label{Hadamardpm}
\end{equation}
Here $V(x,x')$ and $U(x,x')$ are smooth bi-scalar functions of
$x$ and $x'$, and $\sigma(x,x')$ is half the squared length 
of the geodesic connecting $x$ and $x'$. The function
$\Theta_{\pm}(x,x')$ is unity when $x'$ is in the causal future
(past) of the event $x$ for the advanced (retarded) Green's
function, and vanishes otherwise. 

The retarded solution to Eqs.~(\ref{ScalarFieldEquation}) and (\ref{rho_pp}) is
given by 
\bea
\Phi^{ret}&=& q\int d^4 x'\sqrt{-g}\int d \tau G^{ret}(x,x')\delta^4(x',z(\tau)),\nonumber
\\
&=&q\int d\tau G^{ret}(x,z(\tau)).
\label{PhiGreen2}
\eea

Following Quinn, we break the domain of integration into two regions: the
part of the trajectory in the normal neighborhood $C$ (where the Hadamard 
form of the Green's function is valid) and the rest of the trajectory. We
choose the event $x$ to be close enough to the 
trajectory that the events $z(\tau_{adv})$ and
$z(\tau_{ret})$ both lie in $C$, and we denote by $T_\pm$ the proper 
times at which the
trajectory intersects the boundary $\partial C$: 
The past and future intersection points are respectively    
$z(T_-)$ and $z(T_+)$.  The retarded field then takes the form
\beaa
\Phi^{ret}&=&q\int_{T_-}^{T_+}
\Theta_-(x,z(\tau))\left[U(x,z(\tau))\delta(\sigma(x,z(\tau)))-V(x,z(\tau))\theta(-\sigma(x,z(\tau)))\right]d\tau
+q\int_{-\infty}^{T_-}G^{ret}d\tau,
\\
&=&q\int_{T_-}^{0}\left[U\delta(\sigma)-V\theta(-\sigma)\right]d\tau
+q\int_{-\infty}^{T_-}G^{ret}d\tau,
\label{PhiGreen3}
\eeaa
where we have suppressed the arguments of the biscalar functions.
Noting that in the interval $[T_-,0]$, $\sigma(x,z(\tau))=0$ only at $\tau=\tau_{ret}$, and using $d\tau=\dot{\sigma}^{-1}d\sigma$, with $(\dot{\phantom{x}})= d/d\tau$, 
we have,
\begin{equation}
\Phi^{ret}(x)=q\left(\frac{U(x,z(\tau))}{\dot{\sigma}}\right)_{ret}-q\int_{T_-}^{\tau_{ret}(x)}V(x,z(\tau))d\tau+q\int_{-\infty}^{T_-}G^{ret}(x,z(\tau))d\tau.
\label{PhiGreen4}
\end{equation}

The gradient of $\Phi$ with respect to $x$ is given by 
\begin{equation}
\nabla_{\alpha}\Phi^{ret}=q\nabla_{\alpha}\left[\left(\frac{U}{\dot{\sigma}}\right)_{ret}\right]+q V\nabla_{\alpha}\tau_{ret} -q\int_{T_-}^{\tau_{ret}}\nabla_\alpha Vd\tau+q\int_{-\infty}^{T_-}\nabla_{\alpha} G^{ret}d\tau.
\label{NablaPhiGreen}
\end{equation}
Because $\nabla_\alpha V(x,z(\tau))$ and $\nabla_{\alpha} G^{ret}(x,z(\tau))$ are 
vectors in the tangent space at $x$ for all values of $\tau$, the integrals are well defined.  

Noticing that, for $T_-\leq \tau<\tau_{ret}$,
$G^{ret}(x,z(\tau))=-V(x,z(\tau))$, we write
\begin{equation}
\nabla_{\alpha}\Phi^{ret}=q\nabla_{\alpha}\left[\left(\frac{U}{\dot{\sigma}}\right)_{ret}\right]+q V\nabla_{\alpha}\tau_{ret}+q\lim_{h\rightarrow0}\int_{-\infty}^{\tau_--h}\nabla_{\alpha} G^{ret}d\tau.
\label{RetardedForce}
\end{equation}

We can also write the retarded and advanced solutions to the
field equation as
\begin{equation}
\Phi^{ret/adv}=q\left[\frac{U(x,z)}{\dot{\sigma}}\right]_{ret/adv}\pm q\lim_{h\rightarrow0}\int_{\mp\infty}^{\tau_{ret/adv}\mp h}G^{ret/adv}(x,z)d\tau,
\label{Phiretadv}
\end{equation}
and
\begin{equation}
\nabla_\alpha\Phi^{ret/adv}=q\nabla_{\alpha}\left[\frac{U(x,z)}{\dot{\sigma}}\right]_{ret/adv}\pm q V(x,z)\nabla_{\alpha}\tau_{ret/adv}\pm q\lim_{h\rightarrow0}\int_{\mp\infty}^{\tau_{ret/adv}\mp h}\nabla_{\alpha}G^{ret/adv}(x,z)d\tau.
\label{RetAdvForce}
\end{equation}

Now that we have an expression for the retarded and advanced
forces, we need to find expansions of the three bi-scalars, U,
V, and $\sigma$. 

\subsubsection{Expanding the Biscalars, $U(x,z)$, $V(x,z)$, and $\sigma(x,z)$}

The quantities $U(x,z)$ and $V(x,z)$ have the local expansions \cite{DewittBrehme}
\begin{equation}
U(x,z)=1+\frac{1}{12}R_{\alpha'\beta'}\nabla^{\alpha'}\sigma(x,z)\nabla^{\beta'}\sigma(x,z)+O(\epsilon^3),
\label{U(x,z)}
\end{equation}
\begin{equation}
V(x,z)=-\frac{1}{12}R(z)+O(\epsilon),
\label{v(x,z)}
\end{equation}
where $\nabla^{\alpha'}$ is defined to be the contravariant
derivative at the position of the particle $(z)$,
$R_{\alpha\beta}$ is the Ricci Tensor, and $R(z)$ is the Ricci
Scalar.

We review here the computation that expresses $\dot\sigma_{ret/adv}$ in terms of the coordinates 
$x^{\hat{\alpha}}$, and the particle's 4-velocity $u^\alpha$, acceleration $a^\alpha$, and jerk 
$\dot a^\alpha :=u^\beta\nabla_\beta a^\alpha$ at $\tau=0$.  We write 
$\dot\sigma_{ret/adv}= -(u^\alpha y_\alpha)_{ret/adv}$,
where $-y_{\alpha,\,ret}$ and $-y_{\alpha,\,adv}$ are
the gradients with respect to $z$ of $\sigma(x,z)$ at $z_{ret}=z(\tau_{ret})$ and $z_{adv} = z(\tau_{adv})$,
\begin{equation}
y_{\alpha,ret/adv} := - \left(\nabla_{\alpha}\sigma\right)_{ret/adv}.
\label{nablasigma}
\end{equation}
The contravariant vectors $y^\alpha_{ret/adv}$ are tangent to affinely parameterized null geodesics 
from $z(\tau_{ret/adv})$ to $x$.  Solving the geodesic equation iteratively, we find
\begin{equation}
y^{\hat\alpha}_{ret}=\left(x^{\hat{\alpha}} -z^{\hat\alpha}_{ret}\right)
    -\frac{1}{3}R^{\hat{\alpha}}{}_{\hat{\mu}\hat{\nu}\hat{\gamma}}z^{\hat\gamma}_{ret}\left(x^{\hat{\mu}} 
    -z_{ret}^{\hat\mu}\right)\left(x^{\hat{\nu}} -z^\nu_{ret}\right)+O(\epsilon^4).
\label{yalpha}
\end{equation}
For the advanced term, $y^{\hat\alpha}_{adv}$, replace each
subscript ``ret'' by ``adv''.  We next expand
$z^{\hat{\alpha}}(\tau)$ about $\tau=0$:
\begin{equation}
z^{\hat{\alpha}}(\tau_{ret/adv})= z^{\hat{\alpha}}(0)+\left.\partial_\tau z^{\hat{\alpha}}\right|_{\tau=0}  \tau_{ret/adv}+\left.\frac{1}{2}\partial_\tau^2 z^{\hat{\alpha}}\right|_{\tau=0}\tau_{ret/adv}^2+ O(\tau^3).
\label{z1}
\end{equation}
Using the form of the Christoffel symbols in RNC,  
 $\Gamma^{\hat\alpha}{}_{\hat\beta\hat\gamma}= -\frac23R^{\hat\alpha}{}_{(\hat\beta\hat\gamma)\hat\delta}x^{\hat\delta}$, and the index symmetries of the Riemann tensor, we have    
\be
a^{\hat\alpha} = u^{\hat\beta}\na_{\hat\beta} u^{\hat\alpha}|_{\tau=0} =\partial_\tau^2 z^{\hat\alpha}|_{\tau=0},  \qquad
\dot a^{\hat\alpha} = u^{\hat\beta}\na_{\hat\beta} a^{\hat\alpha}|_{\tau=0} = \partial_\tau^3\hat z^\alpha|_{\tau=0},
\ee
whence
\be
z^{\hat{\alpha}}(t_{ret/adv})= u^{\hat{\alpha}}\tau_{ret}+
\frac{1}{2}a^{\hat{\alpha}}\tau_{ret}^2
+\frac{1}{6}\dot{a}^{\hat{\alpha}} \tau_{ret}^3+ O(\tau_{ret}^4),
\label{z2}
\ee
with each coefficient evaluated at $\tau=0$. Now we use the relation $(g_{\alpha\beta}y^\alpha y^\beta)_{ret/adv} =0$
to find $\tau_{ret/adv}$ in terms of $u^{\hat\alpha}$ and $x^{\hat\alpha}$.  
Writing $\tau_{ret/adv}=\tau_1 +\tau_2  +O(\tau^3)$, with $\tau_n = O(\epsilon^n)$, we find  
\begin{equation}
\tau_1=-\left(u_{\hat{\alpha}} x^{\hat{\alpha}} \pm\sqrt{\left(\eta_{\hat{\alpha}\hat{\beta}}+u_{\hat{\alpha}}u_{\hat{\beta}}\right)x^{\hat{\alpha}} x^{\hat{\beta}}} \right),
\label{t1}
\end{equation}
where the $\pm$ corresponds to retarded (+) and advanced (-) solutions and $u_{\hat\alpha}$ is evaluated 
at $\tau=0$.  We denote by 
\be
q_{\alpha\beta}:= g_{\alpha\beta} + u_\alpha u_\beta 
\ee
the projection operator orthogonal to $u_\alpha$ and, with notation 
motivated by Eq.~(\ref{S0hat}) below, write 
$\hat S_0 = q_{\hat\alpha\hat\beta} x^{\hat\alpha}x^{\hat\beta}$,
where $q_{\hat\alpha\hat\beta}$ is evaluated at $z(0)$.  
Then 
\begin{equation}
\tau_1=-\left(u_{\hat{\mu}} x^{\hat{\mu}}\pm\sqrt{\hat{S}_0}\right).
\label{t12}
\end{equation}
Similarly,
\begin{equation}
\tau_2= \pm \frac{a_{\hat{\alpha}} x^{\hat{\alpha}}}{2\sqrt{\hat{S}_0}}\ \tau_1^2.
\label{t2}
\end{equation}

Finally, by substituting Eqs.~(\ref{t1}), and (\ref{t2}) into
Eq.~(\ref{z2}) we obtain an expression for
$z^{\hat{\alpha}}_{ret/adv}$ (and thus $y^{\hat{\alpha}}$) entirely in terms of
$x^{\hat{\alpha}}$ and of $u^{\hat{\alpha}}$ and their derivatives 
at $t=0$. 

We now expand $\dot{\sigma}$ about $\epsilon=0$. To do this,
we focus on $\dot{\sigma}^2$ and pattern our calculation on that 
of \cite{BO1}. Thus, we write,
\begin{equation}
\dot{\sigma}^2_{ret/adv}=(u_{\hat\alpha} y^{\hat\alpha})_{ret/adv}^2 
	= \left(q_{\hat\alpha\hat\beta} y^{\hat{\alpha}}y^{\hat{\beta}}\right)_{ret/adv}.
\label{sigmadot}
\end{equation}
Here $u^{\alpha}$ is the four velocity of the particle at the
retarded or advanced times (we treat this in a similar manner to
the way we treated $z^{\hat{\alpha}}_{ret/adv}$, using a
similar expansion as in Eq.~(\ref{z2})). Since
$y^\alpha_{ret/adv}$ is a null vector, we were able to add the
term
$g_{\hat{\alpha}\hat{\beta}}y^{\hat{\alpha}}y^{\hat{\beta}}=0$.
The reason for this change will soon be clear.

To keep track of the relevant terms in the calculation, we borrow
a term from \cite{BO1}, and then generalize it. We define
$\hat{S}$ by \footnote{It is useful to note that in [3] the use of the hat denoted a quantity evaluated at $\delta r=0$, whereas we use hats to specify that the expression is one found using RNCs. When we need to make a similar evaluation we will denote these quantities with a tilde.}
\begin{equation}
\hat{S}:=\left[q_{\hat{\alpha}\hat{\beta}}(x^{\hat{\alpha}}-z^{\hat{\alpha}}) ( x^{\hat{\beta}}-z^{\hat{\beta}})\right]_{ret/adv}.
\label{shat}
\end{equation} 
With this definition, we can now write 
\begin{equation}
\dot{\sigma}^2_{ret/adv}=S_{ret/adv}+\frac{1}{3}R_{\hat{\alpha}\hat{\gamma}\hat{\beta}\hat{\lambda}} x^{\hat{\alpha}} x^{\hat{\beta}}u^{\hat{\gamma}}u^{\hat{\lambda}}(x^{\hat{\iota}} x_{\hat{\iota}}) +O(\epsilon^5).
\label{sigmadotsquared}
\end{equation}
Here and in the rest of this section, $q_{\hat\alpha\hat\beta}, u^{\hat\alpha}, a^{\hat\alpha}$, and $\dot a^{\hat\alpha}$ will all be assumed to
be evaluated at $\tau=0$. When we
expand $S$ about $\epsilon=0$, we find
\begin{equation}
\hat{S}=\hat{S}_0+\hat{S}_1+\hat{S}_2+...
\label{Sexpand}
\end{equation}
where $\hat{S}_{n}=O(\epsilon^{n+2})$.  Explicitly, we have
\begin{equation}
\hat{S}_0=(\eta_{\hat{\alpha}\hat{\beta}}+u_{\hat{\alpha}}u_{\hat{\beta}}) x^{\hat{\alpha}}x^{\hat{\beta}} ,
\label{S0hat}
\end{equation}
\begin{equation}
\hat{S}_1=\eta_{\hat{\alpha}\hat{\beta}}a_{\hat{\gamma}} x^{\hat{\alpha}} x^{\hat{\beta}} x^{\hat{\gamma}},
\label{S1hat}
\end{equation}
and
\begin{equation}
\hat{S}_2=S_2^{(1)}\pm S_2^{(\pm)}= \left[\Sigma^{(1)}_{\hat{\alpha}\hat{\beta}\hat{\gamma}\hat{\lambda}}\pm\frac{ x^{\hat{\delta}}}{\sqrt{\hat{S}_0}}\Sigma^{(\pm)}_{\hat{\alpha}\hat{\beta}\hat{\gamma}\hat{\lambda}\hat{\delta}}\right] x^{\hat{\alpha}} x^{\hat{\beta}} x^{\hat{\gamma}} x^{\hat{\lambda}},
\label{S2hat}
\end{equation}
 where the quantities $\Sigma^{(1)}_{\hat{\alpha}\hat{\beta}\hat{\gamma}\hat{\lambda}}$ and $\Sigma^{(\pm)}_{\hat{\alpha}\hat{\beta}\hat{\gamma}\hat{\lambda}\hat{\delta}}$ in Eq.~(\ref{S2hat}) are 
\begin{equation}
\Sigma^{(1)}_{\hat{\alpha}\hat{\beta}\hat{\gamma}\hat{\lambda}} 
:= \frac{a^2}{12}q_{\hat\alpha\hat\beta}\left((\eta_{\hat{\gamma}\hat{\lambda}}
         +7u_{\hat{\gamma}}u_{\hat{\lambda}})
         -u_{\hat{\alpha}}u_{\hat{\beta}}u_{\hat{\gamma}}u_{\hat{\lambda}}\right)
-\frac{u_{\hat{\lambda}}\dot{a}_{\hat{\gamma}}}{3}(3\eta_{\hat{\alpha}\hat{\beta}}+2u_{\hat{\alpha}}u_{\hat{\beta}})
\label{Sigma1}
\end{equation}
and
\begin{equation}
\Sigma^{(\pm)}_{\hat{\alpha}\hat{\beta}\hat{\gamma}\hat{\lambda}\hat{\delta}} :=\frac{2}{3}(\eta_{\hat{\alpha}\hat{\beta}}+u_{\hat{\alpha}}u_{\hat{\beta}})(\eta_{\hat{\gamma}\hat{\lambda}}+u_{\hat{\gamma}}u_{\hat{\lambda}})(a^2u_{\hat{\delta}}-\dot{a}_{\hat{\delta}}).
\label{Sigmapm}
\end{equation}
It is also useful to define
\begin{equation}
r_{\hat{\alpha}}:=\frac{1}{2}\nabla_{\hat{\alpha}}\hat{S}_0= \nabla_{\hat{\alpha}}\left(\eta_{\hat{\mu}\hat{\nu}} + u_{\hat{\mu}}u_{\hat{\nu}}\right)x^{\hat{\mu}}x^{\hat{\nu}}=
\left(\eta_{\hat{\alpha}\hat{\mu}}+u_{\hat{\mu}}u_{\hat{\alpha}}\right)x^{\hat{\mu}}.
\label{r_alphadef}
\end{equation}

We now have the information to write the expansion of the first
term in Eq.~(\ref{Phiretadv}) (sometimes called the `direct'
term). We use Eqs.~(\ref{U(x,z)}), (\ref{sigmadotsquared}),
(\ref{Sexpand}), (\ref{S0hat}), (\ref{S1hat}), and (\ref{S2hat})
to expand $\Phi^{ret/adv}$ to the first three orders in $\epsilon$:

\begin{multline}
\Phi^{ret/adv}  = \frac{q}{\sqrt{\hat{S}_0}}\left[1-\frac{\hat{S}_1}{2\hat{S}_0}
   +\frac{3}{8}\left(\frac{\hat{S}_1}{\hat{S}_0}\right)^2-\frac{\hat{S}_2}{2\hat{S}_0}\right] 
   -\frac{q}{6\hat{S}_0^{3/2}}R_{\hat{\alpha}\hat{\gamma}\hat{\beta}\hat{\lambda}}
		u^{\hat{\lambda}}u^{\hat{\gamma}} x^{\hat{\alpha}} x^{\hat{\beta}} x^2
\\
 +\frac{qR_{\hat{\alpha}\hat{\beta}}}{12}\left[ \frac{r^{\hat{\alpha}}r^{\hat{\beta}}
 +S_0 u^{\hat{\alpha}}u^{\hat{\beta}}}{\sqrt{\hat{S}_0}}\pm 2( x^{\hat{\alpha}}u^{\hat{\beta}}
 +u^{\hat{\alpha}}u^{\hat{\beta}}u_{\hat{\gamma}} x^{\hat{\gamma}}) \right] 
 \pm q\lim_{h\rightarrow0}\int_{\mp\infty}^{\tau_{ret/adv}\mp h}G^{ret/adv}(x,z)d\tau
 +O(\epsilon^2),
\label{PhiRetAdvFull}
\end{multline}
where $x^2\equiv x^{\hat{\epsilon}}x_{\hat{\epsilon}}$.

Because the flat spacetime Green's function has support only on the light cone and not 
in its interior, the last term in (35) vanishes in flat spacetime, as do the 
terms involving the Ricci tensor.  On the other hand, the curved spacetime and flat spacetime 
values of $\hat S_0, \hat S_1$ and $\hat S_2$ coincide.  
The flat-space comparison field $\tilde \Phi$ of 
Eq.~(\ref{RegularizedField}) is thus obtained from the first term of Eq.~(\ref{PhiRetAdvFull}) by taking half the sum of its retarded and advanced forms.  
Noting that $\hat S_0$ and $\hat S_1$ are the same for the retarded and 
advanced field and that, by Eq.~(\ref{S2hat}), $\hat S_2$ is replaced by $\hat S_2^{(1)}$, we have   
\be
  \tilde\Phi  = \frac{q}{\sqrt{\hat{S}_0}}\left[1-\frac{\hat{S}_1}{2\hat{S}_0}+\frac{3}{8}\left(\frac{\hat{S}_1}{\hat{S}_0}\right)^2-\frac{\hat{S}_2^{(1)}}{2\hat{S}_0}\right].
\label{eq:Phitilde}\ee
Note that the flat-space contribution $\tilde\Phi$ includes the leading singular 
behavior (the Coulomb part) of $\Phi^{ret}$. 

It is instructive to see Eq.~(\ref{PhiRetAdvFull}) written in
terms of the acceleration and jerk. Using Eqs.~(\ref{S0hat})-(\ref{Sigmapm}), 
we obtain
\begin{multline}
\Phi^{ret/adv}  = \frac{q}{\sqrt{q_{\hat\mu\hat\nu}x^{\hat{\mu}}x^{\hat{\nu}}}}\left[1- \frac{a_{\hat{\gamma}}x^{\hat{\gamma}}x^2}{2q_{\hat\mu\hat\nu}x^{\hat{\mu}}x^{\hat{\nu}}}\left(1- \frac{3}{4} \frac{a_{\hat{\gamma}}x^{\hat{\gamma}}x^2}{q_{\hat\mu\hat\nu}x^{\hat{\mu}}x^{\hat{\nu}}}\right)\right] 
\mp\frac{q}{3}x^{\hat{\delta}}(a^2u_{\hat{\delta}}-\dot{a}_{\hat{\delta}})\\
-\frac{q\left[a^2\left( q_{\hat\alpha\hat\beta} (\eta_{\hat{\gamma}\hat{\lambda}}+7u_{\hat{\gamma}}u_{\hat{\lambda}})-u_{\hat{\alpha}}u_{\hat{\beta}}u_{\hat{\gamma}}u_{\hat{\lambda}}\right)-4u_{\hat{\gamma}}a_{\hat{\lambda}}(3\eta_{\hat{\alpha}\hat{\beta}}+2u_{\hat{\alpha}}u_{\hat{\beta}})\right]}{24\left(q_{\hat\mu\hat\nu}x^{\hat{\mu}}x^{\hat{\nu}}\right)^{3/2}}
x^{\hat{\alpha}}x^{\hat{\beta}} x^{\hat{\gamma}}x^{\hat{\lambda}} 
\\
+\frac{qR_{\hat{\alpha}\hat{\beta}}}{12}\left[ \frac{ x^{\hat{\alpha}} x^{\hat{\beta}}+2 u^{\hat{\alpha}} x^{\hat{\beta}}(u_{\hat{\gamma}} x^{\hat{\gamma}}) +u^{\hat{\alpha}}u^{\hat{\beta}}(x^\epsilon x_{\epsilon}+2(u_{\hat{\gamma}} x^{\hat{\gamma}})^2)}{\sqrt{q_{\hat\mu\hat\nu}x^{\hat{\mu}}x^{\hat{\nu}}}}\pm 2( x^{\hat{\alpha}}u^{\hat{\beta}}+u^{\hat{\alpha}}u^{\hat{\beta}}u_{\hat{\gamma}} x^{\hat{\gamma}})
\right]\\
-\frac{qR_{\hat{\alpha}\hat{\gamma}\hat{\beta}\hat{\lambda}}u^{\hat{\lambda}}u^{\hat{\gamma}} x^{\hat{\alpha}} x^{\hat{\beta}} x^2}{6\left(q_{\hat\mu\hat\nu}x^{\hat{\mu}}x^{\hat{\nu}}\right)^{3/2}}\pm q\lim_{h\rightarrow0}\int_{\mp\infty}^{\tau_{ret/adv}\mp h}G^{ret/adv}(x,z)d\tau
+O(\epsilon^2).
\label{PhiRetAdvFullA's}
\end{multline}
Noting that $\hat{S}_0=r_{\hat{\alpha}}r^{\hat{\alpha}}$, we
write  Eq.~(\ref{PhiRetAdvFullA's}) as

\begin{multline}
\Phi^{ret/adv}=\frac{q}{r}\left[1-\frac{a_{\hat{\gamma}}r^{\hat{\gamma}}x^2}{2r^2}+\frac{3}{8}\left(\frac{a_{\hat{\gamma}}r^{\hat{\gamma}}x^2}{r^2}\right)^2
-\frac{1}{2r^2}
	\left(\frac{a^2}{12}\left(r^4+6r^2(u_{\hat{\alpha}}x^{\hat{\alpha}})^2- (u_{\hat{\alpha}}x^{\hat{\alpha}})^4\right)
	\right)
\right]
\\
-\frac{q}{12r^3} \left[2u_{\hat{\gamma}}x^{\hat{\gamma}}a_{\hat{\mu}}r^{\hat{\mu}}\left(3r^2-(u_{\hat{\sigma}}x^{\hat{\sigma}})^2\right) +2R\indices{_{\hat{\alpha}\hat{\gamma}\hat{\beta}\hat{\delta}}} x^{\hat{\alpha}}x^{\hat{\beta}}u^{\hat{\gamma}}u^{\hat{\delta}}x^2-r^ 2R\indices{_{\hat{\alpha}\hat{\beta}}}\left(r^{\hat{\alpha}}r^{\hat{\beta}}+r^2u^{\hat{\alpha}}u^{\hat{\beta}}\right)\right]
\\
\pm\frac{q}{6}\left[R\indices{_{\hat{\alpha}\hat{\beta}}}r^{\hat{\alpha}}u^{\hat{\beta}}+2x^{\hat{\alpha}}(\dot{a}_{\hat{\alpha}}-a^2u_{\hat{\alpha}})\right]\pm q\lim_{h\rightarrow0}\int_{\mp\infty}^{\tau_{ret/adv}\mp h}G^{ret/adv}(x,z)d\tau+O(\epsilon^2)
\label{PhiRetAdvFull2}
\end{multline}

Therefore, using Eq.~(\ref{RetAdvForce}), we can write the
gradient of the retarded and advanced fields as
\begin{equation}
\nabla_{\alpha}\Phi^{ret/adv}=\nabla_{\alpha}\left[\left(\frac{qU(x,z)}{\dot{\sigma}}\right)_{ret/adv}\right]-\frac{R(z) q}{12}\left(\frac{\nabla_{\alpha}\hat{S}_0}{2\sqrt{\hat{S}_0}}\pm u_{\alpha}\right)\pm q\nabla_{\alpha}\lim_{h\rightarrow0}\int_{\mp\infty}^{\tau_{ret/adv}\mp h}G^{ret/adv}(x,z)d\tau.
\label{NablaPhiRetAdv}
\end{equation}

Writing out the gradient of the scalar field in terms of the
$S_n$'s, we have
\begin{multline}
\nabla_{\hat{\alpha}}\left[\Phi^{ret/adv}\right]=q\left[-\frac{\nabla_{\hat{\alpha}}\hat{S}_0}{2\hat{S}_0^{3/2}}-\frac{1}{2}\left(\frac{\nabla_{\hat{\alpha}}\hat{S}_1}{\hat{S}_0^{3/2}}-\frac{3}{2}\frac{\hat{S}_1\nabla_{\hat{\alpha}}\hat{S}_0}{\hat{S}_0^{5/2}}\right)
-\frac{15}{16}\frac{{\hat{S}_1}^2\nabla_{\hat{\alpha}}\hat{S}_0}{\hat{S}_0^{7/2}}+\frac{3}{4}\frac{\hat{S}_1\nabla_{\hat{\alpha}}\hat{S}_1}{\hat{S}_0^{5/2}}\right]
\\
+q\left[-\frac{1}{2}\left(\frac{\nabla_{\hat{\alpha}}\hat{S}_2}{\hat{S}_0^{3/2}}-\frac{3}{2}\frac{\hat{S}_2\nabla_{\hat{\alpha}}\hat{S}_0}{\hat{S}_0^{5/2}}\right)\pm\frac{1}{6}R\indices{_{\hat{\mu}\hat{\beta}}}u^{\hat{\beta}}\left(\delta\indices{^{\hat{\mu}}_{\hat{\alpha}}}+u^{\hat{\mu}}u_{\hat{\alpha}}\right)\right]
\\
+\frac{qR\indices{_{\hat{\mu}\hat{\nu}}}}{24}\left[\frac{2}{\sqrt{\hat{S}_0}}\left(r^{\hat{\mu}}\left(\delta^{\hat{\nu}}_{\hat{\alpha}}+u^{\hat{\nu}}u_{\hat{\alpha}}\right)
+u^{\hat{\mu}}u^{\hat{\nu}}\nabla_{\hat{\alpha}}\hat{S}_0\right)-\frac{\nabla_{\hat{\alpha}}\hat{S}_0}{\hat{S}_0^{3/2}}\left(r^{\hat{\mu}}r^{\hat{\nu}}+\hat{S}_0u^{\hat{\mu}}u^{\hat{\nu}}\right)\right]
\\
-\frac{q R\indices{_{\hat{\mu}\hat{\gamma}\hat{\nu}\hat{\delta}}}u^{\hat{\gamma}}u^{\hat{\delta}}x^{\hat{\mu}}} {12 \hat{S}_0^{5/2}}\left(4\hat{S}_0x^2\delta^{\hat{\nu}}_{\hat{\alpha}}+4\hat{S}_0x^{\hat{\nu}}x_{\hat{\alpha}}-3x^{\hat{\nu}}x^2 \nabla_{\hat{\alpha}}\hat{S}_0\right)-\frac{qR(z) }{12}\left(\frac{\nabla_{\hat\alpha}\hat{S}_0}{2\sqrt{\hat{S}_0}}\pm u_{\hat\alpha}\right)
\\
\pm q\nabla_{\hat\alpha}\lim_{h\rightarrow0}\int_{\mp\infty}^{\tau_{ret/adv}\mp h}G^{ret/adv}(x,z)d\tau+O(\epsilon^2).
\label{NablaPhiRetAdvS}
\end{multline}

When re-expressed in terms of $a^{\mu},\dot{a}^{\mu},$ and
$r^{\mu}$, $\nabla_{\hat{\alpha}}\Phi^{ret/adv}$ has the 
form
\begin{multline}
\nabla_{\hat{\alpha}}\Phi^{ret/adv}=
q\left[-\frac{r_{\hat{\alpha}}}{r^3}-\frac{1}{2}\left(\frac{a_{\hat{\alpha}}x^2+2a_{\hat{\gamma}}x^{\hat{\gamma}}x_{\hat{\alpha}}}{r^3}-3\frac{a_{\hat{\gamma}}r^{\hat{\gamma}}x^2r_{\hat{\alpha}}}{r^5}\right)+\frac{3}{4}\frac{a_{\hat{\gamma}}r^{\hat{\gamma}}x^2(a_{\hat{\alpha}}x^2+2a_{\hat{\gamma}}x^{\hat{\gamma}}x_{\hat{\alpha}})}{r^5}\right]
\\
+q\left[-\frac{15}{8}\frac{\left(a_{\hat{\gamma}}r^{\hat{\gamma}}x^2\right)^2r_{\hat{\alpha}}}{r^7}
-\frac{a^2}{24r^5}\left(r^4r_{\hat{\alpha}}+12r^4u_{\hat{\gamma}}x^{\hat{\gamma}}u_{\hat{\alpha}} -6r^2\left(u_{\hat{\gamma}}x^{\hat{\gamma}}\right)^2r_{\hat{\alpha}}-4r^2\left(u_{\hat{\gamma}}x^{\hat{\gamma}}\right)^3u_{\hat{\alpha}} 
+3\left(u_{\hat{\gamma}}x^{\hat{\gamma}}\right)^4r_{\hat{\alpha}}\right)\right]
\\
-\frac{q}{2}\left[\left(1-\frac{1}{3}\left(\frac{u_{\hat{\gamma}}x^{\hat{\gamma}}}{r}\right)^2\right)\frac{1}{r^3}\left(u_{\hat{\gamma}}x^{\hat{\gamma}}\dot{a}_{\hat{\beta}}x^{\hat{\beta}}r_{\hat{\alpha}}-r^2\left(u_{\hat{\alpha}}\dot{a}_{\hat{\beta}}x^{\hat{\beta}}+\dot{a}_{\hat{\alpha}}u_{\hat{\beta}}x^{\hat{\beta}}\right)\right)
+\frac{2\dot{a}_{\hat{\beta}}x^{\hat{\beta}}\left(u_{\hat{\gamma}}x^{\hat{\gamma}}\right)^2}{3r^5}\left(r^2u_{\hat{\alpha}}-r_{\hat{\alpha}}u_{\hat{\gamma}}x^{\hat{\gamma}}\right)\right]
\\
+\frac{qR\indices{_{\hat{\mu}\hat{\nu}}}}{12}\left[\frac{1}{r}\left(r^{\hat{\mu}}\left(\delta^{\hat{\nu}}_{\hat{\alpha}}+u^{\hat{\nu}}u_{\hat{\alpha}}\right)
+2u^{\hat{\mu}}u^{\hat{\nu}}r_{\hat{\alpha}}\right)-\frac{r_{\hat{\alpha}}}{r^3}\left(r^{\hat{\mu}}r^{\hat{\nu}}+r^2u^{\hat{\mu}}u^{\hat{\nu}}\right)\right]
\\
-\frac{q R\indices{_{\hat{\mu}\hat{\gamma}\hat{\nu}\hat{\delta}}}u^{\hat{\gamma}}u^{\hat{\delta}}x^{\hat{\mu}}} {6 r^{5}}\left(2r^2x^2\delta^{\hat{\nu}}_{\hat{\alpha}}+2r^2x^{\hat{\nu}}x_{\hat{\alpha}}-3x^{\hat{\nu}}x^2 r_{\hat{\alpha}}\right)-\frac{qR(z) }{12}\left(\frac{r_{\hat{\alpha}}}{r}\right)
\\
\pm \frac{q}{12}\left[4\left(\dot{a}_{\hat{\alpha}}-a^2u_{\hat{\alpha}}\right) +2R\indices{_{\hat{\mu}\hat{\beta}}}u^{\hat{\beta}}\left(\delta^{\hat{\mu}}_{\hat{\alpha}}+ u^{\hat{\mu}}u_{\hat{\alpha}}\right)-R(z)u_{\hat{\alpha}}\right]
\pm q\lim_{h\rightarrow0}\int_{\mp\infty}^{\tau_{ret/adv}\mp h}\nabla_{\hat\alpha}G^{ret/adv}(x,z)d\tau+O(\epsilon^2).
\label{NablaPhiRetAdvA}
\end{multline}
This is a more general expression than is given in Quinn
\cite{Quinn}.  Only when the field point $x$ is chosen to be
along a geodesic orthogonal to the trajectory at $z(0)$ (that is,
when $u_{\hat\alpha} x^{\hat \alpha} =0$) does this match
Quinn's expression.  

\subsection{The Singular Field}

Using the short-distance expansions we have just presented, we can now identify
a singular field, $f^{sing}_\alpha$, satisfying 
$\dis f^{ren}_{\alpha} = \lim_{x\rightarrow z(0)}[ f^{ret}_{\alpha}(x) - f^{sing}_\alpha(x)]$, 
and a singular field, $\Phi^{sing}$, for which 
$f^{sing}_\alpha = \nabla^\alpha \Phi^{sing}$.  
From the explicit form of Eq.~(\ref{NablaPhiRetAdvA}) for $f^{ret}_\alpha$ 
we will quickly show that $f^{sing}_\alpha$ comprises all but the last line 
of Eq.~(\ref{NablaPhiRetAdvA}) for $f^{ret}_\alpha$.  Then, by recalling the 
terms in $\Phi^{ret}$ that lead to these terms, we write $f^{sing}_\alpha$
in the simpler form,
\be
  f^{sing}_\alpha = \nabla_\alpha\Phi^{sing}, 
\ee
with 
\begin{align}
\Phi^{sing} &= \frac{q}{\sqrt{\hat{S}_0}} -\frac{\hat{S}_1}{2\hat{S}_0^{3/2}}+\left\{\frac{q}{\sqrt{\hat{S}_0}}\left[\frac{3}{8}\left(\frac{\hat{S}_1}{\hat{S}_0}\right)^2-\frac{\hat{S}^{(1)}_2}{2\hat{S}_0}\right] -\frac{q}{\sqrt{\hat{S}}_0}\left[\frac{1}{6\hat{S}_0}R_{\hat{\alpha}\hat{\gamma}\hat{\beta}\hat{\delta}}u^{\hat{\gamma}}u^{\hat{\delta}} x^{\hat{\alpha}} x^{\hat{\beta}} x^{\hat{\epsilon}} x_{\hat{\epsilon}}\right]\right.
\nonumber\\
&\hspace{4cm}\left.+\frac{q}{\sqrt{\hat{S}_0}}\frac{1}{12} R_{\hat{\alpha}\hat{\beta}}\left[ r^{\hat{\alpha}} r^{\hat{\beta}}+u^{\hat{\alpha}}u^{\hat{\beta}}\hat S_0
\right]
-\frac{1}{12}q R(z)\sqrt{\hat{S}_0}\right\},
\nonumber\\
& = \Phi^{sing,\rm L} + \Phi^{sing,\rm SL} + \Phi^{sing,\rm SSL},
\label{SingularField1}
\end{align}  
where the grouping into three terms exhibits the field as a sum of leading, 
subleading and sub-subleading terms, of 
order $\epsilon^{-1}$, $\epsilon^0$, and $\epsilon$, respectively.    
Finally, we will check that Eq.~(\ref{SingularField1}) is the expansion through $O(\epsilon)$ of the 
Detweiler-Whiting singular field, 
\begin{equation}
\Phi_{DW}=\frac{1}{2}\left[\left(\frac{U(x,z)}{\dot{\sigma}}\right)_{ret}+\left(\frac{U(x,z)}{\dot{\sigma}}\right)_{adv}\right]+\frac{q}{2}\int_{\tau_{ret}}^{\tau_{adv}}V(x,z)d\tau.
\label{SingularFieldDW}
\end{equation}

% As we discuss in greater detail below, the identification is not unique.  
% In particular, two vector fields $f^{sing}_\alpha$ and 
% ${\frak f}^{sing,\alpha}$ give the same self-force if their difference 
% has vanishing limit at the particle.  

We begin with the identification of $f^{sing}_\alpha$ with the first five lines 
of Eq.~(\ref{NablaPhiRetAdvA}) for $f^{ret}_\alpha = \nabla_{\hat\alpha}\Phi^{ret}$.
The first three lines of the expression are just $\nabla_\alpha \tilde\Phi$, the gradient of 
Quinn's flat-space comparison field.  The effect of the 
angle average on the remaining terms is to remove all terms involving the 
Riemann tensor and its contractions that have an odd number of factors 
of the coordinates $x^{\hat\alpha}$; these are the terms that comprise the 
4th and 5th lines of (\ref{NablaPhiRetAdvA}).
That the angle average of these terms vanishes follows from the 
fact that the terms are odd under the inversion 
$I:x^{\hat\alpha}\rightarrow - x^{\hat\alpha}$, 
while $I$ maps the domain of integration $S_r$ in the angle average to itself.
To make this precise, note that $dS$ is invariant under $I$ and that 
the restriction to $S_r$ of a function odd under $I$ is a function on $S_r$ 
that is odd under $I$.  That is to say, $\int_{S_r} f dS = \int_{I(S_r)} f\circ I dS = -\int_{S_r} f dS$,
when  $f\circ I=-f$. The remaining terms, the last line of Eq.~(42), are 
continuous, and the limit of their angle average is just their value at 
the particle.  
Thus, with $f^{sing}_\alpha$ identified with the 
first five lines of Eq.~(\ref{NablaPhiRetAdvA}), we have 
$\dis f^{ren}_{\alpha} = \lim_{x\rightarrow z(0)}[ f^{ret}_{\alpha}(x) - f^{sing}_\alpha(x)]$, 
as claimed.  

    As we have noted, the flat spacetime comparison field $\tilde\Phi$ of 
Eq.~(\ref{eq:Phitilde}) includes the leading singular behavior of 
$\Phi^{ret}$. To obtain the singular field $\Phi^{sing}$, we add the additional 
$O(\epsilon)$ terms in the short-distance expansion of $\Phi^{ret}$ whose gradient 
provided the terms of $f^{sing}$ involving the Riemann tensor, Ricci tensor, and 
Ricci scalar terms, the terms in the 4th and 5th lines of (\ref{NablaPhiRetAdvA}).  
The Riemann- and Ricci-tensor terms in $f^{sing}_{\hat\alpha}$ are the Riemann- 
and Ricci-tensor terms of $f^{ret}_{\hat\alpha}$ that are odd in $x^{\hat\alpha}$, 
and they are therefore the gradients of the Riemann- and Ricci-tensor terms 
in the expansion (\ref{PhiRetAdvFull}) of $\Phi^{ret}$ that are even in 
$x^{\hat\alpha}$, namely  
\[
 -\frac{q}{6\hat{S}_0^{3/2}}R_{\hat{\alpha}\hat{\gamma}\hat{\beta}\hat{\lambda}}
		u^{\hat{\lambda}}u^{\hat{\gamma}} x^{\hat{\alpha}} x^{\hat{\beta}} x^2
 +\frac{qR_{\hat{\alpha}\hat{\beta}}}{12}\left[ \frac{r^{\hat{\alpha}}r^{\hat{\beta}}
 +S_0 u^{\hat{\alpha}}u^{\hat{\beta}}}{\sqrt{\hat{S}_0}}\right].
\] 
Finally, the Ricci scalar term in $f^{sing}_{\hat\alpha}$ came from the gradient of the 
upper limit of integration in the last term of (\ref{PhiRetAdvFull}): 
It is the part $\nabla_\alpha \tau_{ret}$ that is odd under $I$.
Using Eq.~(\ref{NablaPhiRetAdv}), we can write that term as the gradient of 
\[
   -\frac1{12} q R\sqrt{\hat S_0}.
\]
Then 
\be
 \Phi^{sing} =  \tilde\Phi +  -\frac{q}{6\hat{S}_0^{3/2}}R_{\hat{\alpha}\hat{\gamma}\hat{\beta}\hat{\lambda}}
		u^{\hat{\lambda}}u^{\hat{\gamma}} x^{\hat{\alpha}} x^{\hat{\beta}} x^2
 +\frac{qR_{\hat{\alpha}\hat{\beta}}}{12}\left[ \frac{r^{\hat{\alpha}}r^{\hat{\beta}}
 +S_0 u^{\hat{\alpha}}u^{\hat{\beta}}}{\sqrt{\hat{S}_0}}\right]-\frac1{12} q R\sqrt{\hat S_0},  
\label{SingularFieldx}\ee 
and Eq.~(\ref{SingularField1}) follows.  The explicit form of $\Phi^{sing}$ in terms of $x^{\hat\alpha}$, 
$u^{\hat\alpha}$, and $a^{\hat\alpha}$ can be read off from the expanded form 
(\ref{PhiRetAdvFull2}) $\Phi^{ret}$:  The field $\Phi^{sing}$ comprises the 
first two lines of that equation.    

From Eq.~(\ref{NablaPhiRetAdvA}), we can immediately see that, when the test-particle 
motion is geodesic (when $a^\alpha = 0$), Gralla's version of $f^{ren}_\alpha$ \cite{gralla11} as an angle 
average of $f^{ret}_\alpha$ holds.  Every term in in the first five lines  of $f^{sing}_{\hat\alpha}$  is then odd under $x^{\hat\alpha} \rightarrow - x^{\hat\alpha}$, implying $\langle f^{sing}_{\hat\alpha}\rangle_r = 0$. We thus have  
\be
  f^{ren}_{\hat\alpha} = 
	\lim_{r\rightarrow 0} \langle f^{ret}_{\hat\alpha} - f^{sing}_{\hat\alpha}\rangle_r 
	= \lim_{r\rightarrow 0} \langle f^{ret}_{\hat\alpha} \rangle_r.
\ee 

The singular field $\Phi^{sing}$ is not unique.  Any function $\Phi^s$ defined near the position of 
the particle is equivalent to $\Phi^{sing}$ if it satisfies the conditions 
\be
\lim_{x\rightarrow z(0)} (\Phi^s - \Phi^{sing}) = 0, \qquad 
\lim_{x\rightarrow z(0)} \nabla_\alpha (\Phi^s - \Phi^{sing}) = 0.
\label{equivalence}\ee
That is, two singular fields that are equivalent in this sense give the same values  
of $\Phi^{ren}$ and $f^{ren}_\alpha$.    
We conclude this section by observing that the Detweiler-Whiting singular field, 
$\Phi^{sing}_{DW}$ of Eq.~(\ref{SingularFieldDW}), is equivalent in this sense to $\Phi^{sing}$ 
of Eq.~(\ref{SingularField1}).    

We quickly establish the equivalence as follows.  By comparing Eq.~(\ref{SingularField1}) 
for $\Phi^{sing}$ to Eq.~(\ref{PhiRetAdvFull}) for $\Phi^{ret/adv}$ we have
\be
\Phi^{sing} = \frac12\left(\Phi^{ret}+\Phi^{adv}\right) - \frac1{12}R\sqrt{\hat S_0} 
		-\frac12 q\lim_{h\rightarrow\infty}\left[ \int_{-\infty}^{\tau_{ret}-h}G^{ret}(x,z(\tau)) + \int_{\tau_{adv}+h}^\infty G^{adv}(x,z(\tau))\right] d\tau
\ee 
We next use Eq.~(\ref{Phiretadv}) to write  
\bea
  \frac12\left(\Phi^{ret}+\Phi^{adv}\right) = \frac12 q\left[\left(\frac{U(x,z)}{\dot{\sigma}}\right)_{ret}
	+\left(\frac{U(x,z)}{\dot{\sigma}}\right)_{adv}\right] \nonumber\\
        + \frac12 q\lim_{h\rightarrow\infty}\left[ \int_{-\infty}^{\tau_{ret}-h}G^{ret}(x,z(\tau)) + \int_{\tau_{adv}+h}^\infty G^{adv}(x,z(\tau))\right] .
\eea
From these two equations and the relation $\tau_{adv} - \tau_{ret} = 2\sqrt{\hat S_0} + O(\epsilon^2)$, we 
have the desired result, 
\bea
  \Phi^{sing} &=&  
	\frac12 q\left[\left(\frac{U(x,z)}{\dot{\sigma}}\right)_{ret}
	+\left(\frac{U(x,z)}{\dot{\sigma}}\right)_{adv}\right]
		- \frac 16 qR(\tau_{adv}-\tau_{ret}) +O(\epsilon^2)\nonumber\\
	&=& \frac12 q\left[\left(\frac{U(x,z)}{\dot{\sigma}}\right)_{ret}
			+\left(\frac{U(x,z)}{\dot{\sigma}}\right)_{adv}\right]
		+\frac12 q\int^{\tau_{adv}}_{\tau_{ret}} V d\tau +O(\epsilon^2)\nonumber\\
	&=& \Phi^{sing}_{DW}+O(\epsilon^2). 
\label{DWSingField}
\eea
Finally, $\nabla_\alpha\Phi^{sing}$ differs from $\nabla_\alpha\Phi^{sing}_{DW}$ at 
order $\epsilon$, implying $\lim_{x\rightarrow z(0)} \nabla_\alpha (\Phi^{sing}_{DW} - \Phi^{sing}) = 0$.
\section{Mode-Sum Regularization}
\label{ModeSumReg}

We turn now to mode-sum regularization. We extend previous work to include accelerated trajectories on smooth, globally hyperbolic spacetimes with generic  smooth coordinate systems, and we show the equivalence of mode-sum regularization to the renormalization methods discussed in the previous section.  We begin with a scalar charge and then generalize the results to electromagnetic charges and point masses in the next section.  

In mode-sum regularization one writes the retarded and singular fields as sums of angular harmonics, using the fact that the individual harmonics 
of the retarded field and of the expression for the self-force 
have finite limits on the particle's trajectory. Because the singular part of the retarded field is defined only in a normal neighborhood of the particle, 
its individual angular harmonics are defined only after one extends the field to a thick sphere through a position $z(0)$ of the particle. The singular behavior of the retarded field, however, {\em uniquely determines the large 
$\ell$ behavior of its angular harmonics}:  For a function $f$ on the sphere that is smooth everywhere except at a point $P$, where it has an expansion in powers of the distance to $P$, that short-distance expansion determines the expansion of the angular harmonics of $f$ in powers of $1/\ell$. 

Let $(t,r,\theta,\phi)$ be spherical coordinates related in the usual way to a smooth 
Cartesian chart $(t,x^1,x^2,x^3)$ for which the 2-spheres of constant $t$ and $r$ are in the domain of the chart.  We denote by $\Phi^{sing}$ any smooth extension of the singular field of Eq.~(\ref{SingularField1}) to a thickened 2-sphere on the $t=0$ surface 
through $z(0)$ that includes a finite interval in $r$ about the radial coordinate $r_0$ of $z(0)$. For $\Phi$ representing either $\Phi^{ret}$ or $\Phi^{sing}$, 
each component of the expression  
for the self-force along the Cartesian coordinate basis
has angular harmonics $f^{\ell m}_{\alpha}$ given by 
%\begin{equation}
%f_\alpha^{\ell m}(t,r,\theta,\phi) = q\nabla_{\alpha}\Phi(t,r,\theta,\phi)=q\sum_{\ell=0}^%
%{\infty}\sum_{m=-\ell}^{\ell}f^{\ell m}_{\alpha}(t,r)Y_{\ell m}(\theta,\phi),
%\label{SphericalHarmonicDecomp1}
%\end{equation}
%where
\begin{equation}
f^{\ell m}_{\alpha}(t,r)=q\int d\Omega\, \nabla_\alpha \Phi(t,r,\theta,\phi)\,\bar{Y}_{\ell m}(\theta,\phi).
\label{SphericalHarmonicDecomp2}
\end{equation}
We have seen that the renormalized self-force at $z(0)$ is given by 
\be 
   f^{ren}_\alpha = \lim_{x\rightarrow z(0)} q\nabla_{\alpha}\left(\Phi^{ret}-\Phi^{sing}\right).
\label{frendef}
\ee
To obtain an equivalent mode-sum form of $f^{ren}_\alpha$, we first use the fact that, 
for $r\neq r_0$ on the thickened sphere where $\Phi^{sing}$ is defined, 
$\Phi^{ret}$ and $\Phi^{sing}$ are each smooth; second, that their angular harmonics 
have finite limits as $r\rightarrow r_0^\pm$ (the limits depend whether $r$ approaches 
$r_0$ from above or below); and finally that $\nabla_\alpha\Phi^{ret} - \nabla_\alpha\Phi^{sing}$ is continuous on the 
entire thickened sphere, when its value at $r=r_0$ is taken to be $\lim_{x\rightarrow z(0)}(\nabla_\alpha\Phi^{ret} - \nabla_\alpha\Phi^{sing})$.  We then have 
\begin{align}
  f^{ren}_\alpha/q &=\lim_{r\rightarrow r_0}\nabla_{\alpha}\left(\Phi^{ret}-\Phi^{sing}\right)(t=0,r,\theta_0,\phi_0)   \\
  &=\lim_{r\rightarrow r_0}\sum_{\ell,m} 
   \left[\nabla_{\alpha}\left(\Phi^{ret}-\Phi^{sing}\right)\right]^{\ell m}(t=0,r)
			Y_{\ell m}(\theta_0,\phi_0) \\
 &=\sum_{\ell,m} \lim_{r\rightarrow r_0} 
   \left[\nabla_{\alpha}\left(\Phi^{ret}-\Phi^{sing}\right)\right]^{\ell m}(t=0,r)
			Y_{\ell m}(\theta_0,\phi_0) \\
 &=\sum_{\ell,m} \left[\lim_{r\rightarrow r_0^\pm} 
   \left(\nabla_{\alpha}\Phi^{ret}\right)^{\ell m}(t=0,r)
        -\lim_{r\rightarrow r_0^\pm}\left(\nabla_\alpha\Phi^{sing}\right)^{\ell m}(t=0,r)\right]
			Y_{\ell m}(\theta_0,\phi_0), 
\end{align}
where $r_0$, $\theta_0$, and $\phi_0$ are the angular coordinates of the particle at time $t=0$.

%home
The finite range of the sum over $m$ allows the definitions
\bsube
\begin{align}
f^{ret,\ell\pm}_{\alpha}&
  := q\sum_{m=-\ell}^{\ell}\lim_{r\rightarrow r_{0}^{\pm}}
  \nabla_\alpha	\Phi^{ret, \ell m}(t=0,r)Y_{\ell m}(\theta_0,\phi_0), 
\label{fret_ell}\\
f^{sing,\ell\pm}_{\alpha}&:= q\sum_{m=-\ell}^{\ell}\lim_{r\rightarrow r_{0}^{\pm}}\nabla_\alpha
	\Phi^{sing, \ell m}(t=0,r)Y_{\ell m}(\theta_0,\phi_0), 
\label{SphericalHarmonicSumell}
\end{align} \label{f_ell}
\esube
and the renormalized self-force is then given by 
\begin{equation}
f^{ren}_{\alpha}=\sum_{\ell=0}^{\infty} f^{ren,\ell}_\alpha 
	:= \sum_{\ell=0}^{\infty} \left(f^{ret,\ell\pm}_\alpha-f^{sing,\ell\pm}_\alpha\right).
\label{SphericalHarmonicSumell2}
\end{equation}

% Now, we assume that we have already solved for the retarded field. We would like to be able to write this in the form 
% \begin{equation}
% f^{ren}_{\alpha}=\sum_{\ell=0}^\infty q\left\{\lim_{x\rightarrow z(0)}\left[\nabla_{\alpha}\Phi_{ret}\right]_{\ell}-\lim_{x\rightarrow z(0)}\left[\nabla_{\alpha}\Phi^{sing}\right]\right\},
% \label{Ideal Equation}
% \end{equation}
% but, strictly speaking we cannot separate the self field this way, since the limit of both $\Phi{ret}$ and $\Phi^{sing}$ and their gradients diverges. We avoid this by truncating the sum at large $\ell$, which is to say that we define some $\ell_{max}$ to be a sufficiently large integer such that we can write
% \begin{equation}
% f^{ren}_\alpha\approx q\left\{\sum_{\ell=0}^{\ell_{max}}\lim_{x\rightarrow z(0)}\left[\nabla_\alpha\Phi_{ret}\right]_{\ell}\right\}-q\left\{\sum_{\ell=0}^{\ell_{max}}\lim_{x\rightarrow z(0)}\left[\nabla_\alpha\Phi^{sing}\right]_\ell\right\}.
% \label{TruncatedSum}
% \end{equation}
% Because we are free to choose $\ell_{max}$ we can, in principle, find the self-force to arbitrary precision by increasing the value of $\ell_{max}$.

We will show that $f^{sing,\ell\pm}_{\alpha}$ has the form,
\begin{equation}
f^{sing,\ell\pm}_{\alpha} = \pm A_{\alpha} L +B_{\alpha}.
\label{fsing1}
\end{equation}
where $L\equiv\ell+1/2$, and $A_{\alpha}$ and $B_{\alpha}$ are constants independent of $\ell$. 
This form was obtained for a scalar charge moving on a geodesic in a Kerr background 
by Barack and Ori \cite{bo03} and for a massive particle 
(in a Lorenz gauge) by Barack \cite{barack09}.  We show here 
that the form is valid in our more general context of accelerated motion in 
a smooth, globally hyperbolic spacetime. 

Roughly speaking, functions $g$ on the sphere that diverge as $1/\theta^k$ near 
$\theta = 0$ have angular harmonics $g^\ell$ for which $\sum_{\ell=0}^{\ell_{max}}$ diverges as $\ell_{max}^k$.%
\footnote{Functions of this kind belong to 
Sobolev spaces $H_s$ with $s<0$, and the relation between the singular 
behavior of functions in $H_s$ and that of their angular harmonics is 
described in Appendix B of \cite{sf2}. }  For an expansion in 
$\epsilon$ whose leading term is $\epsilon^{-2}$, one then anticipates 
a singular field for which 
\begin{equation}
f^{sing,\ell}_\alpha= A_{\alpha} L +B_{\alpha}+C_{\alpha}L^{-1}+O(L^{-2}),
\label{f_ell1}
\end{equation} 
again with $A_{\alpha}$, $B_{\alpha}$, and $C_{\alpha}$ constants independent of $\ell$. 
The leading and subleading terms, $A_\alpha L$ and $B_\alpha$, arise from the $1/\epsilon^2$ (Coulomb) behavior of $f^{ret}_\alpha$.  A term $C_\alpha/L$ would yield a logarithmic 
divergence in the sum 
\[
   \sum_{\ell=0}^{\ell_{max}} C_\alpha/L = C_\alpha\log \ell_{max} +O(\ell_{max}^{-1});
\] 
because this would correspond to a (nonexistent) $\log\epsilon$ term in the 
short-distance expansion of $f^{ret}_\alpha$, it cannot be present.  
The argument can be made precise:\footnote{This was pointed out to 
us by Sam Gralla}  After subtracting the leading 
and subleading terms from the singular field, the remainder is defined and 
uniformly bounded everywhere on the sphere except at a point (the position of the 
particle), where it is direction-dependent.  Its angular transform is therefore 
convergent, implying that no term of the form $1/L$ can be present.  Our 
calculation in Sec.~\ref{subsubleading} below explicitly verifies that $C_\alpha=0$.
    
Finally, terms of order $\epsilon^0$ in $f^{sing}_\alpha$ (terms of order $L^{-2}$ 
or higher, including terms falling off faster than any power of $L$) could 
in principle contribute a finite term $\Delta_\alpha$, 
\be
\Delta_\alpha 
   =  \sum_{\ell=0}^{\infty} (f^{sing,\ell}_\alpha - A_{\alpha} L +B_{\alpha}).  
\ee
Following \cite{BO1}, we refer to $A_\alpha, B_\alpha, C_\alpha$ and $\Delta_\alpha$ as `regularization parameters'.%
\footnote{In \cite{BO1}, the term $\Delta_\alpha$ is written as $D_\alpha$.  Other authors, however, reserve the symbol $D_\alpha$ for the coefficient 
of $L^{-2}$ in the expansion of the singular field (see, for example \cite{Heffernan}).  We introduce $\Delta_\alpha$ to avoid confusion.} 
The goals of this section are to show that $\Delta_\alpha$ vanishes and finding the explicit form of  
$A_\alpha$ and $B_\alpha$.
  
Because Eqs.~(\ref{f_ell}) involve sums over all $m$, the values of 
$f^{ret,\ell\pm}_{\alpha}$ and $f^{sing,\ell\pm}_{\alpha}$ are 
invariant under a rotation of the $(\theta,\phi)$ coordinates.  
To evaluate them, it is convenient to choose rotated 
coordinates (that we again denote by $\theta,\phi$) for which the 
particle is on the coordinate axis, $\theta=0$ at $z(0)$.
Using $Y_{\ell m}(\theta=0,\phi)=0$  $\forall$ $m\neq 0$ and Eqs.~(\ref{SphericalHarmonicDecomp2}) and (\ref{SphericalHarmonicSumell}), we can write
\begin{equation}
f^{sing,\ell\pm}_{\alpha} \equiv\left[\nabla_\alpha\Phi^{sing}\right]^{\ell}=\lim_{r\rightarrow r_0^{\pm}}\frac{L}{2\pi}\int d\Omega P_{\ell}(\cos(\theta))\nabla_{\alpha}\Phi^{sing}.
\label{MASTEREQUATION}
\end{equation}
% Combining Eqs.~(\ref{TruncatedSum}) and (\ref{MASTEREQUATION}) we can write,
% \begin{equation}
% f^{sing}_{\alpha}=\sum_{\ell=0}^{\ell_{max}}f^{sing,\ell}_{\alpha}.
% \label{singularForce}
% \end{equation}

To calculate the regularization parameters, we use Eq.~(\ref{MASTEREQUATION}) with $\Phi^{sing}$  given by the expression in Eq.~(\ref{SingularField1}). The singular field is expressed in terms of RNCs, but the integral in Eq.~(\ref{MASTEREQUATION}) is over a sphere that can be arbitrarily large. We therefore need to extend the singular field to the entire sphere. As mentioned above, two extensions that differ by 
a smooth function with support outside a neighborhood of the particle 
do not alter the singular field.     

Because the mode-sum involves spherical harmonics associated with 
a specified coordinate system $(t,r,\theta,\phi)$, we begin by rewriting 
the short-distance expansion Eq.~(\ref{SingularField1}) as an 
expansion in terms of the coordinate distances to the particle.  
To do so, we define Cartesian coordinates $x^\mu$ (termed ``locally 
Cartesian angular coordinates'' in \cite{BO1}) associated with 
these coordinate differences by 
\be 
x^0=t, \quad x^1= x=\rho(\theta)\cos\phi\quad  x^2= y=\rho(\theta)\sin\phi,
\quad
 x^3 =r-r_0,
\label{cartesian}\ee
where $\rho(\theta)=2\sin(\theta/2)$.
In choosing these coordinates -- in particular, choosing $\rho(\theta)$ instead 
of $\sin\theta$ -- and in subsequently discarding terms of 
order $\epsilon^2$, we need to check that different choices give the 
same angular harmonic series up to convergent terms whose sum vanishes at the 
particle.   We can see that this is the case, because two choices of $\rho(\theta)$ 
that differ by terms of order $\theta^3$ and for which the corresponding 
values of $\nabla\rho$ differ by $O(\theta^2)$ give expansions of each 
component $\nabla_\alpha \Phi^{sing}$ that differ by a continuous function 
that is $O(\epsilon)$.  The difference in the angular harmonic series of 
each component $\nabla_\alpha \Phi^{sing}$ is therefore a series that 
converges to zero at the particle.  The values of the regularization 
parameters $A_\alpha$ and $B_\alpha$, regarded as vectors, depend on 
the original coordinate system $(t, r,\theta,\phi)$, but not on the 
locally Cartesian coordinates we use to evaluate them.  Their components, 
of course, depend on the choice of basis.    
 
% The particular choice of $\rho$ is useful for the explicit calculations 
% below.The function $\rho(\theta)$ is an odd function of $\theta$ assumed to be % monotonically increasing along the range $0\leq\theta\leq\pi$ and $\partial_t
% \theta\rho(\theta)|_{\theta=0}=1$. We will use the particular choice .    

The coordinates $x^\mu$ are related to RNCs $x^{\hat{\alpha}}$ by 
\begin{equation}
 x^{\hat{\alpha}}=\partial_\mu x^{\hat{\alpha}} x^\mu+\frac{1}{2}\partial_\mu x^{\hat{\alpha}}\Gamma\indices{^{\mu}_{\epsilon\nu}} x^\epsilon x^\nu+\frac{1}{6}\partial_\mu x^{\hat{\alpha}}\left(\Gamma\indices{^{\mu}_{\nu\gamma}}\Gamma\indices{^{\gamma}_{\epsilon \lambda}}+\partial_{\lambda}\Gamma\indices{^{\mu}_{\nu\epsilon}}\right) x^{\epsilon} x^{\nu} x^{\lambda}+...
\label{CoordinateTransform}
\end{equation}
When we use this relation to replace the RNCs by the coordinates $x^\mu$, the expansion Eq.~(\ref{SingularField1}) retains the same form, with  
$\hat{S}_0$, $\hat{S}_1$, and $\hat{S}_2$ replaced by quantities 
$S_0$, $S_1$, and $S_2$, where 
\begin{equation}
S_0:=q_{\mu\nu} x^\mu x^\nu,
\label{S0}
\end{equation}
\begin{equation} 
S_1:=\left(a_\lambda g_{\mu\nu}+\frac{1}{2}g_{\mu\nu,\lambda}+u_\epsilon u_\lambda \Gamma\indices{^{\epsilon}_{\mu\nu}}\right) x^{\mu} x^\nu x^\lambda=:2\zeta_{\mu\nu\lambda} x^\mu x^\nu x^\lambda,
\label{S1}
\end{equation}
with all quantities in parentheses evaluated at $z(0)$.
We will not use the explicit expression for $S_2$ 
and do not give it here because of its length; we need only the 
fact that it is a homogeneous polynomial of degree 4 in the coordinates 
$ x^\mu$.  
% As we will show, terms in $\Phi^{sing}$ do not contribute to the 
% self-force if they are form, whose numerator 
% has an even number of factors of the coordinates contributes to the self-force.

From Eq.~(\ref{SingularField1}), the singular field's leading order term is 
$O(\epsilon^{-1})$, and the leading-order term in its derivative is 
$O(\epsilon^{-2})$. Recalling Eq.~(\ref{MASTEREQUATION}), we write
\begin{equation}
f^{sing,\ell}_\alpha=f^{L,\ell}_{\alpha}+f^{SL,\ell}_{\alpha}+f^{SSL,\ell}_\alpha,
\label{fOrders}
\end{equation}
where $f^{L,\ell}_{\alpha}$, $f^{SL,\ell}_{\alpha}$, and $f^{SSL,\ell}_{\alpha}$
denote respectively the contributions to $f^{sing}_\alpha$ at leading, subleading, and sub-subleading order.  From Eq.~(\ref{SingularField1}), they are given by the following expressions, 
evaluated on the $t=0$ surface:  
\bsube\bea
f^{L,\ell}_{\alpha}&=&\lim_{r\rightarrow r_0^\pm}q\int d\Omega P_\ell(\cos(\theta))\nabla_\alpha\Phi^{sing,L},  
%=\lim_{r\rightarrow r_0^\pm}\frac{q^2L}{2\pi}\int d\Omega P_\ell(\cos(\theta))\nabla_{\alpha}\left(\frac{1}{\sqrt{S_0}}\right),
\label{LeadingOrderEq}
\\
f^{SL,\ell}_{\alpha}&=&\lim_{r\rightarrow r_0^\pm}q\int d\Omega P_\ell(\cos(\theta))\nabla_\alpha\Phi^{sing,SL},
%=\lim_{r\rightarrow r_0^\pm}\frac{q^2L}{2\pi}\int d\Omega P_\ell(\cos(\theta))\nabla_{\alpha}\left(-\frac{S_1}{2S_0^{3/2}}\right),
\label{SubLeadingOrderEq}
\\
f^{SSL,\ell}_{\alpha}&=&\lim_{r\rightarrow r_0^\pm}q\int d\Omega P_\ell(\cos(\theta))\nabla_\alpha\Phi^{sing,SSL}.
% \nonumber\\
% &&=\lim_{r\rightarrow r_0^\pm}\frac{qL}{2\pi}\int d\Omega P_\ell(\cos(\theta))\nabla_{\alpha}% \left[\Phi^{sing}-q\left(\frac{1}{\sqrt{S_0}}-\frac{S_1}{2S_0^{3/2}}\right)\right].
\label{SSLOrderEq}
\eea\label{fsingell}\esube

In the remainder of this section, we use Eq.~(\ref{fsingell}), with
$\Phi^{sing,\rm L}$, $\Phi^{sing,\rm SL}$, and $\Phi^{sing,\rm SSL}$ given by 
Eq.~(\ref{SingularField1}), to show that the large $\ell$ behavior 
of $f^{sing}_\alpha$ given in Eq.~(\ref{fsing1}) follows from the general 
character of the short-distance form of $\Phi^{sing}$, given in Eqs.~(\ref{PolyForm}) below.
We then find the explicit forms of $A_\alpha$ and $B_\alpha$.  
Denoting by $P^{(k)}(x^\mu)$ a homogeneous polynomial of 
degree $k$ in the coordinates $x^\mu$, we write the leading, subleading, 
and sub-subleading terms of $\Phi^{sing}$ in the form
\bsube\bea
\Phi^{sing,L} &=& \frac{C}{\hat S_0^{1/2}}\\
\Phi^{sing,SL} &=& \frac{P^{(3)}(x^\mu)}{\hat S_0^{3/2}}\\
\Phi^{sing,SSL} &=& \frac{P^{(6)}(x^\mu)}{\hat S_0^{5/2}}.
\label{PolyFormSSL}  
\eea\label{PolyForm} \esube 
For $\Phi^{sing,L}$ and $\Phi^{sing,SL}$, this form is explicit in 
Eq.~(\ref{SingularField1}); for $\Phi^{sing,SSL}$, terms are grouped with 
the common denominator $S_0^{5/2}$.  

That the mode-sum expression (\ref{fsing1}) holds for electromagnetic and 
gravitational perturbations will again follow from the fact that each component
of the corresponding singular fields (the singular parts of the perturbed 
vector potential and metric) satisfies Eq.~(\ref{PolyForm}).
%home

Our treatment of Eqs.~(\ref{SubLeadingOrderEq}) and (\ref{SSLOrderEq}) differs from that of Eq.~(\ref{LeadingOrderEq}). In the former cases, we are allowed to take the limit inside the integral, which simplifies the calculation. In the latter case we cannot do this. The fact that the limit and integral commute follows from the fact 
that, after one writes $d\Omega = d\theta d\phi\sin\theta$, the integrands in Eqs.~(\ref{SubLeadingOrderEq}) and (\ref{SSLOrderEq}) are bounded functions of $\theta$ and $\phi$ and are defined everywhere except at $\theta=0$.%
\footnote{The result is an immediate consequence of Lebesgue's dominated convergence theorem
(see, for example, \cite{phillips}, p. 191):
Let $\{F_n\}$ be a sequence of integrable functions that converges almost everywhere to $F$.
If $|F|< G$, for some integrable function $G$, then $F$ is integrable and 
$\int F d\mu = \lim_{n\rightarrow\infty} \int F_n d\mu$.  For functions of the 
type we consider here, a proof can also be found in \cite{BO1}.} 
We examine these subleading and sub-subleading terms before evaluating the leading term.

Thus far, we have been following the methods of Barack and Ori \cite{BO1} exactly. At this point they used properties of the Schwarschild geometry, and we rephrase the argument 
in a way that holds for a general background spacetime.

\subsection{The Sub-Sub-Leading term}
\label{subsubleading}

The sub-subleading term in the self-force is the easiest to evaluate, and we will see that 
it vanishes.  A function $\Phi^{sing,SSL}$ of the form~(\ref{PolyFormSSL}) has gradient
of the form  
\begin{equation}
\nabla_\alpha\Phi^{sing,SSL} = \frac{P_\alpha^{(7)}( x^\mu)}{S_0^{7/2}},
\label{PolyForm2}
\end{equation}
where each component $P_\alpha^{(7)}$ is a homogeneous polynomial of degree 7.
Because only polynomials in the three coordinates $x^i$, $i=1,\ldots, 3$ survive when $f^{SSL,\ell}_{\alpha}$ is evaluated on the $t=0$ surface, we have 
\begin{equation}
f^{SSL,\ell}_{\alpha}=\lim_{r\rightarrow r_0}\frac{q^2 L}{2\pi}\int d\Omega P_\ell(\cos(\theta))\frac{P_\alpha^{(7)}( x^i)}{S_0^{7/2}}.
\label{SSLOrderPolyForm}
\end{equation}

That a function of the form $P^{(k)}(x^i)/S_0^{k/2}$ is bounded follows immediately 
from the definition (\ref{S0}) of $S_0$ and the fact that the spatial part $q_{ij}$ of 
$q_{\mu\nu}$ is positive definite. As noted above, we can then interchange the order of the 
limit and integration.  To see that the integral over the sphere at $r=r_0$ vanishes, 
we use the fact that $P^{(7)}$ is odd under $I: x^\mu\rightarrow -x^\mu$, while $S_0$ 
is even (see the specific discussion in next section, after Eq.~(\ref{S_0Tilde})). 
From Eq.~(\ref{cartesian}) the restriction of $I$ to the 
$t=0$, $r=r_0$ sphere is the map $\phi\rightarrow\phi+\pi$, implying that 
the sphere itself and the measure $d\Omega$ are invariant under $I$.    
Because the integrand is odd under $I$ and $d\Omega$ is invariant, 
the integral vanishes.  

\subsection{The Subleading Term}

The subleading term of Eq.~(\ref{SubLeadingOrderEq}), 
\be
  f^{SL,\ell}_{\alpha}  
=\lim_{r\rightarrow r_0^\pm}\frac{q^2L}{2\pi}\int d\Omega P_\ell(\cos(\theta))\nabla_{\alpha}\left(-\frac{S_1}{2S_0^{3/2}}\right), 
\ee
is more singular 
than the sub-subleading term by an additional power of $S_0^{1/2}$ in its denominator. 
It has the form
\be
f_\alpha^{SL,\ell} =\lim_{r\rightarrow r_0^\pm}\frac{q^2L}{2\pi}  \int d\theta d\phi \sin\theta P_\ell(\cos\theta)\frac{ P_\alpha^{(2n)}(x^i)}{S_0^{n+1/2} },
\label{fSL}\ee  

To compute the explicit form of $f_\alpha^{SL,\ell}$ and to see that 
$\dis\sin\theta \frac{ P_\alpha^{(2n)}(x^i)}{S_0^{n+1/2} }$ is bounded, we 
begin by noting that, restricted to the $r=r_0, t=0$ sphere, $P_\alpha^{(2n)}$ and 
$S_0$ are given by 
\begin{equation}
 \left. P_\alpha^{(2n)}(x^i)\right|_{r=r_0} =\rho(\theta)^{2n}\left(\sum_{m=0}^{2n}a_{\alpha,m}\sin^m\phi \cos^{2n-m}\phi\right),
\label{SSLNumerator}
\end{equation}
where $a_{\alpha,m}$ is a constant; and 
\begin{equation}
  \tilde{S}_0:= \left.S_0\right|_{r=r_0}  =\rho(\theta)^2\left(q_{xx}\cos(\phi)^2+q_{yy}\sin(\phi)^2\right),
\label{S_0Tilde}
\end{equation}
where we have used the fact that,    
with our rotated $\theta,\phi$ coordinates, $q_{xy}=0$. In effect, this is exactly what Barack and Ori \cite{BO1} do for Schwarzschild, choosing their coordinates such that $u_y=0$, and then relying on the diagonal form of the metric to make $q_{xy}=0$. Then, because the 
eigenvalues of $q_{IJ}$, $I,J=1\ldots 2$, are positive definite, 
$S_0$ can be written as
\begin{equation}
\tilde{S}_0=\rho(\theta)^2 q_{yy}\left(1+\beta^2\cos^2\phi\right),
\label{S_0Tilde2}
\end{equation}
where 
\begin{equation}
\beta^2:=\frac{q_{xx}-q_{yy}}{q_{yy}}.
\label{beta^2}
\end{equation}
From Eqs.~(\ref{SSLNumerator}) and (\ref{S_0Tilde2}), it follows that $S_0^{n+1/2}$ has 
one more power of $\rho(\theta)$ than $P_\alpha^{(2n)}$ and hence that the integrand, 
$\sin\theta P_\ell(\cos\theta)P_\alpha^{(2n)}S_0^{-(n+1/2)}$, is bounded.

We can therefore again bring the limit inside the integral in 
Eq.~(\ref{fSL}).  Substituting the expressions (\ref{SSLNumerator}) and 
(\ref{S_0Tilde2}) for $P_\alpha^{(2n)}$ and $\tilde S_0$ in Eq.~(\ref{fSL}),
we have  
\begin{equation}
f^{SL,\ell}_\alpha 
  =\frac{q^2 L}{2\pi{ q_{yy}}^{n-1/2}}
    \int_0^\pi d\theta \sin\theta \frac{P_\ell (\cos(\theta))}{\rho(\theta)} 
   \sum_{m=0}^{2n}\int_0^{2\pi}\frac{\left(a_{\alpha,m}\sin^m\phi \cos^{2n-1-m}\phi\right)}
              {\left(1+\beta^2\cos^2\phi\right)^{(2n-1)/2}}d\phi.
\label{SSLPolyForm2}
\end{equation}
The integral over $\theta$ has the value
\begin{equation}
\int_0^\pi d\theta \sin\theta \frac{P_\ell (\cos(\theta))}{\rho(\theta)}=\int_0^{\pi} d\theta \sin\theta \frac{P_{\ell}(\cos(\theta))}{\sqrt{2-2\cos(\theta)}}=\frac{1}{L}, 
\label{Bterm2}
\end{equation}
implying $f^{SL,\ell}_{\alpha}$ is independent of $\ell$:   
\begin{equation}
f^{SL,\ell}_{\alpha}=B_{\alpha}. 
\label{Bterm3}
\end{equation}
The integration over $\phi$ involves the  complete elliptic integrals 
\be
  E(w) = \int_0^{\pi/2} (1-w\sin^2\phi)^{1/2}d\phi,\qquad K(w) = \int_0^{\pi/2} (1-w\sin^2\phi)^{-1/2}d\phi,   
\ee
where 
\be
   w := \frac{\beta^2}{1+\beta^2}.
\ee
After a straightforward computation, we find   
\begin{equation}
B_{\alpha}=\frac{2q^2}{3\pi(1+\beta^2)^{3/2}\beta^4q_{yy}^{5/2}}\left(B_{\alpha}^{(E)}E(w)+B_{\alpha}^{(K)}K(w)\right),
\label{Balpha}
\end{equation}
where 
\bsube
\begin{eqnarray}
B_{\alpha}^{(E)}&= &(1+\beta^2)(2+\beta^2)\Lambda_{\alpha XXYY}-2\left[(1+2\beta^2)\Lambda_{\alpha xxxx}+ (1+\beta^2)^2(1-\beta^2)\Lambda_{\alpha yyyy}\right],
\label{BalphaE}
\\
B_{\alpha}^{(K)}&= & (2+3\beta^2)\Lambda_{\alpha xxxx}\nonumber\\
       & &+(1+\beta^2)\left[(2-\beta^2)\Lambda_{\alpha yyyy}-2\Lambda_{\alpha XXYY}\right],
\label{BalphaK}
\end{eqnarray}\esube
with the quantities $\Lambda_{\alpha\beta\gamma\delta\epsilon}$ given in terms of $\zeta_{\beta\gamma\delta}$ of Eq.~(\ref{S1}) by
\begin{equation}
\Lambda_{\alpha\beta\gamma\delta\epsilon}:=3\zeta_{(\alpha\beta\gamma)}q_{\delta\epsilon}-3\zeta_{\beta\gamma\delta}q_{\alpha\epsilon},
\label{LambdaTensor}
\end{equation}
%where $\zeta_{\beta\gamma\delta}$ is given in Eq.~(\ref{S1}) as $\zeta_{\beta\gamma\delta}=1/4\left(2a_\beta g_{\gamma\delta}+g_{\gamma\delta,\beta}+2u_\epsilon u_\beta \Gamma\indices{^{\epsilon}_{\gamma\delta}}\right)$. 
and we define the $\Lambda_{\alpha XXYY}$ as follows;
\begin{eqnarray}
\Lambda_{\alpha XXYY}=\Lambda_{\alpha xxyy}+\Lambda_{\alpha xyxy}+\Lambda_{\alpha xyyx}+x\leftrightarrow y.
\label{LambdaAlphaNScalar}
\end{eqnarray}

In summary, we have shown that the angular harmonic decomposition of the subleading term has only 
a $B$ term, a term independent of $\ell$, whose explicit form is given by Eqs.~(\ref{Balpha})-(\ref{LambdaTensor}). 

These parameters agree with those of Barack and Ori for Schwarzschild \cite{BO1}, and also 
with Warburton and Barack \cite{Warburton} and \cite{Warburton2} in Kerr. 
(In particular note the equivalence of our Eq.~(\ref{LambdaTensor}) with Eqs.~(B5), (B6) 
and (B7) of \cite{Warburton}).

\subsection{Leading Term}
Finally, we turn to the leading term $f^{L,\ell}_\alpha$.  
From Eq.~(\ref{LeadingOrderEq}) and the relation 
$\nabla_\alpha S_0=2 q_{\alpha\beta} x^\beta$, we have 
% \begin{equation}
% f^{L,\ell}_{\alpha\pm}=-\frac{1}{2}\lim_{r\rightarrow r_0^\pm}\frac{L}{2\pi}q^2\int d\Omega P_\ell(\cos(\theta))\frac{\nabla_\alpha S_0}{S_0^{3/2}},
% \label{A1}
% \end{equation}
% which, after we recognize that, allows us to write
\begin{equation}
f^{L,\ell}_{\alpha\pm}=-\frac{L}{2\pi}q^2 q_{\alpha\beta}\tilde{F}^{\beta\ell}_\pm,
\label{A2}
\end{equation}
where
\begin{equation}
\tilde{F}^{\beta\ell}_\pm=\lim_{r \rightarrow r_0^{\pm}}\int d\Omega P_\ell(\cos(\theta))\frac{x^\beta}{S_0^{3/2}}.
\label{Ftilde}
\end{equation}

Because we are working on a $t=0$ surface, we have $\tilde{F}^{0\ell}_\pm =0$.  
To evaluate $\tilde{F}^{i\ell}_\pm$, we follow Barack and Ori \cite{BO1}, dividing the 
$r=$ constant sphere that constitutes the domain of integration into two parts:
the coordinate square ${\cal S}_\epsilon$ for which $|x|<\epsilon$ and 
$|y|<\epsilon$ (some $\epsilon < \pi/2$); and the rest of the sphere, $S^2\backslash{\cal S}_\epsilon$.
The domains are chosen to be symmetric under a rotation 
by $\pi$ about $\theta=0$.  
 
On $S^2\backslash{\cal S}_\epsilon$, the integrand is smooth, and we can 
bring the limit inside the integral, writing 
\beaa
 \lim_{r \rightarrow r_0^{\pm}}\int_{S^2\backslash{\cal S}_\epsilon} d\Omega P_\ell(\cos(\theta))\frac{x^i}{S_0^{3/2}}&=& \int_{S^2\backslash{\cal S}_\epsilon} d\Omega P_\ell(\cos(\theta)) \frac{x^i}{\tilde S_0^{3/2}}.
\eeaa
We immediately see that the contribution to the radial component $\tilde{F}^{1\ell}_\pm$ 
vanishes.  The remaining $x$ and $y$ components of the integral vanish because 
the domain of integration and the function $\tilde S_0$ are invariant 
under a rotation by $\pi$ about $\theta = 0$, while $x$ and $y$ change sign.   

The only contribution to $\tilde{F}^{\beta\ell}_\pm$ is then from the 
integral over ${\cal S}_\epsilon$.  Because $\epsilon$ is arbitrary,
the value of the integral is independent of $\epsilon$, 
determined only by the singular behavior of the integrand at 
$\theta =0$.  
To evaluate the integral, we change integration variables from $(\theta,\phi)$ to 
$(x,y)$.  From Eq.~(\ref{cartesian}), the Jacobian of the transformation is  
\begin{equation}
\frac{\partial(\theta,\phi)}{\partial({x,y})}=\sin\theta,
\label{Jacobian2}
\end{equation}
and we have
\begin{equation}
\tilde{F}^{i\ell}_\pm =\lim_{r\rightarrow r_0^\pm}
		\int_{{\cal S}_\epsilon} dx dy 
		P_\ell(\cos\theta) \frac{x^i}{S_0^{3/2}}
		= \lim_{r\rightarrow r_0^\pm}
		\int_{-\epsilon}^\epsilon  dx \int_{-\epsilon}^\epsilon dy 
		P_\ell(\cos\theta) \frac{x^i}{S_0^{3/2}}.
\label{Ftildexy1}
\end{equation}
Because $P_\ell(\cos\theta)$ differs from its value at $\theta=0$ only 
at $O(\theta^2)$, replacing $P_\ell$ by $1$ does not alter the leading 
singular behavior of the integrand and should therefore not change 
the value of the integral.  To verify this, we write  
\be 
P_\ell(\cos\theta) = 1+h(\theta)\sin^2\theta, 
\ee
where $h$ is smooth on ${\cal S}_\epsilon$.  We then have 
 \beaa
\tilde{F}^{i\ell}_\pm &=& \lim_{r\rightarrow r_0^\pm} 
\int_{{\cal S}_\epsilon} dx dy \frac{x^i}{S_0^{3/2}}  
+ \int_{{\cal S}_\epsilon} dx dy
  \lim_{r\rightarrow r_0^\pm}  \left(h\sin^2\theta \frac{x^i}{S_0^{3/2}}\right)
\equiv (\lim_{r\rightarrow r_0^\pm}  I_1^i) + I_2^i,
\label{Ftildexy2}
\eeaa
where we have used the fact that the function $h\sin^2\theta\ x^i/S_0^{3/2}$ 
is bounded to bring the limit inside the second integral, $I_2^i$. 
Then $I_2^i$ has the form   
\be
I_2^i = \int_{{\cal S}_\epsilon} dx dy \left(h\sin^2\theta \frac{x^i}{\tilde S_0^{3/2}}\right).
\label{eq:I2i}
\end{equation} 
Again the vanishing of $I_2^r$ is immediate, and the symmetry argument 
we have now used twice implies that the remaining components also vanish:  That is, 
from the invariance of ${\cal S}_\epsilon$ and $h\sin^2\theta/\tilde S_0^{3/2}$ 
under a $\pi$ rotation, together with the fact that $x$ and $y$ change sign, 
we have $I_2^x = I_2^y = 0$.  

We are now left with 
\begin{equation}
\tilde{F}^{i\ell}_{\pm}=\lim_{r\rightarrow r_0^\pm}\int_{{\cal S}_\epsilon}\frac{ x^i}{S_0^{3/2}}d x d y.
\label{Ftilde4}
\end{equation} 
We can already see that this integral is independent of $L$, because $P_\ell$ has 
been replaced by $1$. It immediately follows from Eq.~(\ref{A2}) 
that $f^{L,\ell}_\alpha$ is proportional to $L$, and {\em we have thus established 
our central claim, that the singular part of the self-force has the  
form given in Eq. (\ref{eq:fsingAB})}. 

Finally, we evaluate $\tilde{F}^{i\ell}_{\pm}$ to find the explicit form of $A_\alpha$.
We begin by showing that the $x$- and $y$-components can be expressed in terms of the 
third spatial component $\tilde{F}^{r\ell}_{\pm}$.      
From the definition (\ref{S0}) of $S_0$, we have 
\be
\partial_x \frac1{S_0^{1/2}} = -\frac {q_{xx} x + q_{xr}(r-r_0)}{S_0^{3/2}},
\ee
and the $x$-component of Eq.~(\ref{Ftilde4}) takes the form
\be 
\tilde{F}^{x\ell}_{\pm} =-\frac1{q_{xx}}\lim_{r\rightarrow r_0^\pm}\int_{{\cal S}_\epsilon}\left[\partial_x \frac1{S_0^{1/2}} +\frac{q_{xr}}{S_0^{3/2}}(r-r_0) \right]d x d y.
\ee
Using $\dis \int_{-\epsilon}^\epsilon dx\ \partial_x S_0^{-1/2} = 0$, we have
\be
 \tilde{F}^{x\ell}_{\pm}= -\frac{q_{xr}}{q_{xx}}\lim_{r\rightarrow r_0^\pm}\int_{{\cal S}_\epsilon}\frac{r-r_0}{S_0^{3/2}} d x d y 
 = -\frac{q_{xr}}{q_{xx}}\tilde{F}^{r\ell}_{\pm}, 
\label{Fx}\ee
as claimed.  Similarly, 
\be 
\tilde{F}^{x\ell}_{\pm} =-\frac{q_{yr}}{q_{yy}}\tilde{F}^{r\ell}_{\pm}.
\label{Fy}\ee   

To evaluate $\tilde{F}^{r\ell}_{\pm}$, we introduce as integration variables 
\be
  X= \frac x{r-r_0}, \quad Y = \frac y{r-r_0}.
\ee
With $e:\epsilon/(r-r_0)$, we have
\bea
\tilde{F}^{r\ell}_{\pm} &=& \lim_{e\rightarrow \infty}
  \int_{-e}^e d X \int_{-e}^e d Y[q_{xx}X^2 + 2q_{xr}X +q_{yy}Y^2 + 2q_{yr}Y +q_{rr}]^{-3/2}\nonumber\\
&=&  \pm 2\pi(q_{xx} q_{yy} q_{rr} - q_{yy} q_{xr}^2 - q_{xx} q_{yr}^2)^{-1/2}.
\label{Ftilde6}\eea

Finally, using $f^{L,\ell}_{\alpha\pm}=A_\alpha L$, together with Eqs.~(\ref{A2}), (\ref{Fx}), (\ref{Fy}) and (\ref{Ftilde6}), we obtain 

\begin{equation}
A_{\alpha\pm}=\mp q^2 \ 
 \frac{ q_{\alpha r}-q_{\alpha x}q_{xr}/q_{xx}
              -q_{\alpha y} q_{yr}/q_{yy} }
           {(q_{xx} q_{yy} q_{rr} - q_{yy} q_{xr}^2 - q_{xx} q_{yr}^2)^{1/2}}
.
\label{Aalpha}
\end{equation}

It is worth noting that this agrees with the form given in \cite{BO1} and also has the same property that $u_{\alpha}A^{\alpha}=0$.

Thus, as claimed, the regularization parameters for the self force on a point scalar charge 
moving along an arbitrary trajectory through a generic spacetime are given by 
$A_{\alpha}L+B_{\alpha}$, with the terms for a logarithmic divergence ($C_\alpha L^{-1}$) 
and a finite remainder ($\Delta_{\alpha}$) both vanishing. We have given the explicit forms 
of the regularization parameters in the `locally Cartesian angular coordinates,' in 
Eqs.~(\ref{Aalpha}) and (\ref{Balpha}). Their values for the original coordinate system 
are given in Appendix C.

It is important to note that we have recovered the regularization parameters for  $f^{sing,\ell}_{\alpha}$, whose values are not (necessarily) trivially related to those for $f^{sing,\ell,\alpha}$. For now we will just claim that the parameters for the raised indices, the regularization parameters have the form, $A^{\alpha} L + B^{\alpha}$, and postpone the proof to the end of next section, where we can discuss it in the context of extending the four velocity away from the world-line. 

We now turn to the regularization parameters for electromagnetism and gravity.

\section{Generalizing to higher spins}
\label{HigherSpins}

In this section, to distinguish an electromagnetic vector potential from the regularization parameter $A^\alpha$, we use a different font, 
denoting the vector potential by $\texttt{A}^\alpha$.  

We will see that, in a Lorenz gauge, each Cartesian component of the vector potential $\texttt{A}^{\alpha}$ of an electric point charge and of the metric perturbation $h_{\alpha\beta}$ of a point mass has a short-distance expansion similar to that of the field of a scalar charge. We show that this similarity of form implies that the 
mode-sum expression for the singular part of the retarded field is again 
given by Eq.~(\ref{fsing1}).  In particular the term $\Delta_{\alpha}$ again vanishes.
We again rely on the Hadamard expansion of the Green's functions as laid out 
in \cite{Poisson}.

\subsection{Electromagnetic Self-Force}
  In a Lorenz gauge, the electromagnetic vector potential $\texttt{A}^{\alpha}$ of 
a point charge $e$ satisfies
\begin{equation}
\nabla^{\beta}\nabla_{\beta}\texttt{A}^{\alpha}-R\indices{^{\alpha}_{\beta}}\texttt{A}^{\beta}=-4\pi j^{\alpha}, \quad \nabla_{\alpha}\texttt{A}^{\alpha}=0, 
\label{EDfieldEq.}
\end{equation}
with current density 
\begin{equation}
j^{\alpha}(x)=eu^\alpha(x)\int\delta^{(4)}(x,z(\tau))d\tau.
\label{4current}
\end{equation}
The solution to Eq.~(\ref{EDfieldEq.}) has components $\texttt{A}^\mu$ in a global coordinate system given by
\begin{equation}
\texttt{A}_{adv/ret}^{\mu}(x)=\int\left[G^{\mu}_{\nu'}(x,x')\right]_{adv/ret}j^{\nu'}(x')\sqrt{-g}d^4x',
\label{EDsolution}
\end{equation}
where each Green's function satisfies the equation
\begin{equation}
\nabla^{\gamma}\nabla_{\gamma}G\indices{^{\alpha}_{\hat\beta}}(x,x')-R\indices{^{\alpha}_{\beta}}G\indices{^{\beta}_{\beta'}}(x,x')=-4\pi \delta^\alpha_{\beta'}\delta^{(4)}(x,x').
\label{EDGreen'sFunctionEQ}
\end{equation}
Unprimed and primed indices are tensor indices at $x$ and $x'$, respectively, 
and the covariant derivatives are with respect to $x$.   

The expansion of the Green's function in the normal neighborhood $C$ is analogous to that of the scalar field, having the form \cite{Poisson}
\begin{equation}
G\indices{^{\alpha}_{\beta'}}(x,x')=\Theta(x,x')\left[U\indices{^{\alpha}_{\beta'}}(x,x')\delta(\sigma)-V\indices{^{\alpha}_{\beta'}}(x,x')\theta(-\sigma)\right],
\label{LocalEDExpansion}
\end{equation}
where the bi-tensors $U\indices{^{\alpha}_{\beta'}}(x,x')$ and $V\indices{^{\alpha}_{\beta'}}(x,x')$ have in RNC the local expansions
\begin{equation}
U\indices{^{\hat\alpha}_{\hat\beta}}(x,x')
=\delta\indices{^{\hat\alpha}_{\hat{\beta}}}+\frac{1}{12}\left[2R\indices{^{\hat\alpha}_{\hat{\gamma}\hat{\beta}\hat{\delta}}}+\delta^{\hat\alpha}_{\hat\beta}R_{\hat{\gamma}\hat{\delta}}\right]
y^{\hat\gamma}y^{\hat\delta}+O(\epsilon^3)
\label{EDUexpand}
\end{equation}
and
\begin{equation}
V\indices{^{\hat\alpha}_{\hat\beta}}=\frac{1}{2}\left(R\indices{^{\hat\alpha}_{\hat\beta}}-\frac{1}{6}\delta^{\hat\alpha}_{\hat\beta}R\right)+O(\epsilon).
\label{EDVexpand}
\end{equation}
In these expansions, each tensor is evaluated at the point $x'$.

The same steps we followed for the scalar field now give for each component 
of $\texttt{A}^{\alpha}$ essentially the same form as that of the scalar field 
in Eq.~(\ref{Phiretadv}), namely
\begin{equation}
\texttt{A}^{\alpha}_{adv/ret}
 =e\left[\frac{U\indices{^{\alpha}_{\beta'}}u^{\beta'}}{\dot{\sigma}}\right]_{adv/ret}\mp e\lim_{h\rightarrow0^+}\int_{\pm\infty}^{\tau_{adv/ret}\pm h} u^{\nu'}\left[G\indices{^{\alpha}_{\nu'}}\right]_{adv/ret}d\tau.
\label{EDsolution2}
\end{equation}
The force has the formal expression 
\begin{equation}
f^{\alpha}_{EM}=-\nabla_{\beta}T^{\alpha\beta}_{EM}=F^{\alpha\beta}j_{\beta},
\label{EDForce}
\end{equation}
where $F_{\mu\nu}=\nabla_\mu\texttt{A}_{\nu}-\nabla_\nu\texttt{A}_{\mu}$, and the 
expression for the singular part of the force is given in terms of the 
singular part of the vector potential by 
\begin{equation}
f^{sing,\alpha}_{EM}=e u^{\beta}g^{\alpha\sigma}\left[\nabla_{\sigma}\texttt{A}^{sing}_{\beta}-\nabla_{\beta}\texttt{A}^{sing}_{\sigma}\right],
\label{EDSingForce}
\end{equation}
where components of the metric and 4-velocity are evaluated at the position of the particle.  

% Now, we have the form of the solution to the retarded or advanced fields in Eq.~(\ref% {EDsolution2}), and we know that we need to only consider the first derivatives of % 
% this solution to calculate the force. We now show the same for gravity.

\subsection{Gravitational Self-Force}
\label{gravitational}

The test-particle limit of the trajectory of a massive particle moving 
in a curved spacetime is a geodesic.  To consistently compute the self-force 
on a massive particle whose trajectory is accelerated in the test-particle limit, 
one must include whatever additional fields are responsible for the acceleration.
In this section, we find the formal contribution from gravity to the self-force 
on a particle in a generic vacuum spacetime, showing that the form (\ref{fsing1}) holds 
in this general context.    

As noted in the introduction, we find that the expression can be used in a 
mode-sum regularization of a particle with scalar or electromagnetic charge 
moving in a spacetime with a background scalar or electromagnetic field, 
when one works to linear order in the mass and charge (with a fixed ratio $q/m$).  That is, we assume that renormalization can again 
be accomplished by an angle average of the retarded field together with a mass renormalization.  From a short-distance expansion of 
the retarded field, we then obtain the singular field and find that the regularization coefficients have the form $A_\alpha L + B_\alpha$ with $A_\alpha$ the sum of the coefficients for gravity and electromagnetism obtained here, and with $B_\alpha$ having an 
additional contribution from the coupling of the two fields. 

Returning to the task of this section, finding the regularization parameters for gravity, we will write the spacetime metric as  $\tilde{g}_{\alpha\beta}=g_{\alpha\beta}+h_{\alpha\beta}$, where $\tilde{g}_{\alpha\beta}$ is the total metric, $g_{\alpha\beta}$ is the background metric, and $h_{\alpha\beta}$ is the perturbation. We will restrict our discussion to 
background metrics $g_{\alpha\beta}$ that satisfy the vacuum Einstein equation. We 
raise and lower indices with the background metric $g_{\alpha\beta}$ and denote 
by $\nabla_{\alpha}$ the covariant derivative operator of $g_{\alpha\beta}$.

With $\gamma_{\alpha\beta}:=h_{\alpha\beta}-\frac12 g_{\alpha\beta} h$, the 
Lorenz gauge condition is $\nabla_{\alpha}\gamma^{\alpha\beta}=0$, and the linearized Einstein equation has the form
\begin{equation}
\nabla_{\mu}\nabla^{\mu}\gamma^{\alpha\beta}+2R\indices{_{\gamma}^{\alpha}_{\delta}^{\beta}}\gamma^{\gamma\delta}=-16\pi T^{\alpha\beta}.
\label{linearizedEinstein}
\end{equation}
Here, $T^{\alpha\beta}$ is the stress energy tensor of a point particle of mass $m$, 
given by
\begin{equation}
T^{\alpha\beta}=m u^\alpha u^\beta\int\delta^{(4)}\left(x'-z(\tau)\right)d\tau.
\label{perturbedstressenergy}
\end{equation}

As before, we write the solution to the field equation (in this case, Eq.~(\ref{linearizedEinstein})) in terms of a Green's function,
\begin{equation}
\gamma^{\alpha\beta}=4\int G\indices{^{\alpha\beta}_{\gamma'\delta'}}(x,x')T^{\gamma'\delta'}\sqrt{-g'}d^4x',
\label{GravityGreen'sfunction1}
\end{equation}
where $G\indices{^{\alpha\beta}_{\gamma'\delta'}}(x,x')$ satisfies
\begin{equation}
\nabla_{\mu}\nabla^{\mu}G\indices{^{\alpha\beta}_{\gamma'\delta'}}(x,x') +2{{{R_{\gamma}}^{\alpha}}_{\delta}}^{\beta}G\indices{^{\gamma\delta}_{\gamma'\delta'}}(x,x')=-4\pi g\indices{^{(\alpha}_{\gamma'}}g\indices{^{\beta)}_{\delta'}} \delta^{4}(x,x').
\label{GravityGreen'sFunction2}
\end{equation}
As in the spin-0 and spin-1 cases, the Green's function,
$ G\indices{^{\alpha\beta}_{\gamma'\delta'}}(x,x')$, has the form
\begin{equation}
 G\indices{^{\hat\alpha\hat\beta}_{\hat\gamma\hat\delta}}(x,x')=\Theta(x,x')\left[U\indices{^{\hat\alpha\hat\beta}_{\hat\gamma\hat\delta}}(x,x')\delta(\sigma)-V\indices{^{\hat\alpha\hat\beta}_{\hat\gamma\hat\delta}}(x,x')\theta(-\sigma)\right],
\label{GravityHadamard}
\end{equation}
where the bitensors $U\indices{^{\alpha\beta}_{\gamma'\delta'}}$ and $V\indices{^{\alpha\beta}_{\gamma'\delta'}}$ have, in RNC about $x$, the expansions \cite{Poisson}
\bsube\bea
U\indices{^{\hat\alpha\hat\beta}_{\hat\gamma\hat\delta}}(x,x')
   &=&\delta^{(\hat\alpha}_{\hat{\gamma}} \delta^{\hat{\beta})}_{\hat{\delta}}
	+\frac{1}{3}\delta\indices{^{(\hat{\alpha}}_{(\hat{\gamma}}}R\indices{_{\hat{\delta})\hat{\sigma}}^{\hat{\beta})}_{\hat{\mu}}}x^{\hat\sigma}x^{\hat\mu}+O(\epsilon^3),
\label{UGravExpand}\\
V\indices{^{\hat\alpha\hat\beta}_{\hat\gamma\hat\delta}}(x,x')
  &=&R\indices{_{\hat{\gamma}}^{(\hat{\alpha}\hat{\beta})}_{\hat{\delta}}}+O(\epsilon).
\label{VGravExpand}
\eea\esube
When we evaluate the perturbation using Eq.~(\ref{GravityGreen'sfunction1}), we find
\begin{equation}
{\gamma^{\alpha\beta}}_{adv/ret}=4m\left[\frac{u^{\gamma'}u^{\delta'}U\indices{^{\alpha\beta}_{\gamma'\delta'}}}{\dot{\sigma}}\right]_{adv/ret}\mp 4m\lim_{h\rightarrow0^+}\int_{\pm\infty}^{\tau_{adv/ret}\pm h} u^{\gamma'}u^{\delta'}\left[G\indices{^{\alpha\beta}_{\gamma'\delta'}}\right]_{adv/ret}d\tau.
\label{PerturbationSolution}
\end{equation}

Now, solving the perturbed geodesic equation allows us to write
\begin{equation}
f^{\alpha,sing}_{GR}=-m\left(g^{\alpha\delta}+u^{\alpha}u^{\delta}\right)\left(\nabla_{\beta}h^{sing}_{\gamma\delta}-\frac{1}{2}\nabla_{\delta}h^{sing}_{\beta\gamma}\right)u^{\beta}u^{\gamma}.
\label{selfforceGravity}
\end{equation}

Therefore, just as for the scalar charge in Eqs.~(\ref{Phiretadv}) and (\ref{RetAdvForce}), and as for the electric charge in Eqs.~(\ref{EDsolution2}) and (\ref{EDSingForce}), we have expressed the metric perturbation in Eq.~(\ref{PerturbationSolution}) and the expression for the force in Eq.~(\ref{selfforceGravity}).

\subsection{The vanishing C and $\Delta$ terms}

We will now argue that the $C_{\alpha}$ and $\Delta_{\alpha}$ terms vanish in both the electrodynamic and gravitational self-force regularization (computed in a Lorenz gauge). In the scalar case, the singular field was 
given by Eq.~(\ref{DWSingField}) reproduced here
\begin{equation*}
 \Phi^{sing}=\frac12 q\left[\left(\frac{U(x,z)}{\dot{\sigma}}\right)_{ret}
			+\left(\frac{U(x,z)}{\dot{\sigma}}\right)_{adv}\right]
		+q\frac12 \int^{\tau_{adv}}_{\tau_{ret}} V d\tau +O(\epsilon^2).
\end{equation*}

To produce the singular vector potential $\texttt{A}^{\alpha}$ and metric perturbation, $\gamma^{\alpha\beta}$, we will use the same procedure, taking the averaged sum of the advanced 
and retarded solutions of the function ``$U$" (Eqs.~(\ref{EDUexpand}) and (\ref{UGravExpand})) over $\dot{\sigma}$, and add to it the leading order contribution from the function ``$V$'' (Eqs.~(\ref{EDVexpand}) and (\ref{VGravExpand})).

Now, before we continue, it is useful to recall the properties of $\Phi^{sing}$ which we noted gave rise to the vanishing $C_{\alpha}$ and $\Delta_{\alpha}$ terms for a scalar charge. In particular,
\begin{equation*}
\Phi^{sing}=\frac{1}{\sqrt{S_0}}+\frac{\zeta_{\alpha\beta\gamma}x^{\alpha}x^{\beta}x^{\gamma}}{S_0^{3/2}}+\frac{\Lambda_{\alpha\beta\gamma\delta\epsilon\sigma}x^{\alpha}x^{\beta}x^{\gamma}
x^{\delta}x^{\epsilon}x^{\sigma}}{S_0^{5/2}}+O(\epsilon^2).
\end{equation*}

The $A_{\alpha}$ term came from the gradient of the leading term, and the $B_{\alpha}$ term came from the gradient of the sub-leading term. When we considered the mode-sum decomposition of the sub-sub-leading term, we noted that its contribution vanished because, to borrow the expression from Quinn, it had an odd number of unit normal vectors.

Therefore, our aim will be to demonstrate that both $\texttt{A}^{\alpha}_{sing}$ and $\gamma^{\alpha\beta}_{sing}$ (and therefore also $h^{\alpha\beta}_{sing}$) have the same form as $\Phi^{sing}$, and that this will lead to $C_{\alpha}=\Delta_{\alpha}=0$.% We will show it explicitly for the electric charge and then state the result for the point mass.

By comparing with Eq.~(\ref{DWSingField}), we notice that the only term from the integral over ``$V$" will be the leading order term in the expansion multiplied by $\sqrt{S_0}$. That is, the Detweiler-Whiting piece for electromagnetism and gravity will just take on the form of $\sqrt{S_0}\times Const$. The gradient of this term will just give an odd polynomial in $x$ divided by $\sqrt{S_0}$, and so its mode-sum decomposition vanishes. 

Having established this, we can focus on the direct piece of the singular field. The only qualitatively new feature that the arises in the direct part of the field is the presence of the four velocity in the numerator. Consider the explicit expression for the four velocity at the retarded or advanced times:
\begin{equation}
u^{\alpha}_{ret/adv}=u^{\alpha} +a^{\alpha}(\tau_1+\tau_2+...)+\frac{1}{2} \dot{a}^{\alpha}(\tau_1+\tau_2+...)^2+...
\label{uretadv}
\end{equation}

By using Eqs.~(\ref{t12}) and (\ref{t2}), we can rewrite $u^{\alpha}_{ret/adv}$ in terms of the coordinates of $x$ as
\bea
u^{\hat{\alpha}}_{ret/adv}=u^{\hat{\alpha}}-a^{\hat{\alpha}}u_{\hat{\mu}}x^{\hat{\mu}}+\left[a^{\hat{\alpha}}a_{\hat{\mu}}u_{\hat{\nu}}+\frac{1}{2}\dot{a}^{\hat{\alpha}}(q_{\hat{\mu}\hat{\nu}}+u_{\hat{\mu}}u_{\hat{\nu}})\right]x^{\hat{\mu}}x^{\hat{\nu}} \nonumber\\
\pm\left[\frac{x^{\hat{\gamma}}}{2}(a^{\hat{\alpha}}a_{\hat{\gamma}}(q_{\hat{\mu}\hat{\nu}}+u_{\hat{\mu}}u_{\hat{\nu}})+2\dot{a}^{\hat{\alpha}}q_{\hat{\mu}\hat{\nu}}u_{\hat{\gamma}})-a^{\hat{\alpha}}q_{\hat{\mu}\hat{\nu}}
 \right]\frac{x^{\hat{\mu}}x^{\hat{\nu}}}{\sqrt{\hat{S}_0}}
\label{uretadv2}
\eea
%Therefore, we can write $u^{\alpha}_{ret/adv}$ in the form,
%\begin{equation}
%u^{\alpha}_{ret/adv}={ _{(0)}P^{{\alpha}}}+ {_{(1)}P^{\alpha}_{\mu} x^{\mu}}+ {_{(2)}P^{\alpha}_{\mu\nu}x^{\mu}x^{\nu}} \pm\frac{ _{(2)}\bar{P}^{\alpha}_{\mu\nu}x^{\mu}x^{\nu}}{\sqrt{S_0}}\pm\frac{ _{(3)}\bar{P}^{\alpha}_{\mu\nu\gamma}x^{\mu}x^{\nu}x^{\gamma}}{\sqrt{S_0}}+O(\epsilon^3).
%\label{uretadvForm}
%\end{equation}

Now, if we turn to $U\indices{^{\alpha}_{\beta}}$ in Eq.~(\ref{EDUexpand}), and we note that to leading order $y^\alpha=x^{\alpha}-u^{\alpha}\tau_1$, we can write

\bea
U\indices{^{\hat{\alpha}}_{\hat{\beta}}}=\delta\indices{^{\hat{\alpha}}_{\hat{\beta}}}+\frac{\left(-2R\indices{^{\hat{\alpha}}_{(\hat{\gamma}\hat{\delta})\hat{\beta}}} +R_{\hat{\gamma}\hat{\delta}}\delta\indices{^{\hat{\alpha}}_{\hat{\beta}}}\right)}{12}
\left[\delta\indices{^{\hat{\gamma}}_{\hat{\mu}}}\delta\indices{^{\hat{\delta}}_{{\hat{\nu}}}} +u^{\hat{\gamma}}u^{\hat{\delta}}(q_{{\hat{\mu}}{\hat{\nu}}}+u_{{\hat{\mu}}} u_{{\hat{\nu}}})+2u^{\hat{\gamma}}\delta\indices{^{\hat{\delta}}_{{\hat{\nu}}}}u_{\mu}\right]x^{{\hat{\mu}}}x^{{\hat{\nu}}}\nonumber\\
\pm\frac{u^{\hat{\gamma}}\left(-2R\indices{^{\hat{\alpha}}_{(\hat{\gamma}\hat{\delta}){\hat{\beta}}}} +R_{\hat{\gamma}\hat{\delta}}\delta\indices{^{\hat{\alpha}}_{\hat{\beta}}}\right)}{6}\left[u^{\hat{\delta}}u_{{\hat{\nu}}}x^{{\hat{\nu}}}+x^{\hat{\delta}}\right]\sqrt{S_0}
\label{EDUexpand2}
\eea

%Notice, this can also be written in the form  
%\begin{equation}
%[U\indices{^{\alpha}_\beta}]_{ret/adv}={ _{(0)}P\indices{^{\alpha}_\beta}}+ {_{(1)}P\indices{^{\alpha}_{\beta\mu}} x^{\mu}}+ {_{(2)}P\indices{^{\alpha}_{\beta\mu\nu}}x^{\mu}x^{\nu}} \pm\frac{ _{(2)}\bar{P}\indices{^{\alpha}_{\beta\mu\nu}}x^{\mu}x^{\nu}}{\sqrt{S_0}}\pm\frac{ _{(3)}\bar{P}\indices{^{\alpha}_{\beta\mu\nu\gamma}}x^{\mu}x^{\nu}x^{\gamma}}{\sqrt{S_0}}+O(\epsilon^3).
%\label{UEDForm}
%\end{equation}

%In this case, ${_{(1)}P\indices{^{\alpha}_{\beta\mu}}}= { _{(2)}\bar{P}\indices{^{\alpha}_{\beta\mu\nu}}}=0$. By performing the exact same steps, it can be seen that $U^{\alpha\beta}_{\gamma'\delta'}$ in Eq.~(\ref{UGravExpand}) can be written in the same form as Eq.~(\ref{UEDForm}), where each of the $P$s have been given two additional indices, to account for the index structure of $U^{\alpha\beta}_{\gamma'\delta'}$. 
%\begin{multline}

Using Eqs.~(\ref{uretadv2}) and (\ref{EDUexpand2}), we obtain
\bea
\left[U\indices{^{\hat{\alpha}}_{\hat{\beta}}}u^{\hat{\beta}}\right]_{ret/adv}=u^{\hat{\alpha}}-a^{\hat{\alpha}}u_{\hat{\gamma}} x^{\hat{\gamma}}+\left[a^{\hat{\alpha}}u_{\hat{\mu}} a_{\hat{\nu}} +\frac{\dot{a}^{\hat{\alpha}}}{2}(q_{{\hat{\mu}}{\hat{\nu}}}+u_{\hat{\mu}} u_{\hat{\nu}})\right]x^{\hat{\mu}} x^{\hat{\nu}}+\nonumber\\
\frac{(u^{\hat{\alpha}}R_{{\hat{\gamma}}{\hat{\delta}}}-2R\indices{^{\hat{\alpha}}_{({\hat{\gamma}}{\hat{\delta}}){\hat{\beta}}}}u^{\hat{\beta}})}{12}\left[\delta\indices{^{\hat{\gamma}}_{\hat{\mu}}}\delta\indices{^{\hat{\delta}}_{\hat{\nu}}} +u^{\hat{\gamma}}u^{\hat{\delta}}(q_{{\hat{\mu}}{\hat{\nu}}}+u_{{\hat{\mu}}} u_{\hat{\nu}})+2u^{\hat{\gamma}}\delta\indices{^{\hat{\delta}}_{\hat{\nu}}}u_{{\hat{\mu}}}\right]x^{\hat{\mu}}x^{{\hat{\nu}}}\nonumber\\
\pm\frac{x^{\hat{\mu}} x^{\hat{\nu}}}{\sqrt{S_0}}\left[-a^{\hat{\alpha}}q_{{\hat{\mu}}{\hat{\nu}}}+
\frac{x^{\hat{\gamma}}}{6}\left(3a^{\hat{\alpha}}(q_{{\hat{\mu}}{\hat{\nu}}}+u_{\hat{\mu}} u_{\hat{\nu}})a_{\hat{\gamma}}+6 \dot{a}^{\hat{\alpha}}u_{\hat{\gamma}} q_{{\hat{\mu}}{\hat{\nu}}}+(u^{\hat{\alpha}}R_{{\hat{\epsilon}}{\hat{\sigma}}}u^{\hat{\sigma}}-2 R\indices{^{\hat{\alpha}}_{({\hat{\epsilon}}{\hat{\sigma}}){\hat{\beta}}}}u^{\hat{\beta}}u^{\hat{\sigma}})q\indices{^{\hat{\epsilon}}_{\hat{\gamma}}}q_{{\hat{\mu}}{\hat{\nu}}}
\right)\right]\nonumber\\
\label{EDUdotu}
\eea

Now, recalling Eq.~(\ref{sigmadotsquared}) we can write the direct piece of the electromagnetic vector potential,
\bea
\left[\frac{U\indices{^{\hat{\alpha}}_{\hat{\beta}}}u^{\hat{\beta}}}{\dot{\sigma}}\right]
=\frac{u^{\hat{\alpha}}}{\sqrt{S_0}}\left[1-\frac{S_1}{2S_0}+\frac{3S_1^2}{8S_0^2}-\frac{S^{(1)}_2}{2S_0}-\frac{R_{{\hat{\mu}}{\hat{\nu}}{\hat{\epsilon}}{\hat{\delta}}}x^{\hat{\mu}}u^{\hat{\nu}}x^{\hat{\epsilon}}u^{\hat{\delta}}x^2}{6 S_0}\right]- \frac{a^{\hat{\alpha}}u_{\hat{\mu}}x^{\hat{\mu}}}{\sqrt{S_0}}\left(1-\frac{S_1}{2S_0}\right)
\nonumber\\+
\frac{\left[2a^{\hat{\alpha}}u_{\hat{\mu}} a_{\hat{\nu}} +\dot{a}^{\hat{\alpha}}(q_{{\hat{\mu}}{\hat{\nu}}}+u_{\hat{\mu}} u_{\hat{\nu}})\right]x^{\hat{\mu}} x^{\hat{\nu}}}{2\sqrt{S_0}}\nonumber\\
\frac{(u^{\hat{\alpha}}R_{{\hat{\gamma}}{\hat{\delta}}}-2u^{\hat{\beta}} R\indices{^{\hat{\alpha}}_{({\hat{\gamma}}{\hat{\delta}}){\hat{\beta}}}})}{12\sqrt{S_0}}\left[\delta\indices{^{\hat{\gamma}}_{\hat{\mu}}}\delta\indices{^{\hat{\delta}}_{\hat{\nu}}} +u^{\hat{\gamma}}u^{\hat{\delta}}(q_{{\hat{\mu}}{\hat{\nu}}}+u_{{\hat{\mu}}} u_{\hat{\nu}})+2u^{\hat{\gamma}}\delta\indices{^{\hat{\delta}}_{\hat{\nu}}}u_{\hat{\mu}}\right]x^{\hat{\mu}}x^{\hat{\nu}}\pm \frac{a^{\hat{\alpha}} S_1}{2S_0}\mp u^{\hat{\alpha}}\frac{S_2^{(\pm)}}{2S_0^{3/2}} 
\nonumber\\
\pm\frac{x^{\hat{\mu}} x^{\hat{\nu}}}{S_0}\left[-a^{\hat{\alpha}}q_{{\hat{\mu}}{\hat{\nu}}}+
\frac{x^{\gamma}}{6}\left(3a^{\alpha}(q_{{\hat{\mu}}{\hat{\nu}}}+u_{\hat{\mu}} u_{\hat{\nu}})a_{\hat{\gamma}}+6 \dot{a}^{\hat{\alpha}}u_{\hat{\gamma}} q_{{\hat{\mu}}{\hat{\nu}}}+(u^{\hat{\alpha}}R_{{\hat{\epsilon}}{\hat{\sigma}}}u^{\hat{\sigma}}-2 R\indices{^{\hat{\alpha}}_{({\hat{\epsilon}}{\hat{\sigma}})\hat{\beta}}}u^{\hat{\sigma}}u^{\hat{\beta}})q\indices{^{\hat{\epsilon}}_{\hat{\gamma}}}q_{{\hat{\mu}}{\hat{\nu}}}
\right)\right] ,
\nonumber\\
\label{Aretadvfull}
\eea
where we have decomposed $S_2$ into two pieces, $S_2^{(1)}$, which does not change sign when switching from retarded to advanced times, and $S_2^{(\pm)}$, which does.

In the average of the retarded and advanced fields, the contribution from each term in Eq.~(\ref{Aretadvfull}) preceded by $\pm$ vanishes, so we can write the singular vector potential as,
\bea
\frac1e\texttt{A}^{\hat{\alpha}}_{sing}=\frac{u^{\hat{\alpha}}}{\sqrt{S_0}}\left[1-\frac{S_1}{2S_0}+\frac{3S_1^2}{8S_0^2}-\frac{S^{(1)}_2}{2S_0}-\frac{R_{{\hat{\mu}}{\hat{\nu}}{\hat{\epsilon}}{\hat{\delta}}}x^{\hat{\mu}}u^{\hat{\nu}}x^{\hat{\epsilon}}u^{\hat{\delta}}x^2}{6 S_0}\right]- \frac{a^{\hat{\alpha}}u_{\hat{\mu}}x^{\hat{\mu}}}{\sqrt{S_0}}\left(1-\frac{S_1}{2S_0}\right)
\nonumber\\+
\frac{(u^{\hat{\alpha}}R_{{\hat{\gamma}}{\hat{\delta}}}-2u^{\hat{\beta}} R\indices{^{\hat{\alpha}}_{({\hat{\gamma}}{\hat{\delta}}){\hat{\beta}}}})}{12\sqrt{S_0}}\left[\delta\indices{^{\hat{\gamma}}_{\hat{\mu}}}\delta\indices{^{\hat{\delta}}_{\hat{\nu}}} +u^{\hat{\gamma}}u^{\hat{\delta}}(q_{{\hat{\mu}}{\hat{\nu}}}+u_{{\hat{\mu}}} u_{\hat{\nu}})+2u^{\hat{\gamma}}\delta\indices{^{\hat{\delta}}_{\hat{\nu}}}u_{\hat{\mu}}\right]x^{\hat{\mu}}x^{\hat{\nu}}\nonumber\\
+
\frac{\left[2a^{\hat{\alpha}}u_{\hat{\mu}} a_{\hat{\nu}} +\dot{a}^{\hat{\alpha}}(q_{{\hat{\mu}}{\hat{\nu}}}+u_{\hat{\mu}} u_{\hat{\nu}})\right]x^{\hat{\mu}} x^{\hat{\nu}}}{2\sqrt{S_0}}+\frac{6 R\indices{^{\hat{\alpha}}_{\hat{\beta}}}u^{\hat{\beta}}-u^{\hat{\alpha}}R}{12}\sqrt{S_0}.
\nonumber\\
\label{AsingularFull}
\eea
The last term is the `Detweiler and Whiting term' which is $V(0)\sqrt{S_0}$. The first term of Eq.~(\ref{AsingularFull}) is just $u^{\alpha}$ multiplied by the scalar field. The second term contains a sub-leading contribution, which is of the form of a linear term divided by $\sqrt{S_0}$. As we saw in the previous section, terms of this type provide a `B' term. All of the rest of the terms (the final term from the first line, the total of the second line, and the beginning of the third) are terms of order $\epsilon$, and have the form of a polynomial of even degree in $x$ divided by $S_0$ raised to a half integer power. Therefore, the derivatives of these terms will give us polynomials of odd integer powers in the numerator, and thus their mode-sum decomposition will vanish.

Therefore, the regularization parameters for the electrodynamic self-force will also be of the form $A^{\alpha}L+B^{\alpha}$. It is also worth noting that the regularization parameters for the electrodynamic vector potential will just be of the form $B'^{\alpha}$ (we use $B'$ to indicate that this will not have the same value as the $B$ for the self-force).

When we apply the same procedure to Eq.~(\ref{UGravExpand}) and solve for the retarded and advanced $\gamma_{\alpha\beta}$, we find
\bea
\frac{1}{m}\gamma^{ret/adv}_{{\hat{\alpha}}{\hat{\beta}}}=\frac{4u_{\hat{\alpha}}u_{\hat{\beta}}}{\sqrt{S_0}}\left[1-\frac{S_1}{2S_0}+\frac{3S_1^2}{8S_0^2}-\frac{S^{(1)}_2}{2S_0}-\frac{R_{{\hat{\mu}}{\hat{\nu}}{\hat{\epsilon}}{\hat{\delta}}}x^{\hat{\mu}}u^{\hat{\nu}}x^{\hat{\epsilon}}u^{\hat{\delta}}x^2}{6 S_0}\right]-
 8\frac{ u_{({\hat{\beta}}} a_{{\hat{\alpha}})} u_{\hat{\mu}}x^{\hat{\mu}}}{\sqrt{S_0}}\left(1-\frac{S_1}{2S_0}\right)
\nonumber\\
+\frac{4x^{\hat{\mu}}x^{\hat{\nu}}}{\sqrt{S_0}}\left[(a_{\hat{\alpha}}a_{\hat{\beta}} +\dot{a}_{({\hat{\alpha}}}u_{{\hat{\beta}})})(q_{{\hat{\mu}}{\hat{\nu}}}+u_{\hat{\mu}} u_{\hat{\nu}})+2a_{({\hat{\alpha}}}u_{{\hat{\beta}})}a_{{\hat{\mu}}}u_{\hat{\nu}} -\frac{u_{({\hat{\alpha}}}R_{{\hat{\beta}}){\hat{\epsilon}}{\hat{\delta}}{\hat{\sigma}}} u^{{\hat{\delta}}}}{3}(\delta\indices{^{\hat{\epsilon}}_{\hat{\mu}}}\delta\indices{^{\hat{\sigma}}_{\hat{\nu}}}+u^{\hat{\epsilon}}\delta\indices{^{\hat{\sigma}}_{\hat{\mu}}}u_{{\hat{\nu}}})
\right]\nonumber\\
\pm 8 u_{({\hat{\alpha}}}a_{{\hat{\beta}})}\left(1-\frac{S_1}{2S_0}\right)
\pm\frac{8x^{\hat{\mu}}x^{\hat{\nu}}x^{\hat{\delta}}}{S_0}\left[(a_{{\hat{\alpha}}}a_{\hat{\beta}}+\dot{a}_{({\hat{\alpha}}}u_{{\hat{\beta}})})u_{\hat{\delta}} q_{{\hat{\mu}}{\hat{\nu}}}-a_{({\hat{\alpha}}}u_{{\hat{\beta}})}a_{\hat{\delta}}(q_{{\hat{\mu}}{\hat{\nu}}}+u_{\hat{\mu}} u_{\hat{\nu}} )\right]\mp 2\frac{u_{\hat{\alpha}}u_{{\hat{\beta}}} S^{(\pm)}_2}{S_0^{3/2}}
\nonumber\\
-4 u^{\hat{\mu}} u^{\hat{\nu}} R_{{\hat{\mu}}({\hat{\alpha}}{\hat{\beta}}){\hat{\nu}}}(\sqrt{S_0}\mp u_{\hat{\mu}} x^{\hat{\mu}})\nonumber\\
\label{gammaretadvFull}
\eea

Therefore, we can write the singular, trace-reversed, metric perturbation as
\bea
\frac{1}{m}\gamma^{sing}_{{\hat{\alpha}}{\hat{\beta}}}=\frac{4u_{\hat{\alpha}}u_{\hat{\beta}}}{\sqrt{S_0}}\left[1-\frac{S_1}{2S_0}+\frac{3S_1^2}{8S_0^2}-\frac{S^{(1)}_2}{2S_0}-\frac{R_{{\hat{\mu}}{\hat{\nu}}{\hat{\epsilon}}{\hat{\sigma}}}x^{\hat{\mu}}u^{\hat{\nu}}x^{\hat{\epsilon}}u^{{\hat{\sigma}}}x^2}{6 S_0}\right]-
 8\frac{ u_{({\hat{\beta}}} a_{{\hat{\alpha}})} u_{\hat{\mu}}x^{\hat{\mu}}}{\sqrt{S_0}}\left(1-\frac{S_1}{2S_0}\right)+
\nonumber\\
\frac{4x^{\hat{\mu}}x^{\hat{\nu}}}{\sqrt{S_0}}\left[(a_{{\hat{\alpha}}}a_{\hat{\beta}} +\dot{a}_{({\hat{\alpha}}}u_{{\hat{\beta}})})(q_{{\hat{\mu}}{\hat{\nu}}}+u_{\hat{\mu}} u_{\hat{\nu}})+2a_{({\hat{\alpha}}}u_{{\hat{\beta}})}a_{\hat{\mu}}u_{\hat{\nu}} -\frac{u_{({\hat{\alpha}}}R_{{\hat{\beta}}){\hat{\epsilon}}\gamma{\hat{\sigma}}} u^{\gamma}}{3}(\delta\indices{^{\hat{\epsilon}}_{\hat{\mu}}}\delta\indices{^{\hat{\sigma}}_{\hat{\nu}}}+u^{\hat{\epsilon}}\delta\indices{^{\hat{\sigma}}_{\hat{\mu}}}u_{\hat{\nu}})
\right]\nonumber\\-4u^{\hat{\mu}} u^{\hat{\nu}} R_{{\hat{\mu}}({\hat{\alpha}}{\hat{\beta}}){\hat{\nu}}}\sqrt{S_0}.
\nonumber\\
\label{gammasingFull}
\eea

When we write $h^{sing}_{\alpha\beta}=\gamma_{\alpha\beta}-1/2 g_{\mu\nu}\gamma^\mu_\mu$, we will only introduce one new term, which will be from $R_{\alpha\gamma\beta\delta}x^{\gamma} x^\delta (S_0^{-1/2})$ (from the Riemann tensor correction to the metric multiplied by the leading order term in $\gamma^{\mu}_\mu$). This is an order $\epsilon$ term already in the form of an even polynomial divided by $S_0$ to a half integer power. Therefore, looking at the singular field in Eq.~(\ref{gammasingFull}), we can see that it also has the exact same algebraic form as the scalar and electrodynamic singular fields.
Therefore, the mode-sum decomposition of the derivative of this field will have the form $A_{...} L + B_{...}$ (where the `...' represent suppressed indices). However, for the gravitational self-force, we also have terms that are linear in the field, not just the derivatives (since $\nabla\rightarrow\partial+\Gamma$).

When we expand the Cristoffel symbols, then we will get $\Gamma \rightarrow \Gamma_0 +\partial_\mu \Gamma x^\mu+O(\epsilon^2)$. When we apply this to the singular metric perturbation, (recalling that the sub-sub-leading terms vanish, since they are already of order $\epsilon^1$), we will have something of the form
\begin{equation*}
\Gamma^{...}_{...}\gamma_{...}=\frac{C^{(1)}_...}{\sqrt{S_0}}+\frac{C^{2}_{...\mu\nu\delta} x^{\mu}x^{\nu}x^{\delta}}{S_0^{3/2}}+O(\epsilon).
\end{equation*}
The $C^{(n)}_{...}$ are constants. The mode-sum decomposition of the first term gives us a piece independent of $L$ ( another `$B$' term), and the second term is an odd polynomial divided by $S_0$ to an odd integer power, and so this term's mode-sum decomposition vanishes.
 
The expressions for the self-force in an electromagnetic or gravitational context depend on how one extends $g^{\alpha\beta}[z(0)]$ and $u^\alpha[z(0)]$ to a neighborhood of the particle (and there is even this ambiguity in how one defines the scalar self-force with raised indices). If we return to the definition of the scalar, electromagnetic, or gravitational self-force, (Eqs.~(\ref{frendef}), (\ref{EDSingForce}) or (\ref{selfforceGravity}), then we can rewrite them as
\begin{eqnarray}
f^{s=0,sing,\mu}&=&k^{\mu\nu}\nabla_{\nu}\Phi^{sing}=g^{\mu\nu}\nabla_{\nu}\Phi^{sing}\nonumber
\\
f^{s=1,sing,\mu}&=&k\indices{^{\mu\alpha\beta}}\nabla_{\beta}\texttt{A}^{sing}_{\alpha}= \left(\delta^{\mu\beta}u^{\alpha}-\delta^{\mu\alpha}u^{\beta}\right)\nabla_{\beta}\texttt{A}^{sing}_{\alpha}
\nonumber\\
f^{s=2,sing,\mu} &=&k\indices{^{\mu\beta\gamma\delta}}\nabla_{\beta}\gamma_{\gamma\delta}^{sing}=\frac{\left(q^{\beta\mu}\left(q^{\gamma\delta}+u^{\gamma}u^{\delta}\right)-4q^{\delta\mu}u^{\beta}u^{\gamma}\right)}{4}\nabla_{\beta}\gamma^{sing}_{\gamma\delta}.
\label{HigherSpinsForceGeneral}
\end{eqnarray}
In particular, the quantities $k^{\mu...}$ are only properly defined on the trajectory of the particle for $s=1,2$, and we are allowed a choice in how we extend $k^{\mu...}$ away from the world line. One popular way is to use the `fixed extension' \cite{barack09}, in which one defines $k^{\mu...}(x\neq z(0))=k^{\mu...}(x=z(0))$, and is the one we use in this paper, but other choices are available \cite{BO2}. We now show that as long as $k^{\mu...}$ is a smooth function in $x$ then the regularization parameters retain the form $A_\alpha L +B_\alpha$.

Since each component of $\texttt{A}^{sing}_{\alpha}$ and $\gamma^{sing}_{\alpha\beta}$ has the same form as $\Phi^{sing}$, we will consider finding the regularization parameters for $f^{s=0,sing,\mu}$. Denote by $k^{\mu\nu}_{0}$, $\partial_{\gamma} k^{\mu\nu}_0$, and $\partial_{\delta}\partial_{\gamma}k^{\mu\nu}_0$ the values of $k^{\mu\nu}$ and its derivatives at $z(0)$.   For an extension $k^{\mu\nu}[x]$ of 
$k^{\mu\nu}[z(0)]$ the departure of $k^{\mu\nu}\nabla_\nu\Phi^{sing} $ from $k^{\mu\nu}_0\nabla_\nu\Phi^{sing} $
is given by 
\bea
(k^{\mu\nu}-k^{\mu\nu}_0) \nabla_{\nu}\Phi^{sing} 
&=& x^\gamma \partial_k^{\mu\nu}g_0 \nabla_\nu\phi^{sing,L} 
 +\left(x^\gamma\partial_\gamma k^{\mu\nu}_0 \nabla_\nu \phi^{sing,SL} 
+ \frac12 x^\gamma x^\delta \partial_\gamma\partial_\delta k^{\mu\nu} \nabla_\nu \phi^{sing,L}   
\right)\nonumber
\\
& & +O(\epsilon).
\label{SmoothTimesDelPhi}
\eea

The first term on the right has the form $P^{(4)}(x^\mu) S_0^{-5/2}$, and it thus
gives a correction to the $B$ term. The term in parentheses on the right is order 
unity and has the form $P^{(7)}(x^\mu) S_0^{-7/2}$; its contribution to the 
$f^{SSL,\ell}$, given by its contribution to the integral on the right side of 
Eq.~(\ref{SSLOrderPolyForm}) therefore vanishes. Because the remaining part of 
the right side of (\ref{SmoothTimesDelPhi}) is $O(\epsilon)$, its contribution 
to the $f^{sing}_\alpha$ also vanishes.

Therefore, we have demonstrated our claim in Eq.~(\ref{eq:fsingAB}). In doing so, we have shown that to regularize the fields themselves, one needs only subtract of a `$B$' term from the mode-sum of the retarded field, which is to say, for a field $\psi$, $\psi^{sing,\ell}_{...}=B_{...}$. We give the explicit values of the self-force regularization parameters in Appendix A, and the expressions for these parameters in the original coordinate system in Appendix C.

It is important to note that even though the higher order terms in the expansion of the singular field do not contribute to the entire mode-sum, they do contribute mode by mode. That is to say, that when we perform the infinite sum over all modes, the higher order terms vanish, but they do not vanish mode by mode. As we noted in a previous footnote, these terms are important for increasing the speed of convergence in computations \cite{Heffernan}.

\section{The Radial Self-Force on a Static Scalar Charge Outside of a Schwarzschild Black Hole}
\label{RadialForce}

We will write the Schwarzschild metric in the usual manner,
\begin{equation}
ds^2=g_{\mu\nu}dx^\mu dx^\nu=-\left(1-\frac{2M}{r}\right)dt^2+\frac{dr^2}{1-\frac{2M}{r}}+r^2d\Omega^2,
\label{SchildMetric}
\end{equation}
where $M$ is the mass of the black hole. We will now proceed to calculate the regularization parameters for a static charge outside this black hole.

Given a static charge outside a Schwarzschild black hole, we can place the particle at the pole, allowing us to write, $z^{\alpha}(0)=(0,r_0,0,0)$. From the normalization of the four-velocity, $u_{\mu}u^{\mu}=-1$ we can write the four velocity as
\begin{equation}
u_{\mu}=-\left(\sqrt{1-\frac{2M}{r_0}},0,0,0\right).
\label{fourvelocity}
\end{equation}
The four acceleration is given by $a^\mu\equiv u^\nu\nabla_\nu u^\mu$, allowing us to write
\begin{equation}
a_{\mu}=\left(0,\frac{M}{r_0^2-2M r_0},0,0\right).
\label{fouracceleration}
\end{equation}

To use the results of Eqs.~(\ref{Balpha}) and (\ref{Aalpha}), we need to write the line element in terms of our LCACs. This is trivially done for the static charge in Schwarzschild, for which $g_{\delta t\delta t}=g_{tt}$, $g_{\delta r\delta r}=g_{rr}$, and $g_{xx}=g_{yy}=r_0^2$.

With this information we can use Eq.~(\ref{Aalpha}) to write
\begin{equation}
A_{r\pm}L=\mp\frac{Lq^2}{r_0^2\sqrt{1-\frac{2M}{r_0}}}.
\label{ArSchild}
\end{equation}

To calculate the $B_r$ term, we can use Eq.~(\ref{Balpha}), but, since $\beta^2\equiv g_{yy}^{-1}(g_{xx}-g_{yy}+u_x^2)=0$ it is actually easier to return to the integral in Eq.~(\ref{SubLeadingOrderEq}) reproduced below,
\begin{equation}
B_{\alpha}=\lim_{\delta r\rightarrow 0^{\pm}}\frac{q^2 L}{2\pi}\int d\Omega P_{\ell}(\cos(\theta))\left(\frac{3}{4}\frac{S_1\nabla_{\alpha}S_0}{S_0^{5/2}}-\frac{1}{2}\frac{\nabla_{\alpha}S_1}{S_0^{3/2}}\right).
\label{BalphaIntegral}
\end{equation}

From Eqs.~(\ref{S0}), (\ref{S1}), (\ref{fourvelocity}), and (\ref{fouracceleration}), we have
\begin{equation}
\begin{array}{cc}
S_0=\frac{\delta r^2}{1-\frac{2M}{r_0}}+r_0^2\rho(\theta)^2\\
S_1=r_0\delta r \rho(\theta)^2\left[1+\frac{M}{r_0}\frac{1}{1-\frac{2M}{r_0}}\right].
\label{SchildS0S1}
\end{array}
\end{equation}

Since we can bring the limit inside the integral in Eq.~(\ref{BalphaIntegral}), $B_t$, $B_x$. and $B_y$ all vanish, and $B_r$ has the form
\begin{equation}
B_{r}=-\frac{q^2}{2r_0^2}\left[1+\frac{M}{r_0}\frac{1}{1-\frac{2M}{r_0}}\right].
\label{BrSchild}
\end{equation}

Eqs.~(\ref{ArSchild}) and (\ref{BrSchild}) match the corresponding Eqs.~(10.17a) and (10.17b) in Casals et. al. \cite{CasalsEtAl}.

\subsection{The Retarded Field}

From Wiseman \cite{Wiseman}, the retarded field has the form
\begin{equation}
\phi_{ret}=q\frac{\sqrt{1-\frac{2M}{r_0}}}{\sqrt{(r-M)^2-2(r-M)(r_0-M)\cos(\theta)+(r_0-M)^2-M^2\sin^2(\theta)}},
\label{Wiseman}
\end{equation}

Using the relation
\begin{equation}
\frac{1}{\sqrt{a^2+b^2-2abx-(1-x^2)}}=\sum_{\ell=0}^\infty(2\ell+1)P_{\ell}(x)P_{\ell}(a)Q_{\ell}(b),
\label{LegendreTrick}
\end{equation}
we can write Eq.~(\ref{Wiseman}) as
\begin{equation}
\phi_{ret}=\frac{q}{M}\sqrt{1-\frac{2M}{r_0}}\sum_{\ell=0}^{\infty}2L P_{\ell}(\cos(\theta)) P_{\ell}\left(\frac{r_{<}}{M}-1\right)Q_{\ell}\left(\frac{r_{>}}{M}-1\right),
\label{SchildRet}
\end{equation}
where $Q_\ell$ is a Legendre function of the second kind, $r_{>}=\max(r,r_0)$, and $r_<=\min(r,r_0)$.

Now, to make the comparison of the mode sum decomposition of the retarded field and of the singular field easier, we define
\begin{equation}
\begin{array}{cc}
F_{+,\ell}:=\left[\frac{q}{2}(\partial_{r_>}+\partial_{r_<})\phi_{ret}\right]_{r=r_0}\\
F_{-,\ell}:=\left[\frac{q}{2}(\partial_{r_>}-\partial_{r_<})\phi_{ret}\right]_{r=r_0}
\label{Symmetrized}
\end{array}
\end{equation}

Thus, if we want to know the force from the retarded field for $r>r_0$, then we can write $F_{>,\ell}=F_{+,\ell}+F_{-,\ell}$. Similarly $F_{<,\ell}=F_{+,\ell}-F_{-,\ell}$. Notice that this suggests we can write
\begin{equation}
\begin{array}{cc}
F_{+,\ell}=B+F^{(self)}_{\ell}\\
F_{-,\ell}=A_+ L.
\label{FpmAB}
\end{array}
\end{equation}

The anti-symmetric term is the easier one to calculate. Using the formula for the Wronskian of $P_\ell$ and $Q_\ell$ given in Eq.~(8.1.9) in Abromowitz and Stegun \cite{AS}, we can write
\begin{equation}
F_{-\ell}=-\frac{q^2L}{r_0^2\sqrt{1-\frac{2M}{r_0}}}=A_+L.
\label{Fminus}
\end{equation}
Thus, we have seen that our formula in Eq.~(\ref{Aalpha}) has successfully found the leading order term for the static charge in Schwarzschild.

Now, we can write the symmetric term, $F_{+,\ell}$ as
\begin{equation}
F_{+,\ell}=\frac{q^2}{2M}\sqrt{\frac{b-M}{b+M}}\partial_b\left[(2\ell+1)P_{\ell}\left(\frac{b}{M}\right)Q_{\ell}\left(\frac{b}{M}\right)\right],
\label{Fplus}
\end{equation}
where $b\equiv r_0-M$. Now, we will rewrite the $B_r$ term given in Eq.~(\ref{BrSchild}) in a similar form, yielding
\begin{equation}
B_r=\frac{q^2}{2}\sqrt{\frac{b-M}{b+M}}\partial_b\left[\frac{1}{\sqrt{b^2-M^2}}\right].
\label{Br2}
\end{equation}

In the limit that $\ell_{max}\rightarrow\infty$, we can write
\begin{equation}
f^{self}_{r}=\sum_{\ell=0}^{\ell_{max}}\left[(F_{+,\ell}+F_{-,\ell})-(A_+L+B_r)\right]= \sum_{\ell=0}^{\ell_{max}}\left[(-F_{+,\ell}+F_{-,\ell})-(A_-L+B_r)\right].
\label{ScSF}
\end{equation}

Since we showed that the $F_{-,\ell}$ term is exactly canceled by the $A$ term, we can just focus on the difference between the $F_{+,\ell}$ and the $B_r$ term. From Wiseman's result, \cite{Wiseman}, we know that the self-force vanishes. Therefore we need to show that,

\begin{equation}
\frac{q^2}{2}\sqrt{\frac{b-M}{b+M}}\partial_b\sum_{\ell=0}^{\infty}\left[\frac{(2\ell+1)}{M}P_{\ell}\left(\frac{b}{M}\right)Q_{\ell}\left(\frac{b}{M}\right)-\frac{1}{\sqrt{b^2-M^2}}\right]=0.
\label{ExpansionEquals0}
\end{equation}

We will demonstrate that this sum vanishes for the first several orders in $M<<1$. From Arfken \cite{Arfken}
we have the following two equations,
\begin{equation}
P_{\ell}(x)=\sum_{k=0}^{\ell/2}\frac{(-1)^k}{2^n}\frac{(2\ell-2k)!}{(\ell-k)!(\ell-2k)!}\frac{x^{\ell-2k}}{k!},
\label{Pell}
\end{equation}
and
\begin{equation}
Q_{\ell}(x)=2^{\ell}\sum_{s=0}^{\infty}\frac{(\ell+s)!(\ell+2s)!}{(2\ell+2s+1)!s!}x^{-2s-\ell-1}.
\label{Qell}
\end{equation}

Using these equations we can expand each $F_{+,\ell}$ mode in powers of $M$. Doing this, we find
\begin{multline}
\frac{(2\ell+1)}{M}P_{\ell}\left(\frac{b}{M}\right)Q_{\ell}\left(\frac{b}{M}\right)=
\frac{1}{b}+\frac{M^2}{2b^3}\left[1+\frac{1}{(2\ell+3)(2\ell-1)}\right]+\\
\frac{3}{8}\frac{M^4}{b^5}\left[1+\frac{2}{(2\ell+3)(2\ell-1)}+\frac{11}{(2\ell+5)(2\ell+3)(2\ell-1)(2\ell-3)}\right]+O\left(M^6\right).
\label{Mexpand}
\end{multline}

If we focus on the $\ell$ independent pieces of Eq.~(\ref{Mexpand}), then we notice that they are exactly the expansion of $(b^2-M^2)^{-1/2}$. The $\ell$ dependent pieces are each part of a vanishing sum (see Appendix B). Therefore, these terms vanish upon summation, leaving us with no force, as we expect from \cite{Wiseman}.

\section{Conclusions}

For scalar and electromagnetic charges we have extended to accelerated 
trajectories and generic smooth spacetimes and coordinate systems the 
mode-sum regularization developed 
previously for geodesic orbits in a Kerr or Schwarzschild 
background.  In this broader arena and for massive particles 
on geodesic trajectories in generic spacetimes, the singular 
behavior of the retarded self-force in a Lorenz or smoothly 
related gauge retains the form 
\begin{equation*}
 f^{sing,\ell\pm}_\alpha = \pm A_\alpha (\ell+1/2) + B_\alpha,
\end{equation*}
and we have obtained expressions for the regularization parameters 
$A_\alpha$ and $B_\alpha$.

\begin{acknowledgments}
We are indebted to Sam Gralla for helpful conversations and to an anonymous referee for 
corrections and comments.    
This work was supported in part by NSF Grants PHY 1001515, PHY 1307429, and PHY 0970074.  
\end{acknowledgments}
\appendix
\section{The Regularization Parameters for Higher Spins}
\label{RegularizationParametersforHigherSpins}

Here we write the explicit regularization parameters for the self-force on a point electric charge and a point mass (computed in a Lorenz gauge). We directly parallel the approach taken for the scalar charge.

\subsection{Electromagnetic Regularization Parameters}

Until the final equation of this section, we set the charge $e$ to 1. 
 
We begin by writing Eq.~(\ref{AsingularFull}), but we keep only the leading and sub-leading terms
\begin{equation}
\texttt{A}^{sing}_{\hat{\alpha}}=\frac{u_{\hat{\alpha}}}{\sqrt{\hat{S}_0}}-\frac{\left[u_{\hat{\alpha}}\zeta_{\hat{\gamma}\hat{\delta}\hat{\epsilon}} +a_{\hat{\alpha}}u_{\hat{\gamma}}\left(\eta_{\hat{\epsilon}\hat{\delta}}+u_{\hat{\epsilon}}u_{\hat{\delta}}\right)\right] x^{\hat{\epsilon}}x^{\hat{\delta}}x^{\hat{\gamma}}}{\hat{S}_0^{3/2}}.
\label{ASingInRNCs}
\end{equation}
We now transform to our curvilinear coordinates, $v_{\alpha}=\partial_{\alpha}x^{\hat{\mu}}v_{\hat{\mu}}$. Expanding about the position of the particle (which is the origin of both our RNC and our locally Cartesian angular coordinates), we have
\begin{eqnarray}
\partial_{\alpha}x^{\hat{\mu}}=\left(\partial_{\alpha}x^{\hat{\mu}}\right)_0+\left(\partial_{\delta}\partial_{\alpha} x^{\hat{\mu}}\right)_0 x^{\delta}+O(x^2)\nonumber\\
\partial_{\alpha}x^{\hat{\mu}}=\left(\partial_{\alpha}x^{\hat{\mu}}\right)_0+\left(\partial_{\epsilon} x^{\hat{\mu}}\Gamma\indices{^{\epsilon}_{\alpha\delta}}\right)_0x^{\delta}+O(x^2),
\label{TransformExpand}
\end{eqnarray}
where the subscript `$0$' denotes the value of a quantity at the position of the particle 
at time $t=0$.

Applying this coordinate transformation, we find
\be
\texttt{A}^{sing}_{\alpha}
  =\frac{u_{\alpha}}{\sqrt{S_0}}+\frac{\zeta_{\alpha\gamma\delta\epsilon}x^{\gamma}x^{\delta}x^{\epsilon}}{S_0^{3/2}},
\label{EDEffective}
\ee
where 
\begin{equation}
\zeta_{\alpha\gamma\delta\epsilon}:=\left(2u_{\sigma}\Gamma\indices{^{\sigma}_{\alpha\delta}}- a_{\alpha}u_{\delta}\right) q_{\gamma\epsilon}-u_{\alpha}\zeta_{\delta\gamma\epsilon}.
\label{EDzeta}
\end{equation}
To calculate the regularization parameters for electromagnetism we use Eq.~(\ref{EDSingForce}), written as
\begin{equation*}
f^{sing,\alpha}_{EM}=e u^{\beta}g^{\alpha\sigma}\left[\nabla_{\sigma}\texttt{A}^{sing}_{\beta}-\nabla_{\beta}\texttt{A}^{sing}_{\sigma}\right]
=u^{\beta}g^{\alpha\sigma}\left[\partial_{\sigma}\texttt{A}^{sing}_{\beta}-\partial_{\beta}\texttt{A}^{sing}_{\sigma}\right].
\end{equation*}
We now calculate the value of the individual modes of $\partial\texttt{A}^{sing}$ 
in the limit that the field point approaches the source (i.e. as $\epsilon\rightarrow 0$). We then write the regularization parameters for the force as a linear combination of these.

%Now, from Eqs.~(\ref{psiEffective}) and (\ref{UFunctions}), we can write (assuming unit charge),
%\begin{eqnarray}
%\texttt{A}^{\alpha}_{sing}=\frac{u^{\alpha}}{\sqrt{S_0}}+
%\left[\frac{\left(-a^{\alpha} u_{\epsilon}(g_{\beta\gamma}+u_{\beta}u_{\gamma})-{u^{\alpha}}\zeta_{\epsilon\beta\gamma}\right) x^{\epsilon}x^{\beta}x^{\gamma}}{S_0^{3/2}}\right]\nonumber\\
%\texttt{A}^{\alpha}_{sing}=\frac{u^{\alpha}}{\sqrt{S_0}}+\frac{\zeta\indices{^{\alpha}_{\epsilon\beta\gamma}}x^{\epsilon}x^{\gamma}x^{\beta}}{S_0^{3/2}},
%\label{EDEffective}
%\end{eqnarray}
%where we have only kept the pieces which will not vanish trivially. In Eq.~(\ref{EDEffective}) we have defined,
%\begin{equation}
%\zeta\indices{^{\alpha}_{\epsilon\beta\gamma}}=\left(-a^{\alpha} u_{\epsilon}(g_{\beta\gamma}+u_{\beta}u_{\gamma})-{u^{\alpha}}\zeta_{\epsilon\beta\gamma}\right).
%\label{EDzeta}
%\end{equation}

From Eq.~(\ref{EDEffective}), we have 
\begin{equation}
\partial_{\mu}\texttt{A}^{\alpha}_{sing}= - u^{\alpha}\frac{\partial_{\mu}S_0}{S_0^3/2}+ \frac{\Lambda\indices{^\alpha_{\mu\beta\gamma\delta\epsilon}}x^{\beta}x^{\gamma}x^{\delta}x^{\epsilon}}{S_0^{5/2}},
\label{PartialA}
\end{equation}
where 
\begin{equation}
\Lambda\indices{^\alpha_{\mu\beta\gamma\delta\epsilon}}=3\zeta\indices{^{\alpha}_{(\mu\beta\gamma)}}q_{\delta\epsilon}- 3\zeta\indices{^{\alpha}_{\beta\gamma\delta}}q_{\mu\epsilon}.
\label{LambdaTensorED1}
\end{equation}

In Eq.~(\ref{PartialA}), the leading order term is simply the four-velocity multiplied by the leading order term of the scalar field. We can therefore immediately evaluate the mode decomposition of this term,
\begin{eqnarray}
A\indices{^{\alpha}_{\mu}}L&=&\left[u^{\alpha}\frac{-\partial_{\mu}S_0}{S_0^3/2}\right]_{\ell}=u^{\alpha}\lim_{\delta r\rightarrow 0^{\pm}}\frac{L}{2\pi}\int d\cos(\theta) P_{\ell}(\cos(\theta))\int d\phi\left[\frac{-\partial_{\mu}S_0}{S_0^3/2}\right]=u^{\alpha}A^{(scalar)}_{\mu}L\nonumber\\
&=&\mp\frac{L u^{\alpha}}{\sqrt{g_{yy}}}\left[\frac{q_{\mu r}-q_{\mu x}q_{xr}/q_{xx}-g_{\mu y}g_{yr}/g_{yy}}{\sqrt{g_{yy}\tilde{\gamma}^2+\lambda(g_{yy}+\Gamma^2)}}\right],
\label{AalphaEDpotential}
\end{eqnarray}
where we have used Eq.~(\ref{Aalpha}).

Now, we define 
\begin{eqnarray}
\Lambda\indices{^{\alpha}_{\mu XXYY}}=\Lambda\indices{^{\alpha}_{\mu xxyy}}+\Lambda\indices{^{\alpha}_{\mu xyxy}}+\Lambda\indices{^{\alpha}_{\mu xyyx}}+x\leftrightarrow y,\nonumber\\
\label{LambdaTensorED2}
\end{eqnarray}
which we use to write (recalling $w=\beta^2(1+\beta^2)^{-1}$)
\begin{eqnarray}
B\indices{^{\alpha}_{\mu}}=\left[\frac{\Lambda\indices{^\alpha_{\mu\beta\gamma\delta\epsilon}}x^{\beta}x^\gamma x^{\delta} x^{\epsilon}}{S_0^{5/2}}\right]_{\ell}=\lim_{\delta r\rightarrow 0^{\pm}}\frac{L}{2\pi}\int d\cos(\theta) P_{\ell}(\cos(\theta))\int d\phi\left[\frac{\Lambda\indices{^\alpha_{\mu\beta\gamma\delta\epsilon}}x^{\beta}x^\gamma x^{\delta} x^{\epsilon}}{S_0^{5/2}}\right]
\\
=\frac{2}{3\pi(1+\beta^2)^{3/2}\beta^4 q_{yy}^{5/2}}\left(B^{(E),\alpha}_{\mu}\hat{E}(w)+B^{(K),\alpha}_{\mu}\hat{K}(w)\right),\nonumber\\
\label{BalphaEDpotential}
\end{eqnarray}
where we define
\begin{eqnarray}
B\indices{^{(E),\alpha}_{\mu}}=(1+\beta^2)(2+\beta^2)\Lambda\indices{^{\alpha}_{\mu XXYY}}-2\left[(1+2\beta^2)\Lambda\indices{^{\alpha}_{\mu xxxx}}+ (1+\beta^2)^2(1-\beta^2)\Lambda\indices{^{\alpha}_{\mu yyyy}}\right]\nonumber,\\
\label{BalphaEDpotentialE}
\end{eqnarray}
and
\begin{eqnarray}
B\indices{^{(K),\alpha}_{\mu}}=(2+3\beta^2)\Lambda\indices{^{\alpha}_{\mu xxxx}} +(1+\beta^2)\left[(2-\beta^2)\Lambda\indices{^{\alpha}_{\mu yyyy}}-2\Lambda\indices{^{\alpha}_{\mu XXYY}}\right].
\label{BalphaEDpotentialK}
\end{eqnarray}

We have cast Eqs.~(\ref{BalphaEDpotential}), (\ref{BalphaEDpotentialE}) and (\ref{BalphaEDpotentialK}), into forms matching those of Eqs.~(\ref{Balpha}), (\ref{BalphaE}), and (\ref{BalphaK}) for the 
scalar case. The sole differences are the presence of the additional raised index and the additional term in the definition of $\Lambda\indices{^{\alpha}_{\mu\beta\gamma\delta\epsilon}}$. We will see similar symmetries between the scalar field and gravity in the next section.

Now we will write down the regularization parameters in terms of $A\indices{_{\alpha\mu}}$ and $B\indices{_{\alpha\mu}}$.
\begin{eqnarray}
f^{sing,EM\ell}_{\alpha}=\left[u^{\beta}\left(\partial_{\alpha}\texttt{A}_\beta-\partial_{\beta}\texttt{A}_\alpha\right)\right]_{\ell}\nonumber\\ f^{sing,EM\ell}_{\alpha}=u^\beta\left[2A_{[\beta \alpha]}L+2B_{[\beta\alpha]}\right].
\label{EDModeForce}
\end{eqnarray}
Restoring the factors of the charge $e$, we find
\begin{equation}
A^{(EM)}_{\alpha}=2e^2u^\beta A_{[\beta \alpha]}\qquad B^{(EM)}_{\alpha}=2e^2u^\beta B_{[\beta \alpha]}.
\label{EDRPs}
\end{equation}

\subsection{Gravitational Regularization Parameters}

From Eq.~(\ref{gammasingFull}), we can write the singular part of the trace-reversed metric perturbation as
\begin{equation}
\gamma^{sing}_{\hat{\alpha}\hat{\beta}}=\frac{4u_{\hat{\alpha}}u_{\hat{\beta}}}{\sqrt{\hat{S}_0}}-4\frac{\left[2u_{(\hat{\alpha}}a_{\hat{\beta})}u_{\hat{\epsilon}}q_{\hat{\delta}\hat{\gamma}}+ u_{\hat{\alpha}}u_{\hat{\beta}}\zeta_{\hat{\epsilon}\hat{\delta}\hat{\gamma}}\right] x^{\hat{\epsilon}}x^{\hat{\delta}}x^{\hat{\gamma}}}{\hat{S}_0^{3/2}}.
\label{TraceReversedSingularMetricPert}
\end{equation}

We write this in terms of the actual metric perturbation, $h_{\mu\nu}=\gamma_{\mu\nu}-1/2 g_{\mu\nu}\gamma^{\mu}_{\mu}$, and then apply the coordinate transformation to take us from RNCs to our curvilinear coordinates. Upon doing this, we find,
\be
h_{sing}^{\alpha\beta}= 2\frac{g^{\alpha\beta}+2u^{\alpha}u^{\beta}}{\sqrt{S_0}}+\frac{\zeta\indices{^{\alpha\beta}_{\gamma\delta\epsilon}}x^{\gamma}x^{\delta}x^{\epsilon}}{S_0^{3/2}},
\label{EffectivePerturbation}
\ee
where 
\begin{equation}
\zeta\indices{^{\alpha\beta}_{\gamma\delta\epsilon}}:=\left(8u^{(\alpha}a^{\beta)}u_{\gamma}-\partial_{\gamma}g^{\alpha\beta}+4u^{\sigma}u^{(\alpha}\Gamma\indices{_{\sigma}^{\beta)}_{\gamma}}\right)q_{\delta\epsilon}+ (g^{\alpha\beta}+2u^{\alpha}u^{\beta})\zeta_{\gamma\delta\epsilon}.
\label{ZetaGR}
\end{equation}

We now compute $f^{\alpha,sing}_{GR}$ from Eq.~(\ref{selfforceGravity}),  
\begin{eqnarray}
f^{\alpha,sing}_{GR}&=&-m\, q^{\alpha\delta} \left(\nabla_{\beta}h^{(s)}_{\gamma\delta}-\frac{1}{2}\nabla_{\delta}h^{(s)}_{\beta\gamma}\right)u^{\beta}u^{\gamma} 
\nonumber\\
&=& -m\left(g^{\alpha\delta}+u^{\alpha}u^{\delta}\right)u^{\beta}u^{\gamma}\left(
\partial_{\beta}h^{sing}_{\gamma\delta}-\frac{1}{2}\partial_{\delta}h^{sing}_{\beta\gamma}
-\Gamma\indices{^\mu_{\beta\gamma}}h^{sing}_{\mu\delta}
+\Gamma\indices{^\mu_{\delta[\gamma}}h^{sing}_{\beta]\mu}
\right).\nonumber\\
\label{GravForceRPGenerator}
\end{eqnarray}
Therefore, we need to find the leading terms in the mode-sum decomposition of the metric perturbation 
and its derivative. 

We first discuss the mode sum decomposition of the metric perturbation itself. Because the sub-leading term, 
is cubic in the coordinates $x^mu$ and is $O(\epsilon^0)$, its contribution will vanish. This means that the mode-sum decomposition of the metric perturbation evaluated at the position of the mass at time $t=0$, is given by
\begin{eqnarray}
h^{\alpha\beta}_{sing,\ell}=2\lim_{\delta r\rightarrow 0^{\pm}}\frac{L}{2\pi}\int d\cos(\theta) P_{\ell}(\cos(\theta))\int d\phi\left[\frac{g^{\alpha\beta}+2u^{\alpha}u^{\beta}}{\sqrt{S_0}}\right]\nonumber\\
h^{\alpha\beta}_{sing,\ell}=B^{\alpha\beta}_{(h)}=2\left(g^{\alpha\beta}+2u^{\alpha}u^{\beta}\right)\left[\frac{2}{\pi(1+\beta^2)^{1/2}}\hat{K}(w)\right].
\label{hModeSum}
\end{eqnarray}
We use the subscript, $(h)$ to distinguish $B^{\alpha\beta}_{(h)}$ from the quantity $B^{\alpha\beta}$ of  
the electromagnetism section above.  

From Eq.~(\ref{EffectivePerturbation}), we have
\be
\partial_{\mu}h^{\alpha\beta}_{sing}=-(g^{\alpha\beta}+2u^{\alpha}u^{\beta})\frac{\partial_{\mu}S_0}{2S_0^{3/2}}
+\frac{\Lambda\indices{^{\alpha\beta}_{\mu\gamma\delta\epsilon\sigma}}x^{\gamma}x^{\delta}x^{\epsilon}x^{\sigma}}{S_0^{5/2}},
\label{PartialH}
\ee
where 
\begin{equation}
\Lambda\indices{^{\alpha\beta}_{\mu\gamma\delta\epsilon\sigma}}:=\left[3\zeta\indices{^{\alpha\beta}_{(\mu\gamma\delta)}}q_{\epsilon\sigma}-3\zeta\indices{^{\alpha\beta}_{\gamma\delta\epsilon}}q_{\mu\sigma}\right].
\label{LambdaGR}
\end{equation}

The leading order term has the form
\begin{eqnarray}
A\indices{^{\alpha\beta}_{\mu}}L=\left[-(g^{\alpha\beta}+2u^{\alpha}u^{\beta})\frac{\partial_{\mu}S_0}{2S_0^{3/2}}\right]_{\ell}
= - (g^{\alpha\beta}+2u^{\alpha}u^{\beta})\lim_{\delta r\rightarrow 0^{\pm}}\frac{L}{2\pi}\int d\cos(\theta) P_{\ell}(\cos(\theta))\int d\phi\left[\frac{\partial_{\mu}S_0}{2S_0^{3/2}}\right]\nonumber\\
=(g^{\alpha\beta}+2u^{\alpha}u^{\beta})A^{(scalar)}_{\mu}L=
A\indices{^{\alpha\beta}_{\mu}}L=\mp\frac{L (g^{\alpha\beta}+2u^{\alpha}u^{\beta})}{\sqrt{g_{yy}}}\left[\frac{g_{\mu r}+u_\mu u_r-\frac{(g_{\mu x}+u_\mu u_x)(g_{xr}+u_x u_r)}{g_{xx}+U_x^2}-\frac{g_{\mu y}g_{yr}}{g_{yy}}}{\sqrt{g_{yy}\tilde{\gamma}^2+\lambda(g_{yy}+\Gamma^2)}}\right]\nonumber.\\
\label{AalphaGRpartial}
\end{eqnarray}
Now, we define 
\begin{eqnarray}
\Lambda\indices{^{\alpha\beta}_{\mu XXYY}}=\Lambda\indices{^{\alpha\beta}_{\mu xxyy}}+\Lambda\indices{^{\alpha}_{\mu xyxy}}+\Lambda\indices{^{\alpha}_{\mu xyyx}}+x\leftrightarrow y,
\label{LambdaTensorGR}
\end{eqnarray}
which allows us to write, (recalling $w=\beta^2(1+\beta^2)^{-1}$)
\begin{eqnarray}
B\indices{^{\alpha\beta}_{\mu}}=\left[\frac{\Lambda\indices{^{\alpha\beta}_{\mu\sigma\gamma\delta\epsilon}}x^{\sigma}x^\gamma x^{\delta} x^{\epsilon}}{S_0^{5/2}}\right]_{\ell}=\lim_{\delta r\rightarrow 0^{\pm}}\frac{L}{2\pi}\int d\cos(\theta) P_{\ell}(\cos(\theta))\int d\phi\left[\frac{\Lambda\indices{^{\alpha\beta}_{\mu\sigma\gamma\delta\epsilon}}x^{\sigma}x^\gamma x^{\delta} x^{\epsilon}}{S_0^{5/2}}\right]\nonumber
\\
=\frac{2}{3\pi(1+\beta^2)^{3/2}\beta^4 q_{yy}^{5/2}}\left(B\indices{^{(E),\alpha\beta}_{\mu}}\hat{E}(w)+B\indices{^{(K),\alpha\beta}_{\mu}}\hat{K}(w)\right),\nonumber\\
\label{BalphaGRpartial}
\end{eqnarray}
where we define
\begin{eqnarray}
B\indices{^{(E),\alpha\beta}_{\mu}}=-2\left[(1+2\beta^2)\Lambda\indices{^{\alpha\beta}_{\mu xxxx}}+ (1+\beta^2)^2(1-\beta^2)\Lambda\indices{^{\alpha\beta}_{\mu yyyy}}\right]\nonumber\\
+(1+\beta^2)(2+\beta^2)\Lambda\indices{^{\alpha\beta}_{\mu XXYY}},
\label{BalphaGRpartialE}
\end{eqnarray}
and
\begin{eqnarray}
B\indices{^{(K),\alpha\beta}_{\mu}}= (1+\beta^2)\left[(2-\beta^2)\Lambda\indices{^{\alpha\beta}_{\mu yyyy}}-2\Lambda\indices{^{\alpha\beta}_{\mu XXYY}}\right]
+(2+3\beta^2)\Lambda\indices{^{\alpha\beta}_{\mu xxxx}}.
\label{BalphaGRpartialK}
\end{eqnarray}

We can now write the regularization parameters for gravity. From Eqs.~(\ref{GravForceRPGenerator}), (\ref{hModeSum}), (\ref{AalphaGRpartial}), and (\ref{BalphaGRpartial}), we see that only the partial derivatives of the metric 
perturbation contribute to $A^\alpha_{(GR)}$, allowing us to write,
\begin{equation}
A^{\alpha}_{(GR)}=-m\left(g^{\alpha\delta}+u^{\alpha}u^{\delta}\right)u^{\beta}u^{\gamma}\left(
A_{\gamma\delta\beta}-\frac{1}{2}A_{\beta\gamma\delta}\right).
\label{AalphaGR}
\end{equation}
The components $B^\alpha_{(GR)}$ are given by
\begin{equation}
B^{\alpha}_{(GR)}=-m\left(g^{\alpha\delta}+u^{\alpha}u^{\delta}\right)u^{\beta}u^{\gamma}\left(
B_{\gamma\delta\beta}-\frac{1}{2}B_{\beta\gamma\delta}+\Gamma\indices{^{\mu}_{\delta[\gamma}}B^{(h)}_{\beta]\mu}-\Gamma\indices{^\mu_{\beta\gamma}}B^{(h)}_{\mu\delta}
\right).
\label{BalphaGR}
\end{equation}

We have obtained the explicit forms of the regularization parameters for all three spins in Eqs.~(\ref{Aalpha}) and (\ref{Balpha}) (scalar); (\ref{EDRPs}) (electromagnetism); and (\ref{AalphaGR}) and (\ref{BalphaGR}) (gravity). For all three spins, we have given the values in terms of $\zeta$ coefficients, which represent the numerator of the sub-leading terms of the potential (or perturbing metric), and $\Lambda$ coefficients, which represent the numerator of the sub leading terms of the derivative of the potential (or perturbing metric).

\section{Vanishing Sums}

We show the relation  
\begin{equation}
\sum_{\ell=0}^{\infty}\prod_{j=0}^N\frac{1}{(2\ell+1-2m_j)(2\ell+1+2m_j)}=0,
\label{AppBStatement}
\end{equation}
for $N$ and each $m_j$ positive integers with the $m_i$ distinct: $m_i\neq m_j$, $ \forall $ $i\neq j$.  

The product in Eq.~(\ref{AppBStatement}) has a partial fraction decomposition of the 
form 
\begin{equation}
\prod_{j=0}^N\frac{1}{(2\ell+1-2m_j)(2\ell+1+2m_j)}=\sum_{j=0}^N {A_j}\left[\frac1{(2\ell+1-2m_j)}-\frac1{(2\ell+1+2m_j)}\right], 
\label{AppBPartialFrac}
\end{equation}
where 
\begin{equation}
A_i= \left[4m_i\prod_{j\neq i}^N[4(m_i^2-m_j^2)]\right]^{-1}.
\label{AppBCoefficients}
\end{equation}
Eq.~(\ref{AppBCoefficients}) follows quickly from the decomposition 
\mbox{$\displaystyle \frac1{(x-m)(x+m)} = \frac1{4m}\left[\frac1{x-2m}-\frac1{x+2m}\right]$}.  
Because the sum in Eq. (\ref{AppBStatement}) converges absolutely, we can re-order the 
sums over $\ell$ and $j$, writing
\begin{equation}
\sum_{\ell=0}^\infty\prod_{j=0}^N\frac{1}{(2\ell+1-2m_j)(2\ell+1+2m_j)}=\sum_{j=0}^NA_j\sum_{\ell=0}^\infty\left[\frac{1}{2\ell+1-2m_j}-\frac{1}{2\ell+1+2m_j}\right].
\label{AppBDoulbeSum}
\end{equation}
We now show that the sum over $\ell$ vanishes for any positive integer $m_j$.  We start 
by noting that that the first $2m_j$ terms involving $1/(2\ell+1-2m_j)$ separately sum to zero (the terms are antisymmetric about $\ell = m_j-1/2$):  
\bea
    \sum_{\ell=0}^{2m_j-1} \frac{1}{2\ell+1-2m_j} 
   &=& \left(\sum_{\ell=0}^{m_j-1}+\sum_{\ell=m_j}^{2m_j-1} \right)\frac{1}{2\ell+1-2m_j} 
\nonumber\\
   &=& \sum_{\ell=0}^{m_j-1}\frac1{2\ell+1-2m_j}
	-\sum_{\ell'=0}^{m_j-1} \frac1{2\ell'+1-2m_j}
    = 0, 
\eea 
where $\ell' = 2m_j-1-\ell$. 

The remaining terms $1/(2\ell+1-2m_j)$, beginning at $\ell=2m_j$, are now 
identical to, and cancel, the terms $1/(2\ell+1+2m_j)$, beginning at $\ell=0$. 
Denoting by $\Theta(\ell -m_j)$ the step function vanishing for $\ell < m_j$, 
and having the value $1$ for $\ell\geq m_j$, we have  

\bea
\sum_{\ell=0}^\infty\left[\frac{1}{2\ell+1-2m_j}-\frac{1}{2\ell+1+2m_j}\right] 
&=& \sum_{\ell=0}^\infty\left[\frac{\Theta(\ell -m_j)}{2\ell+1-2m_j}-\frac{1}{2\ell+1+2m_j}\right]\nonumber\\
&=& \sum_{\ell=0}^\infty\left[\frac{1}{2\ell+1+2m_j}-\frac{1}{2\ell+1+2m_j} \right]=0.\ \Box \qquad
\label{AppBSingleSum}
\eea
\section{Regularization Parameters in the Original Background Coordinates}

In Sects.~\ref{SelfForceSingField} and \ref{HigherSpins}, the components of the regularization 
parameters are obtained along a basis associated with locally Cartesian angular coordinates (LCAC); and 
the value we obtain for the vector $B_\alpha$ relies on extending the components of $q_{\alpha\beta}$ 
and $u^\alpha$ away from the particle by requiring that their components in the LCAC basis assume the values 
they take at the particle.  For many applications, it is more useful to evaluate the 
components of $A_\alpha$ and $B_\alpha$ in the original coordinate system, as first done 
by Barack and Ori \cite{bo03} and then later explained more completely in an appendix by 
Barack \cite{barack09}. In this appendix, we follow the latter treatment and freeze the components 
of $u^\alpha$ and $q_{\alpha\beta}$ in the original $t,r,\theta,\phi$ coordinates.

%First, it is useful to retrace the steps we took to compute the regularization parameters. We began by considering a particle at an arbitrary position, moving along an arbitrary orbit. We described its motion with some given set of background coordinates, $Z^{\alpha}=(t,r,\theta,\phi)$ (as a concrete example, one could imagine a particle moving around a kerr black hole and then $Z^{\alpha}$ would be the Boyer-Lindquist coordinates.) The particle's position at time $t=0$ would be denoted $Z^{\alpha}_0=(0,r_0,\theta_0,\phi_0)$. To ease the computation of the mode-sum, we rotated our coordinates so that the particle would be at the north pole of our rotated coordinate system, so that $Z'^{\alpha}=(t,r,\theta'\phi')$ (again, in kerr we would think of these as rotated Boyer-Lindquist coordinates).

%Then, due to the singular nature of these coordinates at the pole, we adopted the Locally Cartesian Angular Coordinates, $X^{\alpha}=(\delta t=t, \delta r=r-r_0, x,y)$, where $x=\rho(\theta')\cos(\phi')$ and $y=\rho(\theta')\sin(\phi')$, using the freedom in orientation of $\theta'$ and $\phi'$ to diagonalize $q_{\alpha\beta}$. Finally, a fourth coordinate system is $\tilde x^{\alpha}=(\delta t=t, \delta r, \delta \theta=\theta-\theta_0,\delta\phi=\phi-\phi_0)$, is the same system as that defined by $Z^{\alpha}$, but the coordinates are specified by the coordinate differences between the event and the particle's position at $t=0$.
We define $(\tilde x^\alpha)=(\delta t=t,\delta r=r-r_0,\delta \theta=\theta-\theta_0,\delta\phi=\phi-\phi_0)$,
so that $\tilde x^\mu$ agrees up to a constant with the original $t,r\theta,\phi$ coordinates; we continue 
to denote the locally Cartesian coordinates by $x^{\alpha}=(\delta t,\delta r, x,y)$.
We denote by $\widetilde W^{\mu\dots\nu}_{\sigma\dots\tau}$ the components of a quantity $W^{\dots}_{\dots}$, evaluated using the coordinate system $x^\mu$.  Note that the quantities $\zeta_{\mu\nu\lambda}$ 
and $\Lambda_{\mu\ldots\nu}$ involve partial derivatives of metric components and do 
not transform as tensors.      

From the definitions of $S_0$, $S_1$, and the derivative of our singular field, (Eqs.~(\ref{S0}), (\ref{S1}), and (\ref{NablaPhiRetAdvS}) respectively), we can write the components of the singular force in the original coordinates as
\begin{equation}
q^{-2}\tilde{f}^{sing}_{\mu}=-\frac{\tilde q_{\mu\nu}\tilde x^{\nu}}{\widetilde S_0^{3/2}}+\frac{3 \tilde\zeta_{\gamma\delta\epsilon}q_{\mu\nu}-(2\tilde\zeta_{\mu\gamma\delta}+\tilde\zeta_{\gamma\delta\mu})\tilde q_{\nu\epsilon}}{\widetilde S_0^{5/2}}\tilde x^{\nu}\tilde x^{\epsilon}\tilde x^{\gamma}\tilde x^{\delta}+O(\epsilon^0).
\label{FsingBL}
\end{equation}

We still want to use the LCAC to simplify our integrations, 
retaining the $\tilde x^\mu$ components $\widetilde W^{\mu\dots\nu}_{\sigma\dots\tau}$  of each quantity, 
but expressing them in terms of the LCAC. To do so, we write
\begin{eqnarray}
\tilde x^3 = \delta\theta= x^3+\frac{1}{2}\cot(\theta_0) (x^4)^2+O(\epsilon^3)\nonumber\\
\tilde x^4 = \delta\phi=\sin(\theta_0)^{-1}\left(x^4-\cot(\theta_0) x^3 x^4\right)+O(\epsilon^3)
\label{deltaBL}
\end{eqnarray}
(equivalent to Eq.~(A.17) of \cite{barack09}).
Then  
\begin{equation}
\tilde x^{\alpha}=a_\beta^\alpha x^{\beta}+c^{\alpha}_{\beta\gamma}x^{\beta}x^{\gamma}+O(\epsilon^3), 
\label{BoyerLindquistToLCAC}
\end{equation} 
where $\dis a^\alpha_\beta = \left.\partial_\beta\tilde x^\alpha\right|_0,$ and $ 
c^{\alpha}_{\beta\gamma} = \left.\partial_\beta\partial_\gamma\tilde x^\alpha\right|_0$. 
By the arguments laid down before, it is clear that the higher order terms will give contributions to 
the self-force that either vanish at the particle or contribute to an order-unity term that vanishes upon integration over $\phi$.   
Note that, at linear order, the transformation (\ref{BoyerLindquistToLCAC}) just replaces each occurrence of $\tilde x^4$ by $x^4/\sin\theta_0$.

% In Eq.~(\ref{FsingBL}), the sub-leading term is then changed only by the replacements $\zeta_{\alpha\beta\gamma}\rightarrow\tilde{\zeta}_{\alpha\beta\gamma}$, $q_{\alpha\beta}\rightarrow\tilde{q}_{\alpha\beta}$ and $\tilde x\rightarrow X$.

The leading term acquires a first order correction: 
\begin{eqnarray}
\tilde f^{sing,L}_{\mu}  
=-\frac{\tilde{q}_{\mu\nu} a_\lambda^\nu x^\lambda}
{(\tilde{q}_{\alpha\beta}a^\alpha_\sigma a^\beta_\tau x^{\sigma}x^{\tau})^{3/2}}
 +\frac{\left(3\tilde{q}_{\mu\nu}\tilde{q}_{\iota\kappa}-\tilde{q}_{\mu\iota}\tilde{q}_{\nu\kappa}\right)c^\iota_{\sigma\tau}x^\nu x^\kappa x^\sigma x^\tau}{(\tilde{q}_{\alpha\beta}a^\alpha_\sigma a^\beta_\tau x^{\sigma}x^{\tau})^{5/2}}
\label{FLeadSingBL}
\end{eqnarray}

We take the mode-sum expansion of the force and evaluate these individual modes in the limit that $\epsilon\rightarrow 0$. The leading term will now give us the $A_{\alpha}$ term as before, and in the original coordinates we merely pick up an additional factor of $\sin\theta_0$;
\begin{equation}
\widetilde A_{\alpha\pm}=\mp\sin\theta_0\ q^2 \ 
 \frac{ \tilde q_{\alpha r}-\tilde q_{\alpha \theta}\tilde q_{\theta r}/\tilde q_{\theta\theta}
              -\tilde q_{\alpha \phi} \tilde q_{\phi r}/\tilde q_{\phi\phi} }
           {(\tilde q_{\theta\theta} \tilde q_{\phi\phi} \tilde q_{rr} - \tilde q_{\phi\phi} \tilde q_{\theta r}^2 - \tilde q_{\theta\theta} \tilde q_{ \phi r}^2)^{1/2}}.
\label{Aalphas}
\end{equation}

For $\widetilde B_{\alpha}$, we evaluate the integral
\begin{equation}
\widetilde B_{\alpha}=\frac{q^2}{2\pi}\widetilde P_{\alpha\mu\nu\sigma\tau}\widetilde I^{\mu\nu\sigma\tau}, 
\label{BCoordinateBasis}
\end{equation}
where
\begin{equation}
\widetilde I^{\mu\nu\sigma\tau}= \lim_{\delta r \rightarrow 0}\int_{0}^{2\pi}d\phi\left[\frac{a^\mu_\alpha a^\nu_\beta a^\sigma_\gamma a^\tau_\delta x^{\alpha}x^{\beta}x^{\gamma}x^{\delta}}{(\tilde q_{\kappa\lambda}a^{\kappa}_{\epsilon}a^{\lambda}_{\iota}x^{\epsilon}x^{\iota})^{5/2}}\right],
\label{BOIntegral}
\end{equation}
and
\begin{equation}
\widetilde P_{\alpha\mu\nu\gamma\delta}=3\tilde q_{\alpha\delta}\tilde\zeta_{\mu\nu\gamma}-\tilde q_{\gamma\delta}\left(2\tilde\zeta_{\alpha\mu\nu}+\tilde\zeta_{\mu\nu\alpha}\right)+\left(3\tilde q_{\alpha\mu}\tilde q_{\epsilon\nu}-\tilde q_{\alpha\epsilon}\tilde q_{\mu\nu}\right)c\indices{^{\epsilon}_{\gamma\delta}},
\label{BarackOriPScalar}
\end{equation}
where $c\indices{^{\epsilon}_{\gamma\delta}}$ is defined in Eq.~(\ref{BoyerLindquistToLCAC}), whose only non-vanishing components are $c\indices{^{\theta}_{\phi\phi}}=4^{-1}\sin(2\theta_0)$ and $c\indices{^{\phi}_{\theta\phi}}=c\indices{^{\phi}_{\phi\theta}}=-2^{-1}\cot(\theta_0)$.

Notice that this equation is identical to Eq.~(58) from \cite{barack09}, with the sole exception that we have included the acceleration in our $\tilde\zeta_{\alpha\beta\gamma}$. The limit in Eq.~(\ref{BOIntegral}) means that the integral $I^{\mu\nu\gamma\delta}$ vanishes except when the indices only run over the $(\theta,\phi)$ coordinates. Adopting the notation from \cite{barack09}, we let lowercase roman indices run over only $\theta$ and $\phi$. Barack writes down the solutions to these integrals in Eqs.~(48-57) \cite{barack09}, which we reproduce below. First, we define
\begin{equation}
\alpha=\sin^2(\theta_0)\tilde q_{\theta\theta}/\tilde q_{\phi\phi}-1,\qquad \tilde{\beta}=2\sin(\theta_0)\tilde q_{\theta\phi}/\tilde q_{\phi\phi}.
\label{BarackAlpha}
\end{equation}
Then, $I^{abcd}$ is given by
\begin{equation}
I^{abcd}=\frac{\sin(\theta_0)^{5-N}}{(\alpha^2+\tilde{\beta}^2)^2(4\alpha+4-\tilde{\beta}^2)^{3/2}(Q/2)^{1/2}}\left[QI^{(N)}_K \hat{K}(\omega)+I_{E}^{(N)}\hat{E}(\omega)\right],
\label{BarackIMasterEquation}
\end{equation}
where
\begin{equation}
Q=\alpha+2-(\alpha^2+\tilde{\beta}^2)^{1/2}, \qquad \omega=\frac{2(\alpha^2+\tilde{\beta}^2)^{1/2}}{\alpha+2+(\alpha^2+\tilde{\beta}^2)^{1/2}},
\label{BarackQ}
\end{equation}
and $N=\delta^{a}_{\phi}+\delta^{b}_{\phi}+\delta^{c}_{\phi}+\delta^{d}_{\phi}.$

The ten quantities $I_{K}^{(N)}$ and $I_{E}^{(N)}$ are given by
\begin{eqnarray}
I^{(0)}_{K}=4\left[12\alpha^3+\alpha^2(8-3\tilde{\beta}^2)-4\alpha\tilde{\beta}^2+\tilde{\beta}^2(\tilde{\beta}^2-8)\right],\nonumber\\
I^{(0)}_{E}=-16\left[8\alpha^3+\alpha^2(4-7\tilde{\beta}^2)+\alpha\tilde{\beta}^2(\tilde{\beta}^2-4)-\tilde{\beta}^2(\tilde{\beta}^2+4)\right],
\label{I(0)}
\end{eqnarray}
\begin{eqnarray}
I^{(1)}_{K}=8\tilde{\beta}\left[9\alpha^2-2\alpha(\tilde{\beta}^2-4)+\tilde{\beta}^2\right],\nonumber\\
I^{(1)}_{E}=-4\tilde{\beta}\left[12\alpha^3-\alpha^2(\tilde{\beta}^2-52)+\alpha(32-12\tilde{\beta}^2)+\tilde{\beta}^2(3\tilde{\beta}^2+4)\right],
\label{I(1)}
\end{eqnarray}
\begin{eqnarray}
I^{(2)}_{K}=-4\left[8\alpha^3-\alpha^2(\tilde{\beta}^2-8)-8\alpha\tilde{\beta}^2+\tilde{\beta}^2(3\tilde{\beta}^2-8)\right],
\nonumber\\
I^{(2)}_{E}=8\left[4\alpha^4+\alpha^3(\tilde{\beta}^2+12)+\alpha(\tilde{\beta}^2-4)(3\tilde{\beta}^2-2\alpha)+2\tilde{\beta}^2(3\tilde{\beta}^2-4)\right],
\label{I(2)}
\end{eqnarray}
\begin{eqnarray}
I^{(3)}_{K}=8\tilde{\beta}\left[\alpha^3-7\alpha^2+\alpha(3\tilde{\beta}^2-8)+\tilde{\beta}^2\right],\nonumber\\
I^{(3)}_{E}=-4\tilde{\beta}\left[8\alpha^4-4\alpha^3+\alpha^2(15\tilde{\beta}^2-44)+4\alpha(5\tilde{\beta}^2-8)+\tilde{\beta}^2(3\tilde{\beta}^2+4)\right],
\label{I(3)}
\end{eqnarray}
\begin{eqnarray}
I^{(4)}_{K}=-4\left[4\alpha^4-4\alpha^3+\alpha^2(7\tilde{\beta}^2-8)+12\alpha\tilde{\beta}^2-\tilde{\beta}^2(\tilde{\beta}^2-8)\right],\nonumber\\
I^{(4)}_{E}=16\left[4\alpha^5+4\alpha^4+\alpha^3(7\tilde{\beta}^2-4)+\alpha^2(11\tilde{\beta}^2-4)+(2\alpha+1)\tilde{\beta}^2(\tilde{\beta}^2+4)\right].
\label{I(4)}
\end{eqnarray}
\subsection{The Regularization Parameters for Electromagnetism and Gravity}
First, recall Eq.~(\ref{HigherSpinsForceGeneral}), reproduced below:
\begin{eqnarray}
f^{s=1,sing}_{\mu}&=& \left(\delta_\mu^{\beta}u^{\alpha}-\delta_{\mu}^{\alpha}u^{\beta}\right)\nabla_{\beta}A^{sing}_{\alpha}
\nonumber\\
f^{s=2,sing}_{\mu}&=&\left(q^{\beta}_{\mu}\left(q^{\gamma\delta}+u^{\gamma}u^{\delta}\right)-4q_{\mu}^{\delta}u^{\beta}u^{\gamma}\right)\nabla_{\beta}\frac{\gamma^{sing}_{\gamma\delta}}{4}.\nonumber
\end{eqnarray}
Since we have shown that only the leading and subleading terms in the singular vector potential and metric perturbation will give a non-vanishing contribution to the mode-sum when evaluated at the particle, this allows us to write the expressions for the singular vector potential and metric perturbation in a very convenient form, (taking the charge and mass to be unity)
\begin{eqnarray}
\texttt{A}^{sing}_{\hat{\alpha}}&=&\left[u_{\hat{\alpha}}-a_{\hat{\alpha}}u_{\nu}x^{\nu}+O(\epsilon^1)\right] \Phi^{sing}\nonumber\\
\frac{1}{4} \gamma^{sing}_{\hat{\alpha}\hat{\beta}}&=& \left[u_{\hat{\alpha}}u_{\hat{\beta}}-2a_{(\hat{\alpha}}u_{\hat{\beta})}u_{\nu}x^{\nu}+O(\epsilon^1)\right]\Phi^{sing}.
\label{HigherSpinsSingularField}
\end{eqnarray}
We transform from the RNC basis to the coordinate basis using Eq.~(\ref{CoordinateTransform}), and plug in our expression for $\Phi^{sing}=S_0^{-1/2}-S_1(2S_0^{3/2})^{-1}+O(\epsilon^1)$, we find that the singular force for spin $s=0,1,2$ can be written as
\begin{equation}
\tilde{f}^{s,sing}_{\alpha}=(-1)^s (q_s)^2\left[-\frac{\tilde{q}_{\alpha\nu}\tilde{x}^{\nu}}{\tilde{S}_0^{3/2}}+\frac{\tilde{P}^{s}_{\alpha\mu\nu\gamma\delta}\tilde{x}^{\mu}\tilde{x}^{\nu}\tilde{x}^{\gamma}\tilde{x}^{\delta}}{\tilde{S}_0^{5/2}}+O(\epsilon^0)\right],
\label{SpinSelfForce}
\end{equation}
where $q_s$ is $q,e,m$ for $s=0,1,2$ respectively, and $P^{s}_{\alpha\mu\nu\gamma\delta}$ is given by
\begin{equation}
\tilde{P}^{s}_{\alpha\mu\nu\gamma\delta}=\left(\delta_{s,0}\delta^{\beta}_{\alpha}+\tilde{q}^{\beta}_{\alpha}(1-\delta_{s,0})\right)\left(\tilde{P}_{\beta\mu\nu\gamma\delta}+s^{2}\tilde{a}_{\beta}\tilde{q}_{\mu\nu}\tilde{q}_{\gamma\delta}+s\tilde{q}_{\beta\gamma}\tilde{u}^{\lambda}\tilde{u}^{\rho}\partial_{\delta}\tilde{g}_{\lambda\rho}\tilde{q}_{\mu\nu}\right),
\label{PgeneralSpin}
\end{equation}
 where $P_{\beta\mu\nu\gamma\delta}$ is defined in Eq.~(\ref{BarackOriPScalar}). Thus, we can write the regularization parameters for spins 0,1, and 2:
\begin{equation}
\tilde{A}^{s}_{\alpha\pm}=\mp\sin(\theta_0) q_s^2 (-1)^s \ 
 \frac{ \tilde{q}_{\alpha r}-\tilde{q}_{\alpha \theta}\tilde{q}_{\theta r}/\tilde{q}_{\theta\theta}
              -\tilde{q}_{\alpha \phi} \tilde{q}_{\phi r}/\tilde{q}_{\phi\phi} }
           {(\tilde{q}_{\theta\theta} \tilde{q}_{\phi\phi} \tilde{q}_{rr} - \tilde{q}_{\phi\phi} \tilde{q}_{\theta r}^2 - \tilde{q}_{\theta\theta} \tilde{q}_{ \phi r}^2)^{1/2}},
\label{AalphaSpins}
\end{equation}
and
\begin{equation}
\tilde{B}^{s}_{\alpha}=(-1)^s \frac{q_s^2}{2\pi} \tilde{P}^{s}_{\alpha\mu\nu\gamma\delta}\tilde{I}^{\mu\nu\gamma\delta},
\label{BalphaSpins}
\end{equation}
where $I^{\mu\nu\gamma\delta}$ is given in Eq.~(\ref{BOIntegral}). 

Eqs.~(\ref{AalphaSpins}) and (\ref{BalphaSpins}) simplify exactly to Eqs.~(39-44) given in \cite{barack09}, when we take the geodesic limit, and specialize to a Kerr geometry.

\bibliography{Biblio}
\end{document}